\documentclass[twocolumn]{aastex63}

\usepackage{amsmath}
\usepackage{enumitem}
\usepackage{multirow}

\defcitealias{Raithel:2018}{R18}
\defcitealias{Kalirai:2008}{K08}
\defcitealias{Mroz:2017}{M17}
\defcitealias{Mroz:2019b}{M19}
\defcitealias{Sumi:2016}{SP16}
\defcitealias{Penny:2019}{P19}

\received{}
\revised{}
\accepted{}
\submitjournal{}

\shorttitle{Astrometric Microlensing} 
\shortauthors{Lam, Lu, Hosek}

\begin{document}

\title{PopSyCLE: A New Population Synthesis Code for Compact Object Microlensing Events}

\author[0000-0002-6406-1924]{Casey Y. Lam} 
\correspondingauthor{Casey Y. Lam} 
\email{casey$\_$lam@berkeley.edu}
\affiliation{University of California, Berkeley, Department of Astronomy, Berkeley, CA 94720}

\author[0000-0001-9611-0009]{Jessica R. Lu}
\affiliation{University of California, Berkeley, Department of Astronomy, Berkeley, CA 94720}

\author[0000-0003-2874-1196]{Matthew W. Hosek, Jr.}
\affiliation{University of California, Los Angeles, Department of Astronomy, Los Angeles, CA 90095}

\author[0000-0003-0248-6123]{William A. Dawson}
\affiliation{Lawrence Livermore National Laboratory, 7000 East Ave., Livermore, CA 94550}

\author[0000-0003-2632-572X]{Nathan R. Golovich}
\affiliation{Lawrence Livermore National Laboratory, 7000 East Ave., Livermore, CA 94550}

\begin{abstract}
We present a new Milky Way microlensing simulation code, dubbed \texttt{PopSyCLE} (Population Synthesis for Compact object Lensing Events). 
\texttt{PopSyCLE} is the first resolved microlensing simulation to include a compact object distribution derived from numerical supernovae explosion models and both astrometric and photometric microlensing effects.
We demonstrate the capabilities of \texttt{PopSyCLE} by investigating the optimal way to find black holes (BHs) with microlensing.
Candidate BHs have typically been selected from wide-field photometric microlensing surveys, such as OGLE, by selecting events with long Einstein crossing times ($t_E>120$ days).
These events can be selected at closest approach and monitored astrometrically in order to constrain the mass of each lens; \texttt{PopSyCLE} predicts a BH detection rate of $\sim$40\% for such a program.
We find that the detection rate can be enhanced to $\sim 85\%$ by selecting events with both $t_E>120$ days and a microlensing parallax of $\pi_E<0.08$. 
Unfortunately, such a selection criterion cannot be applied during the event as $\pi_E$ requires both pre- and post-peak photometry.
However, historical microlensing events from photometric surveys can be revisited using this new selection criteria in order to statistically constrain the abundance of BHs in the Milky Way.
The future \emph{WFIRST} microlensing survey provides both precise photometry and astrometry and will yield individual masses of $\mathcal{O}(100-1000)$ black holes, which is at least an order of magnitude more than is possible with individual candidate follow-up with current facilities. 
The resulting sample of BH masses from \emph{WFIRST} will begin to constrain the shape of the black hole present-day mass function, BH multiplicity, and BH kick velocity distributions.
\end{abstract}

\section{Introduction}
\label{sec:Introduction}

The Milky Way is predicted to contain $10^8 - 10^9$ stellar mass black holes \citep{Agol:2002}; however, the exact number is still very uncertain.
Only a few dozen black holes (BHs) have been detected to date in X-ray binaries or BH-BH mergers \citep{Remillard:2006, Abbott:2016}; no \emph{isolated} stellar mass BHs have yet been definitively and unambiguously detected. 
Although BHs likely form after the death of a massive star, uncertainties in the final stages of massive star evolution have made quantitative predictions of BH masses and numbers difficult \citep{Heger:2003}.
Whether a massive star will form a BH, neutron star, or compact remnant at all is not only a function of the initial stellar mass, but other factors such as metallicity, stellar winds and the core structure at time of collapse. 
Additionally, it is still unknown whether BH binary systems form directly from stellar binaries, dynamically from two isolated BHs, or if there are multiple formation channels. 

Detecting and characterizing a sample of isolated stellar mass BHs would provide insight into these unsolved problems by constraining the number of BHs in the Milky Way and the present-day mass function, binary fraction, and spatial and velocity distributions.
This in turn will place constraints on stellar evolution models and improve understanding of supernova physics \citep{Janka:2012}.
It would also allow better interpretation of recent LIGO results. 
For example, is the ``mass gap" between the heaviest neutron stars and the lightest BHs real, or is it an artifact of the observational bias resulting from only having detected binary BHs (from an observational perspective, see \citet{LIGO:2018, Ozel:2010, Farr:2011}; from a theoretical perspective, see \citet{Belczynski:2012, Fryer:2012})? 

Gravitational microlensing is a technique particularly well suited for detecting dark isolated objects in the Milky Way such as BHs, as properties of the lens can be inferred from changes in the source image, without having to directly observe the lens itself \citep{Paczyniski:1986}.
For example, the MACHO \citep{Alcock:1993}, EROS \citep{Aubourg:1993}, and OGLE \citep{Udalski:1994} collaborations used gravitational microlensing to determine that the majority of dark matter in the Galactic halo is not due to ordinary baryonic matter \citep{Alcock:2000, Tisserand:2007, Wyrzykowski:2011}. 
However, after the LIGO BH-BH merger detection, there has been a major resurgence in interest in BHs (and in particular primordial BHs) as a viable dark matter candidate \citep{Bird:2016, Sasaki:2016}.
Additionally, for the past 15 years microlensing has been used to detect extrasolar planets; for a review see \cite{Gaudi:2012}.
Currently, microlensing is being used to search for isolated BHs by combining information from the photometric brightening and parallax signal and the astrometric shift of the source image to constrain the mass of the lens \citep{Lu:2016, Rybicki:2018}.
As all current microlensing surveys are ground-based photometric surveys, a substantial development is space-based survey telescopes such as \emph{Gaia} and \emph{WFIRST}, which will provide astrometric measurements for microlensing events \citep{Gaia:2016, Rybicki:2018, Spergel:2015, Penny:2019}.

Previous theoretical work modeled microlensing in the Milky Way due to a population of stars, brown dwarfs, and stellar remnants \citep{Han:2003, Wood:2005}.
However, these models use a heavily simplified model for compact object population synthesis.
\citet{Oslowski:2008} investigated compact object lensing by using the \texttt{StarTrack} population code \citep{Belczynski:2008} to generate isolated BHs and neutron stars via two different formation channels.
However, in all these models a realistic extinction map, which varies significantly across the sky and affects both the optical depth and events rates, was lacking.
\citet{Dai:2015} followed \citet{Wood:2005} to specifically investigate lensing by neutron stars, but incorporated updated kinematic information and extinction via a variable luminosity function, showing that most neutron star lensing events occur on much shorter timescales than previously though, and that the steepness of the luminosity function has a strong effect on the timescale distribution.
However, they did not have updated modeling for BHs.
Several new simulations have been recently developed to investigate optical depths and event rates of Galactic microlensing involving more sophisticated Galactic models and realistic extinction \citep{Penny:2013, Awiphan:2016}.
However, the population synthesis in these models lack high-mass stellar remnants as they are primarily focused on lower mass stars and exoplanets.

In this paper, we will describe new a code we have developed called Population Synthesis for Compact object Lensing Events (\texttt{PopSyCLE}).\footnote{https://github.com/jluastro/PopSyCLE}
White dwarfs, neutron stars, and BHs are synthesized and injected into a stellar model of the Milky Way using stellar evolution models and an initial-final mass relation. 
Each individual object is then propagated in time to perform a synthetic microlensing survey.
Cuts are then made on observational quantities.
Since the simulation is resolved, all microlensing parameters are known for each individual event.
We can then begin to investigate microlensing events due to BH lenses.

The rest of the paper is organized as follows.
In \S\ref{sec:PopSyCLE Ingredients}, the input models used to perform both stellar and compact object population synthesis in \texttt{PopSyCLE} are described.
The details of the compact object population synthesis are given in \S\ref{sec:Population synthesis}, and the procedure for selecting microlensing events is given in \S\ref{sec:Identifying Microlensing Events}.
Simulation parameters are presented in \S \ref{sec:PopSyCLE Simulations}, then synthetic \texttt{PopSyCLE} surveys are compared to observations and theory in \S\ref{sec:PopSyCLE Comparison}.
In \S\ref{sec:Results} we consider strategies for BH microlensing candidate selection, search, and verification, both from the ground and with the upcoming \emph{WFIRST} mission.
A discussion of \texttt{PopSyCLE} compared to other work is presented in \S\ref{sec:Discussion}, along with comments on further developments to be made to the code. 
Lastly, in \S\ref{Conclusions} we present our conclusions. 

\section{PopSyCLE Ingredients}
\label{sec:PopSyCLE Ingredients}

\begin{figure*}[t]
    \centering
    \includegraphics[scale=0.125]{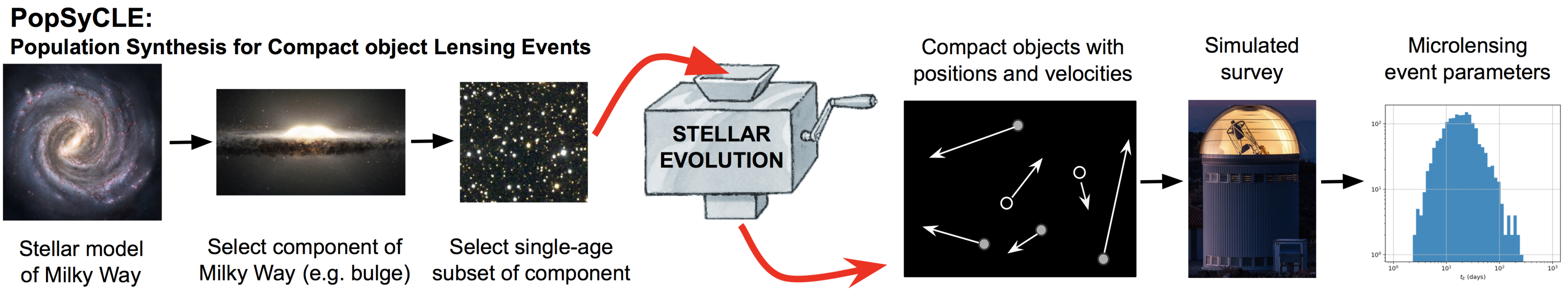}
    \caption{A schematic of the \texttt{PopSyCLE} pipeline. 
    The first three panels are described in \S \ref{sec:Generating Stars with Galaxia}. The next two panels are described in \S \ref{sec:Generating Compact Objects} and \ref{sec:Population synthesis}.
    The last two panels are described in \S \ref{sec:Identifying Microlensing Events}.
    Image credits: Universe Today (Panel 1), ESO/NASA/JPL-Caltech/M. Kornmesser/R. Hurt (Panel 2), NOAO/AURA/NSF (Panel 3), Simply scheme: introducing computer science, Harvey and Wright (Panel 4), OGLE survey (Panel 6).}
\end{figure*}
\label{fig:pipeline}

To calculate microlensing event rates and parameters, knowledge of the masses, positions, and velocities of stars and compact objects in the Milky Way is required.
Photometric information to estimate quantities such as blending and survey depth limits is also required.
We utilize existing models of the Milky Way to describe the stars (\S\ref{sec:Generating Stars with Galaxia}) and inject compact objects according to our own population synthesis estimates and an initial-final mass relation for BHs, neutron stars, and white dwarfs (\S\ref{sec:Generating Compact Objects}).

\subsection{Generating Stars with \texttt{Galaxia}}
\label{sec:Generating Stars with Galaxia}

\texttt{Galaxia} is a synthetic stellar survey of the Milky Way \citep{Sharma:2011}. 
Given a survey area and location, \texttt{Galaxia} will generate a stellar model of the Milky Way for a circular field of sky, which corresponds to a 3-D conical volume. 
\texttt{Galaxia} is a resolved simulation; that is, for every single star in the field, it will return a position, velocity, age, mass, photometry in several filters, 3-D extinction, metallicity, surface gravity, and more.\footnote{http://galaxia.sourceforge.net/Galaxia3pub.html}
\texttt{Galaxia} implements the Besan\c{c}on analytic model for the Milky Way \citep{Robin:2003}, with a modified version of the disk kinematics that adjusts the velocity in the azimuthal direction \citep{Shu:1969}.
For a detailed description of the implementation and galactic model parameters see \cite{Sharma:2011} and references therein; for a summary of the model, see Tables 1-3 in \cite{Sharma:2011}.
We also perform additional modifications to \texttt{Galaxia}'s bulge kinematics, as they are known to produce some inconsistencies with observations.
This is described in more detail in Appendix \ref{Appendix:Galaxia Galaxtic Model}.
In the rest of the main text, whenever reference is made to \texttt{Galaxia}, we specifically mean \texttt{Galaxia} with the modified bulge, unless otherwise mentioned.
Additionally, there is also an option to implement the N-body stellar halo model of \cite{Bullock:2005} instead of the smooth analytic stellar halo model from the Besan\c{c}on model. 
We choose to use the smooth analytic model instead of the N-body model as we are primarily considering lensing in the disk and bulge; stellar halo substructure is not a significant concern.

\texttt{Galaxia} includes a 3-D extinction map; however, the built-in extinction law from \cite{Schlegel:1998} is not well suited to the high-extinction regions towards the inner Galactic bulge. 
Unfortunately, the newest 3-D extinction maps (e.g. \citet{Green:2018}) do not cover significant fractions of the Galactic Bulge region.
Similarly, the extinction maps of \citet{Nataf:2013}, which were constructed using OGLE, VVV, and 2MASS data, only cover the OGLE-III bulge fields.
Instead, we modify the Galaxia output in order to adopt the reddening law of \cite{Damineli:16} as described in Appendix \ref{Appendix:Extinction}.

\subsection{Generating Compact Objects}
\label{sec:Generating Compact Objects}

\texttt{PyPopStar} is a software package that generates single-age, single-metallicity populations (i.e.~star clusters) using adjustable initial mass functions (IMFs), multiplicity distributions, stellar evolution and atmosphere grids, and extinction laws; for a more detailed description see \S 4 of \cite{Hosek:2019}.\footnote{https://github.com/astropy/PyPopStar} 
In summary, IMFs are parametrized as piecewise functions consisting of multiple power-laws and generated according to the prescription from \cite{Pflamm-Altenburg:2006}.
We adopt the MESA Isochrones and Stellar Tracks (MIST) v1.2 stellar evolution models \citep{Choi:2016, Dotter:2016}, which are calculated using the Modules for Experiments in Stellar Astrophysics (MESA) code \citep{Paxton:2011, Paxton:2013, Paxton:2015}. 
These models provide theoretical isochrones with stellar temperatures and surface gravities as a function of mass at a given stellar age. 
Model atmosphere grids are then used to assign an atmosphere to each star.
The stellar temperature determines which grid is used; an ATLAS9 grid is used for stars with T$_{eff}$ $>$ 5500 K \citep{Castelli:2004}, a PHOENIX grid is used for 3800 K $<$ T$_{eff}$ $<$ 5000 K \citep[version 16;][]{Husser:2013}, and a BTSettl grid is used for 1200 K $<$ T$_{eff}$ $<$ 3200 K \citep{Baraffe:2015}. 
For temperatures at a transition region between grids (e.g. 5000 K -- 5500 K), an average atmosphere between the two model grids is used.
For all models, solar metallicity is assumed.
This assumption will not have a large effect on the BH population synthesis; see \S \ref{sec:Metallicity} for justification. 
Although the MIST stellar evolution models do not evolve stars all the way down to neutron stars and BHs, it does have some partial support for white dwarfs; a full discussion of stellar evolution with white dwarfs is presented in \S \ref{sec:White Dwarfs}.

The \texttt{PyPopStar} software as described in \S4 of \cite{Hosek:2019} would discard stars that had evolved into compact objects.
To be able to perform population synthesis, we have updated \texttt{PyPopStar} with an initial-final mass relation (IFMR) that will properly account for compact objects. 
Given the zero-age main sequence (ZAMS) mass of a star, the IFMR describes the type and mass of compact object formed. 
The IFMR is an active area of research, as the ZAMS mass does not solely determine the evolutionary path of a star.  
Factors such as rotation (which induces mixing in the star) and mass loss due to stellar winds (which are a function of metallicity) are important in the final stages of stellar evolution.
In particular, the pre-supernova core structure of the star strongly determines whether there is an explosion or not, which in turn affects the type of compact remnant formed (\citet{Sukhbold:2018} and references therein).
We modify \texttt{PyPopStar} to implement a combination of IFMRs taken from recent simulations to build a population of white dwarfs, black holes, and neutron stars as described in more detail below.

Note that \texttt{PyPopStar} has support for stellar multiplicity. 
However, as \texttt{Galaxia} only has a treatment of single stars, this functionality was not used. 
A discussion of this is presented in \S \ref{sec:Binarity and Mergers}.

\subsubsection{Neutron Stars and Black Holes}
\label{sec:Neutron Stars and Black Holes}

For neutron stars (NSs) and BHs, we have adopted the IFMR presented in \cite{Raithel:2018}, hereafter \citetalias{Raithel:2018}, which is calculated by combining observational data of BH and NS masses along with 1-D neutrino-driven supernova simulations \citep{Sukhbold:2016}. 
We have made several minor modifications, described in Appendix \ref{Appendix:Initial-Final Mass Relation}.
The IFMR of \citetalias{Raithel:2018} is only a function of the ZAMS mass; other factors that influence a BH or NS outcome such as small differences in core structure are taken into account by making the IFMR probabilistic. 
For a given ZAMS mass, the probability of forming a BH or a NS is given in Table \ref{tab:Compact Object Formation Probabilities}; the values come from \citetalias{Raithel:2018}.
There are then two separate IFMRs for BHs and NSs.
The BH IFMR is a piecewise function (Figure \ref{fig:mzams_vs_mbh}), while we make the assumption that the NS IFMR is a constant (justification in Appendix \ref{Appendix:Initial-Final Mass Relation}.)
Thus, because of the probabilistic nature of the total IFMR, one ZAMS mass can be mapped to two remnant masses, of either a BH or NS.

The maximum BH mass generated by the BH IFMR is about $16 M_\odot$. 
This appears to run contrary to the recent results from gravitational wave astronomy experiments, which has found 15 out of 20 pre-merger black holes with masses above $16 M_\odot$ \citep{LIGOVIRGO:2018}.
As discussed in detail in \S 6 of \citetalias{Raithel:2018}, the models of \cite{Sukhbold:2016} are all solar metallicity, while the formation of LIGO-mass BHs from single-star evolution requires lower metallicity. 
Additionally, the models are of single stars, and do not consider formation of BHs via dynamical assembly in dense stellar clusters nor strong magnetic fields, other proposed mechanism for obtaining LIGO-mass BHs.
See \S \ref{sec:Metallicity} and \S \ref{sec:Binarity and Mergers} for a further discussion.

\subsubsection{White Dwarfs}
\label{sec:White Dwarfs}
One complicating factor in white dwarf (WD) population synthesis is that the MIST models do not include the late stages of WD cooling.
Thus not all WDs are generated, in particular the oldest and faintest ones.
This means a WD IFMR is also required. 
For all white dwarfs not included in a set of evolutionary models, we use the empirically determined IFMR of \cite{Kalirai:2008}, hereafter \citetalias{Kalirai:2008}, given by
\begin{equation}
\label{eq:wd_ifmr}
    M_{WD} = (0.109\; M_{ZAMS} + 0.394)\; M_\odot.
\end{equation}
The data in \citetalias{Kalirai:2008} range in initial mass from $1.16 M_\odot < M_{ZAMS} < 6.5 M_\odot$.
However, to have continuous coverage in $M_{ZAMS}$ for the IFMR, we extend the domain to $M_{ZAMS}$ to $0.5 M_\odot < M_{ZAMS} < 9 M_\odot$, where the lower mass limit was chosen the match the MIST models' lower mass limit, and the upper mass limit was chosen to match where the IFMR for NSs and BHs begin. 
The low-mass extrapolation covers all possible WDs formed within the age of the Universe.
The high-mass extrapolation ensures that that there is an IFMR for all stellar masses and the maximum WD mass would then be 
\begin{equation*}
    M_{WD}(M_{ZAMS} = 9 M_\odot) = 1.375 M_\odot \lesssim M_{ch}
\end{equation*}
where $M_{ch} = 1.4 M_\odot$ is the Chandrasehkar mass.

In particular for the MIST models, WDs are tracked only so far down the cooling curve and the coolest/oldest WDs are usually dropped. For these objects, the WD IFMR (Equation \ref{eq:wd_ifmr}) is used to produce \emph{dark} WDs and their luminosity is neglected. 
This is justified as these old WDs have absolute magnitudes of M$\sim$18 in red-optical filters and they contribute negligible flux at typical distances for lenses and sources \citep[e.g.][]{Campos:2016}. 
Thus a \texttt{PopSyCLE} population will contain both luminous WDs from the MIST models and dark WDs from the \citetalias{Kalirai:2008} IFMR. 

\begin{deluxetable}{rclccc}
\tablecaption{Compact Object Formation Probabilities 
    \label{tab:Compact Object Formation Probabilities}}
\tablehead{
    \multicolumn{3}{c}{Mass Range ($M_\odot$)} \hspace{0.2in} &
    \colhead{$P_{WD}$} \hspace{0.2in} & 
    \colhead{$P_{NS}$} \hspace{0.2in} & 
    \colhead{$P_{BH}$} \hspace{0.2in}
}
\startdata
     0.5  & $<$ M $<$ & 9 & 1.000 & 0 & 0 \\
     9    & $<$ M $<$ & 15 & 0 & 1.000 & 0 \\
     15   & $<$ M $<$ & 17.8 & 0 & 0.679 & 0.321 \\
     17.8 & $<$ M $<$ & 18.5 & 0 & 0.833 & 0.167 \\
     18.5 & $<$ M $<$ & 21.7 & 0 & 0.500 & 0.500 \\
     21.7 & $<$ M $<$ & 25.2 & 0 & 0 & 1.000 \\
     25.2 & $<$ M $<$ & 27.5 & 0 & 0.652 & 0.348 \\ 
     27.5 & $<$ M $<$ & 60 & 0 & 0 & 1.000 \\
     60   & $<$ M $<$ & 120 & 0 & 0.400 & 0.600 \\
\enddata
\tablecomments{Given the ZAMS mass (the first column), what are the probabilities of forming a white dwarf, neutron star, or black hole. 
The NS and BH columns come from Table 3 of \citetalias{Raithel:2018}.}
\end{deluxetable}

\section{Population synthesis}
\label{sec:Population synthesis}

First, \texttt{Galaxia} is used to generate a synthetic stellar survey of a circular area on the sky. 
Only stars that are within a distance of 20 kpc to Earth are kept, as these are the ones most likely to be observable lenses and sources. 
In addition to mass, age, 6-D kinematics, and photometry, \texttt{Galaxia} also tags each star by its population type (thin disk, thick disk, bulge, stellar halo.) 
The stars generated are sorted by population to preserve age and kinematic information. 
The thin disk is a multi-age population ranging from less than 0.15 Gyr up to 10 Gyr; we split the thin disk population into sub-populations of 29 finer age bins. 
The thick disk, bulge, and halo are all single-age populations of 11, 10, and 14 Gyr, respectively.
Some of the youngest stars in the thin disk need to have their ages reassigned because the youngest age available for the MIST isochrones is $10^5$ years.
This is only necessary for a very small number of stars; for example, in a survey pointed toward the Galactic center, less than 0.001 percent of stars were younger than $10^5$ years.
Additionally, the age of the 14 Gyr halo star population are reassigned to be 13.8 Gyr, to be within the age of the universe. 

Each of these (sub)-populations is then approximated as a group of single-age stars. 
If necessary, these age groups are further broken up into smaller sub-groups of 2 million stars each, to keep the population synthesis calculation manageable. 

Knowledge of the initial mass of each age group of stars is needed to perform the compact object population synthesis.
The present-day mass of the age group can be calculated with knowledge of the ZAMS mass of each star.
The initial age group mass can be found by multiplying the present-day group mass by an initial-current mass ratio factor; the calculation of this ratio is described in Appendix \ref{Appendix:Initial-Final Group Mass Ratio}. 
With the initial mass of the age group of stars in hand, we can then generate a similar group of stars of the correct mass and age with \texttt{PyPopStar}.

Next, compact objects are generated to inject back into the stellar population using \texttt{PyPopStar}. 
A Kroupa IMF $\xi(m) = dN/dm \propto m^{-\alpha}$,
\begin{equation}
\alpha = 
\begin{cases}
    -1.3, & 0.1\;M_\odot < m \leq 0.5\;M_\odot \\
    -2.3, & 0.5\;M_\odot < m \leq 120\; M_\odot
\end{cases}
\end{equation}
is assumed with the lower mass limit set to match the MIST models' lower mass limit, and the upper mass limit set to match the IFMR's upper mass limit \citep{Kroupa:2001}. 
Any object that evolves off the isochrone becomes a compact object.
As described in \S \ref{sec:Generating Compact Objects}, the IFMRs of \citetalias{Raithel:2018} and \citetalias{Kalirai:2008} are used to obtain masses for BHs, NSs, and WDs.

We assume the positions and velocities of the compact objects follows that of the stellar population in \texttt{Galaxia}; they are assigned using kernel density estimation (KDE). 
The 6-D kinematic data is simultaneously fit using a Gaussian kernel and Euclidean metric with a bandwidth of 0.0001.
Such a small bandwidth means the data is extremely unsmoothed.
However, this is desired as we want to randomly draw from the existing distribution. 
For the BHs and NSs, a tuneable birth kick is also applied in a random direction in addition to the stellar velocity assigned from KDE.
NSs have been observed to have birth velocities of $\sim 200 - 500$ km/s, and up to 1000 km/s; it is thought that the initial kick velocity is due to asymmetry in the supernova explosion \citep{Lai:2001, Janka:2012}. 
BHs should also receive kicks; by conservation of momentum, BH kicks should be smaller than NS kicks (although see \citet{Janka:2013, Repetto:2012} which suggests they will be of similar magnitude).

All the compact objects produced with the IFMRs are assumed to be dark, while luminous WDs produced with the MIST models are assigned the same color excess $E(B-V)$ as the star nearest to them in 3-D space.

For easier processing and parallelization, the stars from \texttt{Galaxia} and the compact objects generated from \texttt{PyPopStar} are sorted into a grid of Galactic latitude and longitude bins as seen from the solar system barycenter. 
The size of these latitude and longitude bins $ \Delta \theta_{bin}$ must be sufficiently large such that most stars do not cross over multiple bins during the survey duration, but not so large that finding source-lens pairs is computationally infeasible (described in \S\ref{sec:Event Selection Algorithm}). 
Assuming that typical microlensing surveys are of $\sim 5$ years, and proper motions of $>$5 arcsec/year are rare, this gives a lower limit on $\Delta \theta_{bin}$ of $25''$.
We have chosen the default bin sizes to be roughly $30'' \times 30''$; however, the bin size is adjustable by the user. 

The above process of generating compact objects and sorting them into bins for each (sub)-population group is then repeated. 
Thus, the generated compact objects preserve the correlations between population age, structure, and dynamics that are present in the stellar population, with the addition of tunable birth kicks.

\section{Identifying Microlensing Events}
\label{sec:Identifying Microlensing Events}

Microlensing occurs when a foreground star or compact object (the lens) passes near a background star (the source) as projected on the sky. 
However, not all close approaches may produce detectable or significant microlensing events. 
Therefore, microlensing event candidates are first identified, then detailed properties of the lensing event are calculated and used to decide whether the candidate is significant enough to classify as a microlensing event. 
In \S\ref{sec:Candidate Selection Parameters} and \ref{sec:Other Selection Parameters}, basic definitions for microlensing event parameters are presented. 
The process by which candidates are identified and converted into a final microlensing list is described in \S\ref{sec:Event Selection Algorithm} and \ref{sec:Survey implementation}. 
Going forward, the point source point lens (PSPL) regime of microlensing is assumed.

\subsection{Candidate Selection Parameters}
\label{sec:Candidate Selection Parameters}

The scale of the microlensing event is set by the the angular \emph{Einstein radius}
\begin{equation}
\label{eq:theta_E}
    \theta_E = \sqrt{\frac{4GM}{c^2} \Bigg( \frac{1}{d_L} - \frac{1}{d_S} \Bigg)},
\end{equation}
where $d_S$ is the observer-source distance, $d_L$ is the observer-lens distance, and $M$ is the mass of the lens. 
Note that in microlensing, $\theta_E$ is unresolved, unlike in strong lensing.
For a ``typical" microlensing event consisting of a stellar bulge source at $d_S = 8$ kpc and a stellar disk lens at $d_L = 4$ kpc with a mass $M = 0.5 M_\odot$, the Einstein radius is $\theta_E = 0.71$ mas.
For context, the highest-resolution images from the largest ground-based or space-based telescopes (with a filled aperture) is 50 mas in the optical or infrared.
However, in some cases Einstein radii of this scale are interferometrically resolvable \citep{Dong:2019}, moving from the regime of microlensing to strong lensing.

The \emph{Einstein crossing time $t_E$}, the time it takes for the lens to traverse the Einstein radius, can be inferred from the photometric light curve. 
The magnitude of the relative source-lens proper motion $\mu_{rel}$ relates the Einstein crossing time and Einstein radius via
\begin{equation}
    \theta_E = \mu_{rel} t_E.
\end{equation}

The dimensionless source-lens separation vector normalized in units of the Einstein radius is
\begin{equation}
    \vec{u} = \frac{\vec{\theta}_S - \vec{\theta}_L}{\theta_E},
\end{equation}
where $\vec{\theta}_L$ and $\vec{\theta}_S$ denote the angular position of the lens and source on the sky, respectively. 
The magnitude of the separation $|\vec{u}| = u$ as a function of time, neglecting higher-order effects such as parallax, can be written as 
\begin{equation}
    u(t) = \sqrt{u_0^2 + \Big( \frac{t - t_0}{t_E} \Big)^2},
\end{equation}
where $t_0$ is the time of closest approach and $u_0$ is the separation at $t_0$; $u_0$ is also known as the \emph{impact parameter}.

The \emph{photometric amplification} of the source is given by
\begin{equation}
    A(u) = \frac{u^2 + 2}{u \sqrt{u^2 + 4}},
\label{eq:amplification}
\end{equation}
and is maximized at $u = u_0$. 
It can be seen from this formula that a smaller $u_0$ produces a larger amplification.
Most microlensing surveys define a ``microlensing event" as an event that has $u_0$ less than 1 or 2.

The \emph{source flux fraction}, which is sometimes called the blend parameter, or confusingly, the blend fraction, is the fraction of source flux over total flux in the telescope's photometric extraction aperture 
\begin{equation}
\label{eq:source flux fraction}
    b_{SFF} = \frac{F_S}{F_S + F_L + F_N},
\end{equation}
where $F_S$, $F_L$, and $F_N$ are the fluxes due to the (unlensed) source, lens, and any neighboring stars. 
From this definition, a highly blended event will have a source flux fraction of $b_{SFF} \approx 0$ and an event with no blending (i.e. only flux from the source) will have $b_{SFF} = 1$.

The photometric extraction aperture is typically proportional to the spatial resolution of the images; we adopt an aperture diameter of the full-width half-maximum (FWHM), which is set by seeing for ground-based surveys like OGLE and the diffraction limit for space-based surveys such as \emph{WFIRST} (these surveys are described in \S \ref{sec:PopSyCLE Simulations}). 
To give concrete numbers, the median seeing for the OGLE survey is roughly 1.3'', which we take to be the FWHM.
\emph{WFIRST}, a planned 2.4 meter space telescope observing at 1630 nm (H-band), will have FWHM $\sim 0.17''$.
These FWHM values for OGLE and \emph{WFIRST} would correspond to aperture radii of roughly $0.65''$ and $0.09''$, respectively. 

The \emph{bump magnitude}, $\Delta m$, is the difference between the baseline magnitude $m_{base}$ and the peak magnitude $m_{peak}$
\begin{equation}
\label{eq:bump magnitude}
    \Delta m = -2.5 \mathrm{log}_{10} \Bigg( \frac{A(u_0) F_S + F_L + F_N}{F_S + F_L + F_N} \Bigg).
\end{equation}
Note that although the magnification of the source flux is achromatic, the bump magnitude and the source flux fraction are dependent on wavelength and the photometric extraction aperture \citep{DiStefano:1995}.
Thus, when events are selected on $\Delta m$ or $b_{SFF}$, we explicitly note the filter and aperture used.

\subsection{Other Selection Parameters}
\label{sec:Other Selection Parameters}

In addition to the selection parameters described in \S \ref{sec:Candidate Selection Parameters}, there are a number of other measureable microlensing quantities. 
In particular, the astrometric and microlensing parallax signals are useful for identifying possible black hole microlensing events. 

The \emph{astrometric shift $\vec{\delta}_c$} of the source image centroid, assuming no blending, is given by
\begin{equation} 
\label{eq:delta_c}
    \vec{\delta}_c = \frac{\theta_E}{u^2 + 2} \vec{u}.
\end{equation}
In contrast to the photometric amplification, the maximum astrometric shift occurs at $u = \pm \sqrt{2}$.
Thus, if $u_0 < \sqrt{2}$, the maximum astrometric shift will occur before and after the maximum photometric amplification. 
However, if $u_0 \geq \sqrt{2}$, the time at which maximum astrometric shift and photometric amplification occur will coincide.
It should be noted that $\delta_c$ is the maximum possible astrometric shift for a given geometry (i.e. for given $M$, $d_L$, $d_S$), as it does not include blending. 

Blending due to a luminous lens modifies the astrometric signal to
\begin{equation}
\label{eq:delta_c_LL}
    \vec{\delta}_{c, LL} = \frac{\theta_E}{1 + g} \frac{1 + g(u^2 - u \sqrt{u^2 + 4} + 3)}{u^2 + 2 + gu\sqrt{u^2 + 4}} \vec{u},
\end{equation}
where $g = F_L / F_S$ \citep{Dominik:2000}. 
Blending due to the lens and other neighboring stars dilutes the astrometric signal further to 
\begin{equation}
    \vec{\delta}_{c, LLN} = \frac{\tilde{A}(u) F_S \vec{\theta}_S + F_N \vec{\theta}_N}{A(u) F_S + F_L + F_N} - \frac{F_S \vec{\theta}_S + F_N \vec{\theta}_N}{F_S + F_L + F_N},
\end{equation}
where $\vec{\theta}_S$ and $\vec{\theta}_N$ are the angular positions of the source and the centroid of all the on-sky neighbors contributing to blending, relative to the lens position; $A(u)$ is given by Equation \ref{eq:amplification} and $\tilde{A}(u)$ is defined
\begin{equation}
    \tilde{A}(u) = \frac{u^2 + 3}{u\sqrt{u^2 + 4}}.
\end{equation}

For long-duration events ($t_E \gtrsim 3$ months), the motion of Earth orbiting the Sun will modify the otherwise symmetric light curve. 
This signal is called the \emph{microlensing parallax $\pi_E$} and is given by
\begin{equation}
\label{eq:pi_E}
    \pi_E = \frac{\pi_{rel}}{\theta_E},
\end{equation}
where $\pi_{rel}$ is the \emph{relative parallax}
\begin{equation}
\label{eq:pi_rel}
    \pi_{rel} = 1 \mathrm{AU} \bigg( \frac{1}{d_L} - \frac{1}{d_S} \bigg).
\end{equation}

Combining Equations \ref{eq:theta_E} and \ref{eq:pi_rel} and solving for the lens mass $M$ yields
\begin{equation}
\label{eq:M}
    M = \frac{\theta_E}{\kappa \pi_E},
\end{equation}
where $\kappa \equiv \frac{4 G}{1 \, AU \cdot c^2}$. 
If both the photometric magnification $A(u)$ and the astrometric shift $\delta_c(u, \theta_E)$ can be measured, the Einstein radius $\theta_E$ can be deduced. 
Along with a measurement of the microlensing parallax $\pi_E$, the mass of the lens $M$ can then be derived using Equation \ref{eq:M}. 
Thus, by having both astrometric and photometric measurements, the degeneracies between lens mass, lens distance, and source distance can be broken.
\footnote{In the case of more complicated microlensing scenarios (e.g. binary sources or lenses that may also be rotating), additional information might be needed to break new degeneracies introduced, such as incorporating e.g. spectroscopic measurements \citep{Smith:2002}. 
Fitting and extracting masses from these types of events is beyond the scope of this paper, although we do discuss binary effects in \S \ref{sec:Binarity and Mergers}.}
The lens mass can then be constrained without having to resort to assumptions about the lens and source distances.
Many microlensing studies only measure the photometric signal and use a Galactic model to weight different source distances in order to derive the lens mass; however, this is a significant assumption.
By combining both the photometric and astrometric microlensing signal, the lens mass can be determined without having to resort to modeling distances. 
Additionally, the astrometric signal peaks after the photometric signal when $u_0 < \sqrt{2}$.

The method just described can be used to measure the masses of BHs.
Since long-duration microlensing events are more likely to be BHs, probable BH microlensing events can be selected from ground-based photometry, then followed up astrometrically \citep{Lu:2016}.

\subsection{Event Selection Algorithm}
\label{sec:Event Selection Algorithm}

With population synthesis complete, we can specify a duration and sampling cadence to perform a synthetic microlensing survey. 
Typical microlensing survey lengths are on the order of years, over which celestial motions are linear for nearly all lenses and sources. 
Thus, with an initial position and velocity, the position of any object at some later time in the survey can be calculated.

First, a grid of latitude and longitude bins is overlaid across the entire survey area. 
\begin{figure*}[t!]
    \centering
    \includegraphics[scale=0.125]{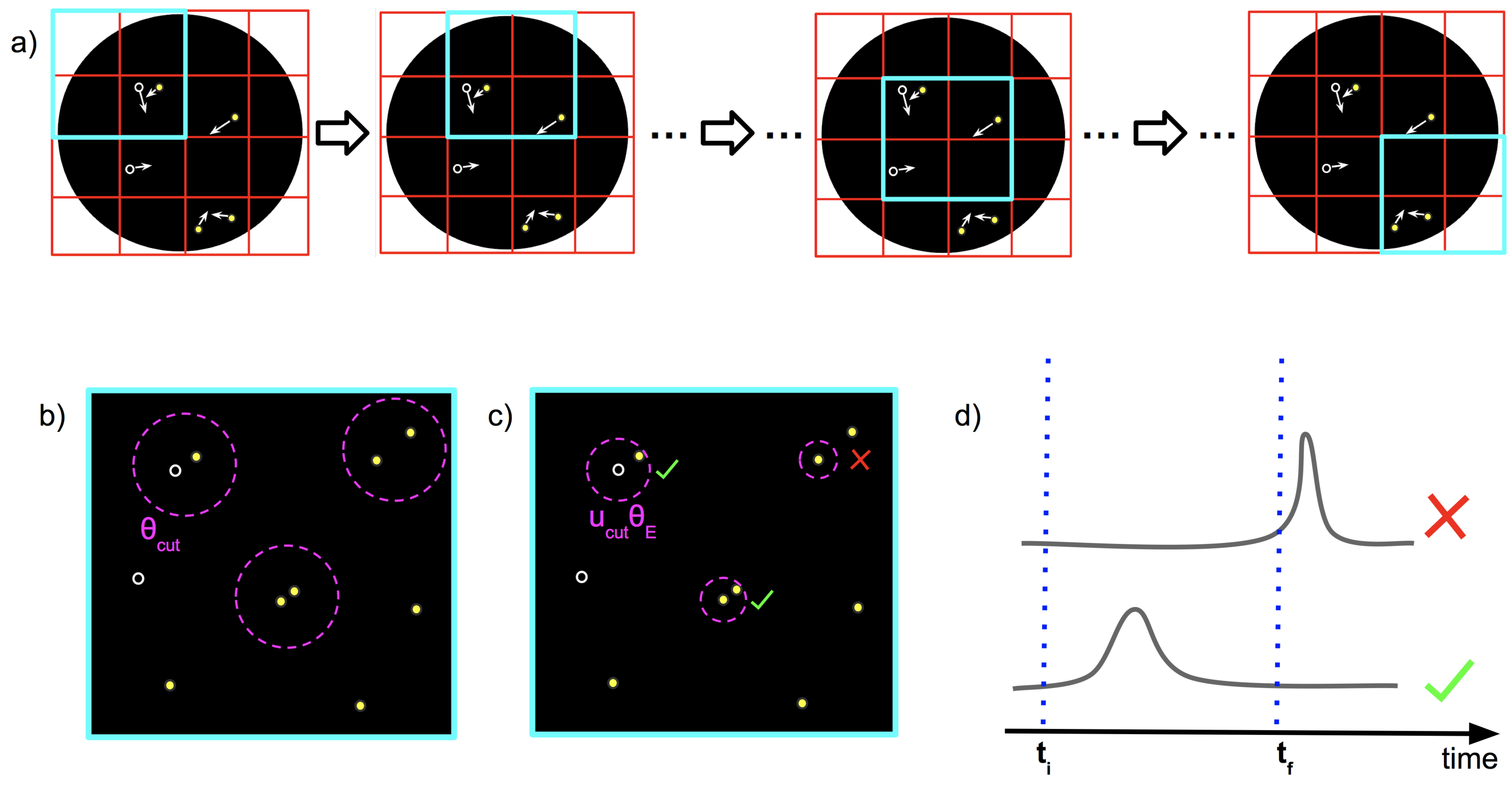}
    \caption{a) A schematic of the binning algorithm described in the second paragraph of \S \ref{sec:Event Selection Algorithm}. 
    Each field is divided into a grid ({\em red lines}) and then the four adjacent boxes at each grid vertex are binned ({\em blue square}) to search for candidate microlensing events and contaminating neighbor stars. 
    Within each blue bin, event selection criteria are applied (subfigures b through d).
    The process is then repeated on the next blue bin until the entire grid has been covered.
    Finally, results are aggregated and duplicates are removed.
    b) A schematic of Cut 1, which selects all object pairs within some on-sky separation of $\Delta \theta < \theta_{cut}$. 
    Microlensing parameters are computed for all of these object pairs.
    c) A schematic of Cut 2, which selects all of the objects pairs left after Cut 1 that satisfy the condition $\Delta \theta / \theta_E < u_{cut}$.
    Note that unlike in Cut 1, the circles have different radii, as $\theta_E$ depends on the lens mass and the source and lens distances.
    The pairs that pass this selection have green checks next to them, while those that do not pass the selection have a red ``x".
    d) A schematic of Cut 3, which selects the source-lens pairs whose point of closest approach on-sky occurs within the survey window.
    This is equivalent to selecting lightcurves that peak photometrically within the survey window.
    Note: none of these cartoons are to scale.}
\end{figure*}
\label{fig:algorithm}
Next, four adjacent latitude and longitude bins of objects generated from population synthesis are read in and combined to make one large bin over which microlensing events are searched for at the sampled times within the survey window.
This combination ensures that microlensing events that occur across the edges of two smaller bins are not missed.
The algorithm for finding event candidates has four ``cuts" as follows: 
\begin{enumerate}
    \item Find the nearest \emph{on-sky} neighbor for each object and calculate the separation $\Delta \theta$.
    Pairs with separations of $\Delta\theta > \theta_{cut}$ are rejected, where $\theta_{cut}$ is a user-specified separation.
    Analogous to the bin size selection, choosing too small a value for $\theta_{cut}$ will mean some events are missed. 
    Choosing too large a value for $\theta_{cut}$ is less problematic; it will merely slow down subsequent calculations by considering pairs that will not produce a detectable microlensing signal.
    \item Calculate the Einstein radius $\theta_E$ of the lenses from these initial microlensing candidate pairs. 
    Only keep the pairs where $\Delta\theta/\theta_E < u_{cut}$, where $u_{cut}$ is another user-specified value that sets the maximum impact parameter. 
    For this further refined list of candidates, record all the stars that fall within the seeing disk radius $\theta_{blend}$, which will be used to calculate blending later on. 
    The seeing disk radius is also a user-specified parameter.
    \item From the new list of microlensing candidate pairs, calculate the time of closest approach $t_0$ when $u = u_0$. 
    Events are kept if $t_0 \in [t_i, t_f]$ where $t_i$ and $t_f$ are the start and end times for the survey. 
    \item Remove unphysical ``events'' where the source is a dark object. 
\end{enumerate}
This yields the list of microlensing event candidates.
The procedure is then repeated for all the bins. 
This algorithm will generate many repeats, due to the multiple overlaps in the bins, and also the multiple evaluation times. 
Only unique events are counted, that is, a lens-source pair is only counted once; the event parameters that get recorded are the ones corresponding to when the source and lens are closest to each other.
A cartoon schematic of this entire process is shown in Figure \ref{fig:algorithm}. 

With complete information about the 6-D kinematics, masses, and photometry, any microlensing parameters of interest can be calculated for all event candidates.

It should also be noted that the choice of sampling cadence $t_{obs}$, defined as the time between consecutive observations, will change the number of event candidates.
For example, $t_{obs} = 1$ day corresponds to a very dense sampling, while $t_{obs} = 100$ days corresponds to a very sparse sampling.
There will be a loss of sensitivity to events with $t_E < t_{obs}$ since shorter events will fall between observations.
It should also be emphasized that the sampling cadence of \texttt{PopSyCLE} has no relation to the observational survey cadence of real microlensing surveys. 
For example, as shown in Figure \ref{fig:samp_cad}, \texttt{PopSyCLE} run with a sampling cadence of 100 days finds nearly all events with $t_E \gtrsim$ 30 days since objects only have to approach each other within $u_{cut} \theta_E$ to be detected.

\begin{figure}[t!]
    \centering
    \includegraphics[scale=0.55]{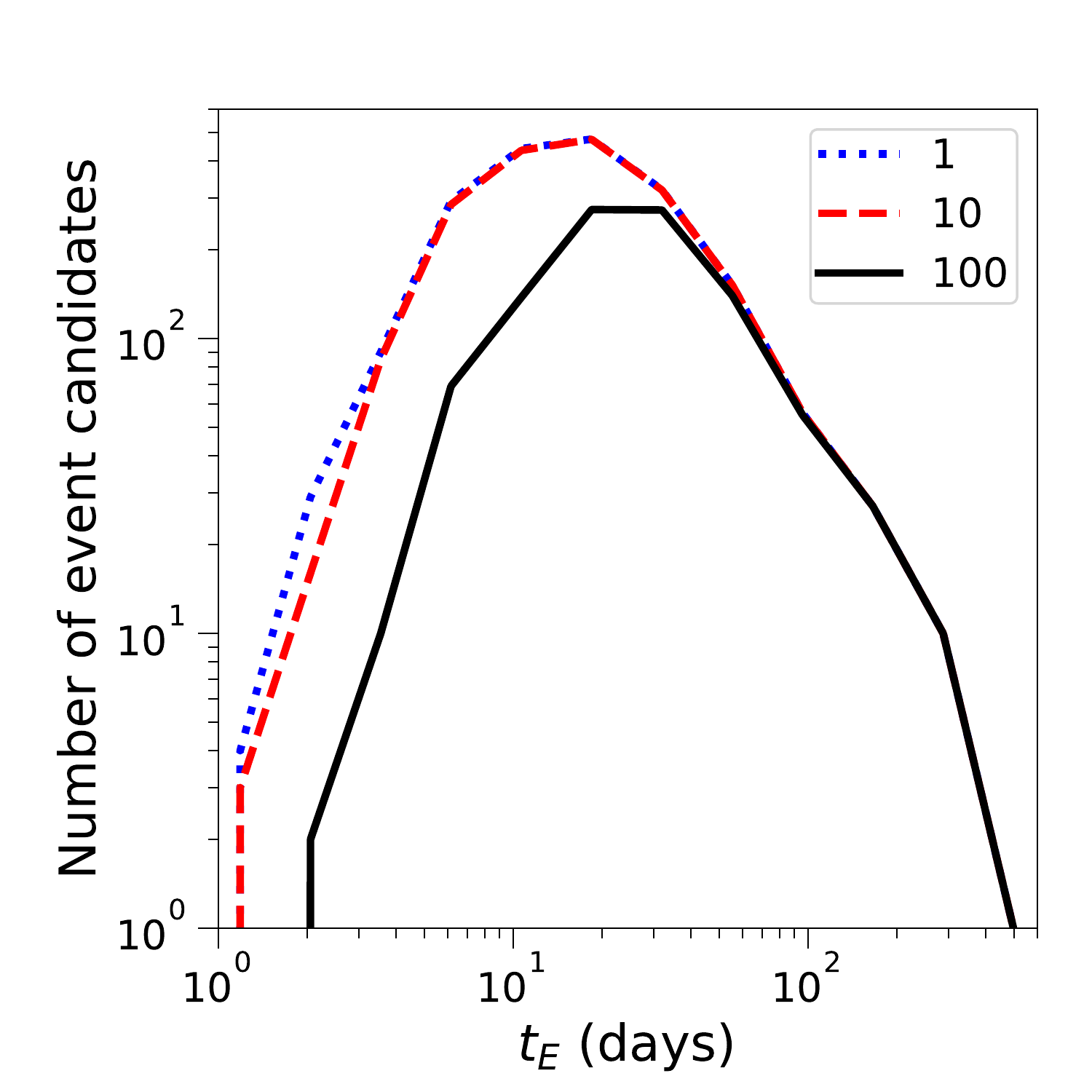}
    \caption{The $t_E$ distribution for all event candidates (i.e. no observational cuts) from field F01, but with three different survey cadences.
    The number in the legend corresponds to the sampling cadence, in days. 
    The long timescale end above 30 days is insensitive to the sampling cadence. 
    Cadences of 10 days or less provide an accurate description of the peak of the $t_E$ distribution.} 
\end{figure}
\label{fig:samp_cad}

\subsection{Survey Implementation}
\label{sec:Survey implementation}

Many of the microlensing events reported by \texttt{PopSyCLE} as described in \S \ref{sec:Event Selection Algorithm} would be undetectable by a real survey.
For example, there will be events too faint to be observed.
Thus, we consider the list of events generated in \S \ref{sec:Event Selection Algorithm} to be event candidates. 
The user can select from this list of event candidates those which satisfy some observational criteria to generate a final list of microlensing events.

\section{PopSyCLE Simulations}
\label{sec:PopSyCLE Simulations}

All the event candidates generated by \texttt{PopSyCLE} are microlensing events, as there are no other types of transient phenomena (e.g. variable stars, telescope artifacts) in the simulation.
Nonetheless, observational cuts are applied to replicate what a real survey would see.
As different microlensing surveys use different telescopes, selection criteria, reduction pipelines and methods, etc. the cuts are different for each survey.
We simulate several current and future microlensing surveys, as described below. 

\begin{deluxetable}{lp{1.5in}}
\tabletypesize{\small}
\tablecaption{Common Simulation Parameters
    \label{tab:Common Simulation Parameters}}
\tablehead{
  \colhead{Parameter} & 
  \colhead{Value}
}
\startdata
NS kick velocity & 350 km/s \\
BH kick velocity & 100 km/s \\
Duration & 1000 days \\
Cadence & 10 days \\
$\theta_{cut}$ & $2''$ \\
$u_{cut}$ & 2 \\
Extinction Law & \cite{Damineli:16} \\
Area Per Field & 0.34 deg$^2$ \\
\enddata
\tablecomments{As described in \S\ref{sec:Population synthesis}, a birth kick is added to the progenitor star's velocity in a random direction for  black holes and neutron stars.
As described in \S\ref{sec:Event Selection Algorithm}, duration describes the length of synthetic survey, and cadence specifies the sampling frequency. 
$\theta_{cut}$ is used to set the maximum separation between potential microlensing event candidates. Considering relatively extreme values to maximize $\theta_E$, such as $d_L = 10$ pc, $d_S = 20$ kpc, and $M = 15 M_\odot$ correspond to $\theta_E = 0.11''$; thus, the choice of $2''$ will capture all observable microlensing events in the simulation.
$u_{cut}$ defines the maximum allowed $u_0$.}
\end{deluxetable}

\begin{deluxetable}{lrrl}
\tabletypesize{\small}
\tablecaption{\texttt{PopSyCLE} Fields
    \label{tab:PopSyCLE Fields}}
\tablehead{
  \colhead{Name} & 
  \colhead{$l$} &
  \colhead{$b$} &
  \colhead{Alt. Name}
}
\startdata
F00 & -0.93$^\circ$ & -7.70$^\circ$ & OGLE-IV-BLG547 \\
F01 &  8.81$^\circ$ & -3.64$^\circ$ & OGLE-IV-BLG527 \\
F02 & -2.75$^\circ$ & -3.32$^\circ$ & OGLE-2017-BLG-0019$^a$ \\
F03 &  3.51$^\circ$ & -3.17$^\circ$ & MOA-II-gb14\\
F04 &  9.62$^\circ$ & -2.93$^\circ$ & MOA-II-gb21 \\
F05 &  1.83$^\circ$ & -2.52$^\circ$ & OGLE-2015-BLG-0029$^a$ \\
F06 &  0.65$^\circ$ & -1.86$^\circ$ & MOA-II-gb5 \\
F07 & -7.91$^\circ$ & -1.85$^\circ$ & OGLE-IV-BLG672 \\
F08 &  1.25$^\circ$ & -1.38$^\circ$ & OGLE-2014-BLG-0613$^a$ \\
F09 &  1.00$^\circ$ & -1.03$^\circ$ & OGLE-IV-BLG500 \\
F10 & -3.61$^\circ$ &  1.86$^\circ$ & OGLE-2015-BLG-0211$^a$ \\
F11 &  0.33$^\circ$ &  2.82$^\circ$ & OGLE-IV-BLG611 \\
F12 &  7.81$^\circ$ &  4.81$^\circ$ & OGLE-IV-BLG629 \\
F13 & -4.21$^\circ$ &  4.96$^\circ$ & OGLE-IV-BLG617 \\
\enddata
\tablecomments{Fields labeled with $^a$ are centered on individual microlensing events from the OGLE EWS.}
\end{deluxetable}

\begin{deluxetable*}{l || lllll}
\tabletypesize{\small}
\tablewidth{0pt}
\tablecaption{Simulation Parameters \label{tab:Simulation Parameters}}
\tablehead{
    \colhead{} &
    \multicolumn{5}{c}{Simulation Name} \\
    \colhead{Parameter/Criteria} &    
    \colhead{Mock Sumi} &
    \colhead{Mock EWS} & 
    \colhead{Mock Mr\'{o}z19} &
    \colhead{Mock Mr\'{o}z17} & 
    \colhead{Mock WFIRST} 
}
\startdata
Filter & I & I  & I & I & H \\
Seeing disk radius, $\theta_{blend}$ & --$^a$ & $0.65''$ & --$^b$ & $0.65''$ & $0.09''$\\ 
Minimum source magnitude, $m_S$ [mag] & $\leq 20$ & -- & $\leq 21$ & $\leq 22$ & -- \\
Minimum baseline magnitude, $m_{base}$ [mag] & -- & $\leq 21$ & -- & -- & $\leq 26$ \\
Maximum impact parameter, $u_0$ & $\leq 1$ & $\leq 2$ & $\leq 1$ & $\leq 1$ &  $\leq 2$ \\
Removal of highly-blended events, $b_{SFF,m}$ & -- & -- & -- & $\geq 0.1$ & -- \\
Einstein crossing time range, $t_E$ [days] & $[0.3, 200]$ & -- & $[0.5, 300]$ & $[0.1, 316]$ & -- \\
Removal of low-amplitude events, $\Delta m$ [mag] & -- & $\geq 0.1$ & -- & -- & $\geq 0.1$ \\
\enddata
\tablecomments{
The selection parameters for Mock Sumi, Mr\'{o}z19, and Mr\'{o}z17 come from \citetalias{Sumi:2016, Mroz:2019b, Mroz:2017} respectively.
For Mock EWS, we choose values based on numbers reported on the OGLE EWS website for the 2016-2018 seasons (described in more detail in \S \ref{sec:Selecting BH Candidates with OGLE}).
For Mock WFIRST, the minimum baseline magnitude comes from \citetalias{Penny:2019}, and the remaining parameter cuts are based on Mock EWS. \\
$^a$ There is no cut involving a parameter dependent on $\theta_{blend}$ (such as $b_{SFF}$ or $\Delta m$) in the Mock Sumi simulation; however, the median seeing at the MOA site is approximately $2.0''$ \citepalias{Sumi:2016}, which would roughly correspond to $\theta_{blend} = 1.0''$. \\
$^b$ Similarly to $^a$, there is no cut involving a parameter dependent on $\theta_{blend}$; however, the seeing is the same as for \citetalias{Mroz:2017}. \\}
\end{deluxetable*}

First, we provide background information on current surveys we will emulate. The Optical Gravitational Lensing Experiment (OGLE) \citep{Udalski:1992} and the Microlensing Observations in Astrophysics (MOA) \citep{Muraki:1999} collaborations run dedicated long-term photometric surveys monitoring the Milky Way bulge (and other targets) for microlensing events from small ground-based, seeing-limited telescopes.
For instance, OGLE-IV regularly observes $\gtrsim 150$ $\mathrm{deg}^2$ of the Galactic bulge for 8 to 9 months out of the year, from February to October.
The observing cadence typically ranges from 20 min to 2 days, depending on the sky location, and events are detected at $I \lesssim 22$ and the median seeing ranges from 1.25'' to 1.35'' \citep{Udalski:2015}.
OGLE also posts alerts about phometric microlensing events through the Early Warning System (EWS).
The OGLE EWS reduces photometry in real time, providing information about potential microlensing candidates.
With the advent of OGLE-IV, roughly 2000 microlensing events are published each season on the EWS website.\footnote{http://ogle.astrouw.edu.pl/ogle4/ews/ews.html}

We compare \texttt{PopSyCLE} simulation results to three different papers that use data gathered from these two surveys: \citet{Sumi:2016}, hereafter \citetalias{Sumi:2016}, which used MOA-II data, and \citet{Mroz:2019b} and \citet{Mroz:2017}, hereafter \citetalias{Mroz:2019b} and \citetalias{Mroz:2017}, which used OGLE-IV data.
Note that \citetalias{Sumi:2016} is an updated version of \citet{Sumi:2013} with corrected event rates, due to a correction in the stellar number count.
Observational surveys must correct for the ability to detect a given microlensing event.
This is quantified by the efficiency correction $\varepsilon$ which is a function of a multitude of parameters, such as the Einstein crossing time, source flux fraction, unlensed source magnitude, survey duration and cadence, and stellar confusion. 
These three papers calculate efficiency corrections to convert the directly observed distribution of microlensing events to efficiency corrected event rates and $t_E$ distributions; we compare our \texttt{PopSyCLE} simulations to these efficiency corrected quantities.

We also run \texttt{PopSyCLE} simulations to explore the possibilities of future microlensing surveys.
The Wide Field Infrared Survey Telescope (\emph{WFIRST}) is a 2.4 meter space telescope scheduled to launch in 2025.
As part of the program, a microlensing survey searching for exoplanets will observe 1.97 deg$^2$ of the Galactic bulge down to $H < 26$ every 15 minutes over 6 observing seasons, where each season is 72 days long \citep[][hereafter \citetalias{Penny:2019}]{Penny:2019}. 
However, the exact way in which the 6 seasons will be distributed across the survey lifetime of 4.5 - 5 years is not yet determined.

\texttt{PopSyCLE} simulations are run in two parts.
We define a \emph{field} as the projected area on the sky over which \texttt{PopSyCLE} is run, and a \emph{simulation} is defined to be the particular parameters used and observational criteria imposed on the \texttt{PopSyCLE} run for some field.
Table \ref{tab:Common Simulation Parameters} lists the parameters common to all simulations presented in this paper.
In summary, for the population synthesis portion of the simulation, we assume that NSs and BHs receive birth kicks of 350 km/s and 100 km/s in a random direction, respectively.
Microlensing events with an impact parameter $u_0 < 2$ are then selected from a 0.34 deg$^2$ area survey of 1000 days in length with observations made every 10 days.
Table \ref{tab:PopSyCLE Fields} contains a list of all fields run in all simulations.
These fields were selected from the OGLE and MOA surveys' Galactic bulge fields, and picked to sample a variety of latitude and longitudes. 
For \emph{WFIRST}, only three of these fields are utilized (F06, F08, F09) from the \texttt{PopSyCLE} simulations, as they are representative samples of the \emph{WFIRST} Cycle 7 design field of view (see Figure 7 of \citetalias{Penny:2019}). 

Note that a survey duration of 1000 continuous days is used in the simulations (Table \ref{tab:Common Simulation Parameters}), which does not match those of any of the surveys described above.
Survey duration and structure have an effect on the types of events detectable.
For example, having gaps between observing seasons can cause a lack of data from one side of the lightcurve, making it difficult to fit for event parameters.
Additionally, some events will be undetectable, simply because they are too long to be observed in a particular observing window, or fall entirely within an observational gap. 
We assume that for published event rates and $t_E$ distributions (e.g. \citetalias{Sumi:2016, Mroz:2017, Mroz:2019b}) efficiency corrections take these effects into account.
However, this statement does not apply to results from the OGLE EWS, as it is only used for alerting events in real time and does not have any type of correction.
Similarly for \emph{WFIRST}, when trying to determine how many BHs are detectable over the duration of the survey, we assume no efficiency correction.
The effect that the season spacing will have on the number of detectable BH events is not so obvious, as the exact details of the season distribution have not been set.
The effect this has on our estimate of the number of BH masses that \emph{WFIRST} can measure will be discussed further in \S \ref{sec: BH Hunting with WFIRST}.

Five different simulations are run, named Mock Sumi, Mock Mr\'{o}z19, Mock Mr\'{o}z17, Mock EWS, and Mock WFIRST.
The first three attempt to mimic the microlensing event selection criteria of \citetalias{Sumi:2016}, \citetalias{Mroz:2019b}, and \citetalias{Mroz:2017} respectively, after taking into account detection efficiency.
These criteria are selected to match the particular magnitude limit, impact parameter, and $t_E$ range of the survey's efficiency correction.
Mock EWS is designed to mimic the detection capabilities of the OGLE EWS, while Mock WFIRST is designed to mimic \emph{WFIRST}'s microlensing survey based on ``reasonable" criteria expected for the mission.

Table \ref{tab:Simulation Parameters} lists the corresponding selection criteria for these five simulations. 

\section{PopSyCLE Comparison}
\label{sec:PopSyCLE Comparison}

\subsection{Number of black holes and neutron stars in the Milky Way}
\label{sec:Number of black holes in the Milky Way}

We first use our population synthesis model to calculate the number of BHs and NSs in the Milky Way. 
First, \texttt{Galaxia} generates a random fraction of the entire Milky Way.
We then perform a simplified version of the compact object population synthesis described in \S \ref{sec:Population synthesis} by ignoring the spatial distribution and kinematics of the compact objects, instead only returning the masses of the compact objects.
We then scale the number of compact objects in accordance to the fraction of stars generated to obtain the actual number.

With this method, we estimate that there are $2.2 \times 10^8$ black holes and $4.4 \times 10^8$ neutron stars in the entire Milky Way. 
Our findings compare well with previous theoretical estimates that predict around $10^8$ to $10^9$ stellar-mass black holes in the Milky Way, with a similar estimate for neutron stars, using methods based on the number of microlensing events toward the Galactic Bulge \citep{Agol:2002} and supernova explosion and Galactic chemical evolution models \citep{Timmes:1996}.

Of the $2.2 \times 10^8$ black holes produced by \texttt{PopSyCLE}, $8.5 \times 10^7$ have masses greater than $10 M_\odot$. 
This is comparable the findings of \cite{Elbert:2018}, where they estimate there should be around $10^8$ black holes with $M > 10M_\odot$ in a galaxy with a stellar mass equal to that of the Milky Way ($M_{*, MW} \approx 6 \times 10^{10} M_\odot$). 
Note that the models of \cite{Elbert:2018} include metallicity and BH-BH mergers, which \texttt{PopSyCLE} currently does not include. 
The mass distribution of BHs is shown in Figure \ref{fig:BHMF} and a discussion of the distribution is presented in \S \ref{sec:Milky Way Present-Day BH Mass Function}.

\subsection{Microlensing event rate}
\label{sec:Microlensing event rate}

\begin{deluxetable}{lcccccc}
\tablecaption{Comparing \texttt{PopSyCLE} to surveys \label{tab:Comparing PopSyCLE to surveys}}
\tablehead{
    \colhead{Field} &
    \multicolumn{2}{c}{$n_s$ ($10^6$/deg$^2$)} &
    \multicolumn{2}{c}{$\Gamma$ ($10^{-6}$/star/yr)} &
    \multicolumn{2}{c}{$\langle t_E \rangle$ (days)} \\
    \colhead{} &
    \colhead{Obs.} &
    \colhead{Mock} &
    \colhead{Obs.} &
    \colhead{Mock} &
    \colhead{Obs.} &
    \colhead{Mock}}
\startdata
F00$^{M19}$ & 2.62 & 2.52 & 1.3 $\pm$ 0.8 & 5.97 & 32.6 & 20.4 \\
F01$^{M19}$ & 4.54 & 2.92 & 5.5 $\pm$ 0.9 & 8.09 & 39.5 & 25.0 \\
F03$^S$ & 3.64 & 3.97 & $14.0^{+2.9}_{-2.4}$ & 22.74 & 25.5 & 18.5 \\
F04$^S$ & 1.11 & 1.12 & $3.5^{+2.6}_{-1.7}$ & 7.66 & 17.0 & 28.5 \\
F06$^S$ & 2.80 & 5.77 & $36.6^{+4.9}_{-4.4}$ & 50.67 & 17.4 & 18.4 \\
F07$^{M19}$ & 1.75 & 2.09 & 5.6 $\pm$ 1.4 & 4.11 & 51.8 & 39.5 \\
F09$^{M19}$ & 4.84 & 4.75 & 23.9 $\pm$ 2.0 & 43.69 & 18.8 & 17.4 \\
F11$^{M19}$ & 4.95 & 5.7 & 16.2 $\pm$ 1.3 & 32.2 & 21.8 & 17.0 \\
F12$^{M19}$ & 3.26 & 2.07 & 3.4 $\pm$ 1.1 & 7.25 & 36.7 & 16.7 \\
F13$^{M19}$ & 4.51 & 3.44 & 5.2 $\pm$ 1.1 & 11.25 & 30.8 & 21.9 \\
\enddata
\tablecomments{Fields with $^{M19}$ indicate the observed values come from \citetalias{Mroz:2019b}, while those with $^S$ are from \citetalias{Sumi:2016}.}
\end{deluxetable}

We next compare stellar densities, event rates, and Einstein crossing times from \texttt{PopSyCLE} simulations with those from \citetalias{Mroz:2019b} and \citetalias{Sumi:2016}. 
A summary is presented in Table \ref{tab:Comparing PopSyCLE to surveys}.

For each field listed in Table \ref{tab:PopSyCLE Fields}, we apply the observational cuts from Tables \ref{tab:Common Simulation Parameters} and \ref{tab:Simulation Parameters} to generate a final list of detectable events as described in \S\ref{sec:PopSyCLE Simulations}. 
In order to convert to an event rate in units of [events star$^{-1}$ year$^{-1}$], we also calculate the total number of detectable stars within the magnitude limits for the corresponding survey.
The microlensing event rate, $\Gamma$, varies between different observational surveys and different fields. 
The rates for Mock Sumi and Mock Mr\'{o}z19 are presented in Table \ref{tab:Mock Simulations}. 

The event rates from \citetalias{Sumi:2016} for fields F03, F04, and F06 (corresponding to MOA-II-gb14, gb21, and gb5) are 14.0, 3.5, and 36.6 events star$^{-1}$ year$^{-1}$, respectively.
For those same fields, the Mock Sumi \texttt{PopSyCLE} simulation gives event rates of 22.74, 7.66, and 50.67 events star$^{-1}$ year$^{-1}$. 
The event rates from \texttt{PopSyCLE} are roughly between 1.4 to 2.2 times higher than the ones reported in \citetalias{Sumi:2016}. 

The event rates from \citetalias{Mroz:2019b} for fields F00, F01, F07, F09, F11, F12, and F13 (corresponding to OGLE-IV-BLG547, 527, 672, 500, 611, 629, and 617) are 1.3, 5.5, 5.6, 23.9, 16.2, 3.4, and 5.2 events star$^{-1}$ year$^{-1}$, respectively.
For those same fields, the Mock Mr\'{o}z19 \texttt{PopSyCLE} simulation gives event rates of 5.97, 8.09, 4.11, 43.69, 32.2, 7.25, 11.25 events star$^{-1}$ year$^{-1}$.
The event rates from \texttt{PopSyCLE} are generally a factor of 2 times higher than the ones reported in \citetalias{Mroz:2019b}, although for one field it is nearly 5 times higher and in one field it is only 0.7 of the observed rate.

Our event rate estimates from \texttt{PopSyCLE} use a total number of stars that is 100\% complete down to the selected magnitude, as we have not imposed any confusion due to the finite beam size of a real telescope.
We assume that the reported event rates in \citet{Sumi:2016, Mroz:2017, Mroz:2019b} use completeness-corrected stellar number counts.

\begin{deluxetable}{llcccc}
\tablecaption{Mock Simulations \label{tab:Mock Simulations}}
\tabletypesize{\scriptsize}
\tablehead{
    \colhead{Field} &
    \colhead{Sim} &
    \colhead{$n_s$ ($10^6$)} & 
    \colhead{$\Gamma$ ($10^{-6}$)} &
    \colhead{$ \langle t_E \rangle$} &
    \colhead{Med($t_E$)} \\
    \colhead{} &
    \colhead{} &
    \colhead{(deg$^{-2}$)} &
    \colhead{(star$^{-1}$ yr$^{-1}$)} &
    \colhead{(days)} &
    \colhead{(days)} 
}
\startdata
F00 & S & 1.46 & 5.87 & 17.3 & 11.2 \\
 & M19 & 2.52 & 5.97 & 20.4 & 14.4 \\
\hline
F01 & S & 1.52 & 6.37 & 25.1 & 14.8 \\
 & M19 & 2.92 & 8.09 & 25.0 & 15.5 \\
\hline
F02 & S & 2.70 & 19.09 & 20.2 & 14.2 \\
 & M19 & 6.11 & 25.66 & 19.5 & 13.9 \\
\hline
F03 & S & 3.97 & 22.74 & 18.5 & 13.2 \\
 & M19 & 7.92 & 30.80 & 19.7 & 14.5 \\
\hline
F04 & S & 1.12 & 7.66 & 28.5 & 21.6 \\
 & M19 & 2.37 & 9.96 & 19.6 & 17.3 \\
\hline
F05 & S & 5.38 & 38.57 & 17.5 & 13.4 \\
 & M19 & 11.05 & 43.65 & 19.5 & 13.9 \\
\hline
F06 & S & 5.77 & 50.67 & 18.4 & 12.4 \\
 & M19 & 12.53 & 59.68 & 18.9 & 12.8 \\
\hline
F07 & S & 0.99 & 4.33 & 40.9 & 28.1 \\
 & M19 & 2.09 & 4.11 & 39.5 & 22.9 \\
\hline
F08 & S & 2.07 & 33.68 & 20.1 & 16.0 \\
 & M19 & 5.02 & 40.04 & 18.1 & 13.1 \\
\hline
F09 & S & 2.06 & 33.84 & 20.5 & 16.0 \\
 & M19 & 4.75 & 43.69 & 17.4 & 12.7 \\
\hline
F10 & S & 0.33 & 19.69 & 13.4 & 13.0 \\
 & M19 & 0.64 & 33.32 & 24.5 & 15.1 \\
\hline
F11 & S & 2.05 & 21.53 & 16.6 & 10.1 \\
 & M19 & 5.70 & 32.20 & 17.0 & 11.4 \\
\hline
F12 & S & 1.11 & 5.80 & 18.1 & 10.1 \\
 & M19 & 2.07 & 7.25 & 16.7 & 9.1 \\
\hline
F13 & S & 1.68 & 6.40 & 12.3 & 7.9 \\
 & M19 & 3.44 & 11.25 & 21.9 & 13.1 \\
\enddata
\tablecomments{For each field simulation (S = Mock Sumi, M19 = Mock Mr\'{o}z19) we list the density of stars $n_s$ brighter than the survey magnitude limit (no correction for blending or confusion), the event rate $\Gamma$, average Einstein crossing time $\langle t_E \rangle$, and the median Einstein crossing time Med($t_E$).}
\end{deluxetable}

\subsection{Einstein Crossing Time Distribution}
\label{sec:Einstein Crossing Time}

The average Einstein crossing time $\langle t_E \rangle$ is a commonly reported parameter for microlensing distributions.
Most authors refer to both the raw $\langle t_E \rangle$ derived from the uncorrected distribution obtained directly after making all observational cuts, and an efficiency corrected $\langle t_E \rangle_\varepsilon$, obtained after correcting for the detection efficiency.
In this section, we compare the reported $\langle t_E \rangle_\varepsilon$ from \citetalias{Sumi:2016} and \citetalias{Mroz:2019b} to those obtained from \texttt{PopSyCLE} simulations.
For brevity, whenever the notation $\langle t_E \rangle$ is used, it is understood to be efficiency corrected, unless otherwise stated.

The $\langle t_E \rangle$ from \citetalias{Sumi:2016} for fields F03, F04, and F06 are 25.5, 17.0, and 17.4 days, respectively.
For these same fields, the Mock Sumi \texttt{PopSyCLE} simulation gives $\langle t_E \rangle$ of 18.5, 28.5, 18.4 days; these values range between factors of 0.7 to 1.7 times the observed timescales. 

The $\langle t_E \rangle$ from \citetalias{Mroz:2019b} for fields F00, F01, F07, F09, F11, F12, and F13 are 32.6, 39.5, 51.8, 18.8, 21.8, 36.7, and 30.8 days, respectively.
For these same fields, the Mock Mr\'{o}z19 simulation gives $\langle t_E \rangle$ of 20.4, 25.0, 39.5, 17.4, 17.0, 16.7, and 21.9 days; these values are all shorter than the observed timescales by factors of 0.5 to 0.9.

Values for $\langle t_E \rangle$ for the events in \citetalias{Mroz:2017} cannot be calculated as they do not report nor present individual event parameters (only the values binned into the histogram are given).
Thus we compare the entire $t_E$ distribution to \texttt{PopSyCLE}, and in particular the ``peak" of the $t_E$ distribution ($t_{E,peak}$), defined as the location of the maximum of the $t_E$ distribution when the individual timescales are binned \emph{logarithmically}.
The peak of a logarithmically binned distribution is not the best quantity to compare, and we advocate performing a fit to the $t_E$ distribution instead.
However, for the moment we use $t_{E,peak}$ as a proxy for some measure of central tendency (note that $t_{E,peak}$ does not correspond to the mean, median, nor mode of the $t_E$ values). 

From the efficiency-corrected $t_E$ distributions presented in several papers, $t_{E,peak}$ is located around 15-20 days (see Figure 13 of \cite{Sumi:2013}, Figure 17 of \citet{Wyrzykowski:2015}, and Figure 2 of \citetalias{Mroz:2017}). 
As can be seen in Figure \ref{fig:mroz_real_vs_mock}, $t_{E,peak}$ from \texttt{PopSyCLE} falls in the lower end of that range. 

\begin{figure}[t!]
    \centering
    \includegraphics[scale=0.55]{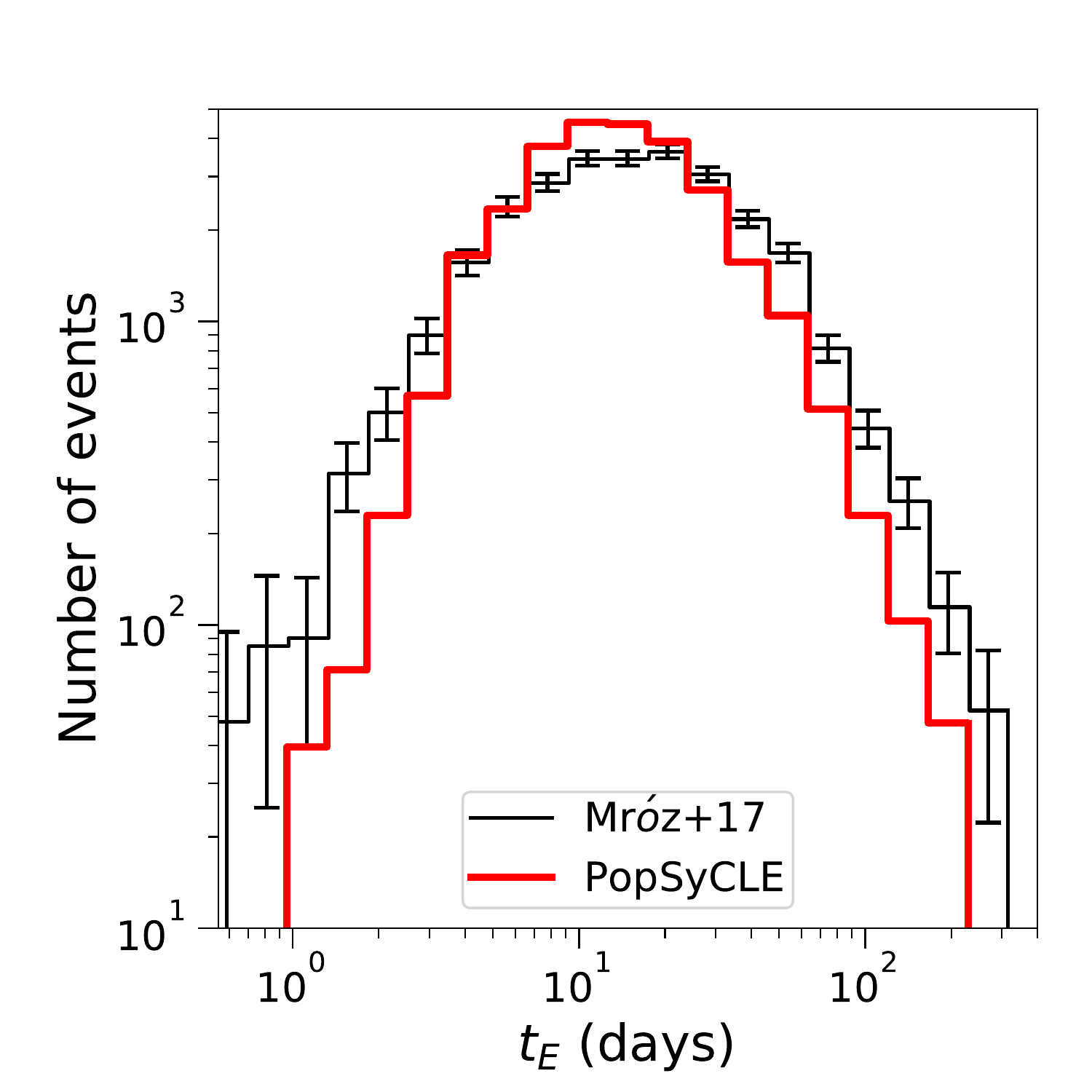}   
    \caption{The $t_E$ distribution in black comes from Figure 2 of \citetalias{Mroz:2017}.
    Overplotted in red is the Mock Mr\'{o}z17 \texttt{PopSyCLE} simulation, scaled such that the number of events are the same.} 
\end{figure}
\label{fig:mroz_real_vs_mock}

\section{Results}
\label{sec:Results}

\subsection{Milky Way Present-Day BH Mass Function}
\label{sec:Milky Way Present-Day BH Mass Function}

The BH present-day mass function (PDMF) encodes information about the BH IFMR and stellar IMF, and to a lesser degree the star formation history (SFH).
It also provides information about BH binaries and their formation channel(s). 
With a sufficiently large sample of BH mass measurements from both LIGO (extragalactic) and microlensing (Galactic), the BH PDMF can be measured and the IFMR can be constrained. 

We use \texttt{PopSyCLE} to generate the Milky Way BH PDMF which is shown in Figure \ref{fig:BHMF}.
The \citetalias{Raithel:2018} IFMR clearly shows the ``mass gap", as the lowest mass BH is around $5 M_\odot$, which is greater than the largest possible NS mass $M_{NS,max} \sim 2 - 3 M_\odot$ (\citet{Ozel:2016}, but also see \citet{Margalit:2017} which suggests an upper limit closer to $\sim 2.2 M_\odot$).
As discussed in \S\ref{sec:Neutron Stars and Black Holes}, there are no $\gtrsim 30 M_\odot$ black holes as the \citetalias{Raithel:2018} IFMR assumes only solar metallicity progenitor stars and \texttt{PopSyCLE} currently implements only single BHs.
For the Milky Way, the SFH (according to the Besan\c{c}on model) does not influence the BH PDMF very much, as most stars are over $10^9$ years old. 
The minimum ZAMS mass for a star to form a black hole is $\sim15 M_\odot$, and the corresponding main-sequence lifetime of such a star is $\sim10^7$ years.
Thus, the vast majority of BHs produced when their progenitor star dies will have already been formed.

\begin{figure}[t!]
    \centering
    \includegraphics[scale=0.55]{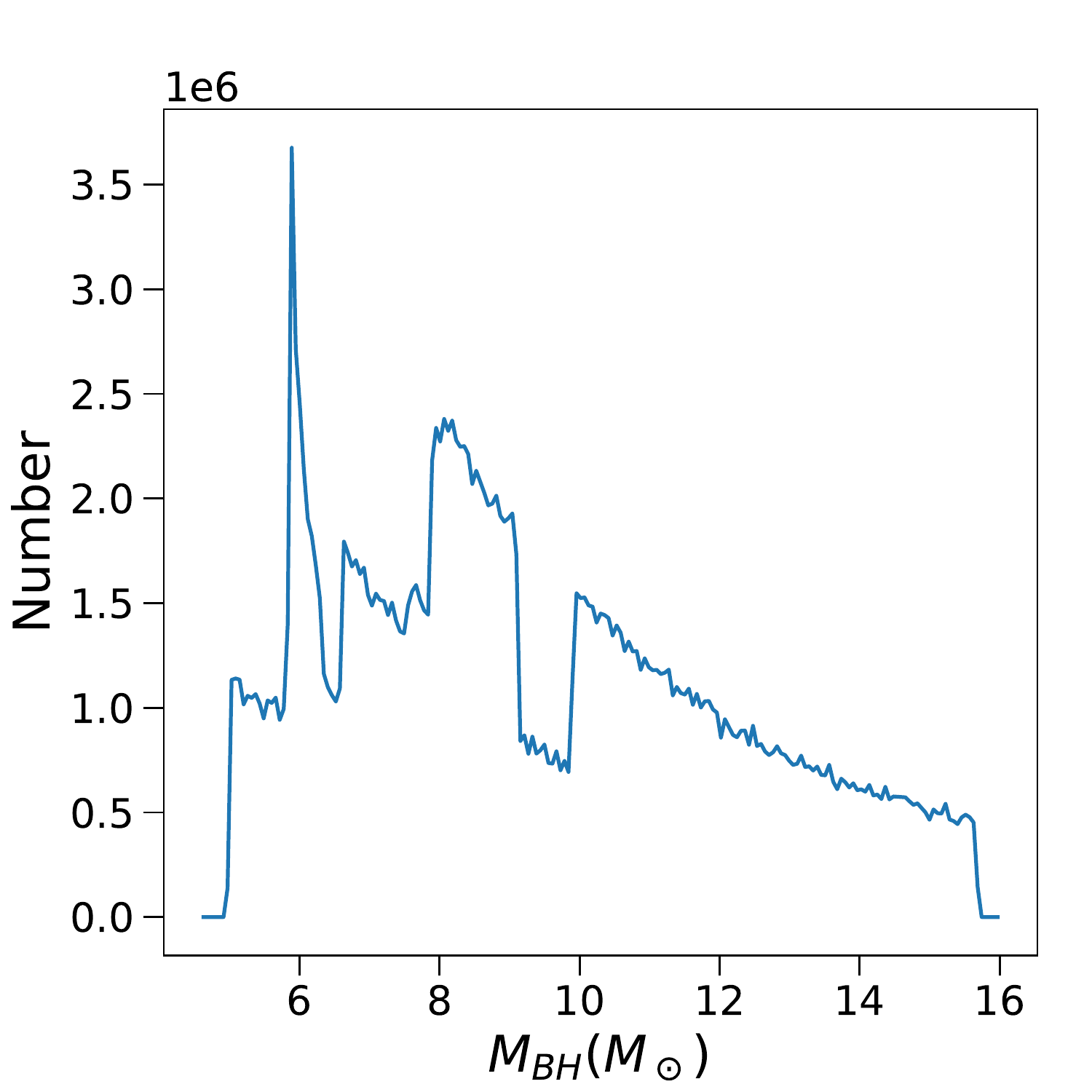} 
    \caption{The Milky Way BH present-day mass function.
    The description of BH population synthesis used to produce this distribution is described in \S \ref{sec:Number of black holes in the Milky Way}.
    Note that this is the theoretical underlying distribution for the \emph{entire} Milky Way, not just those observable via microlensing; no observational constraints or limitations have been taken into account.
    The BH IFMR comes from \citetalias{Raithel:2018}.}
\end{figure}
\label{fig:BHMF}

\begin{figure}[t!]
    \centering
    \includegraphics[scale=0.55]{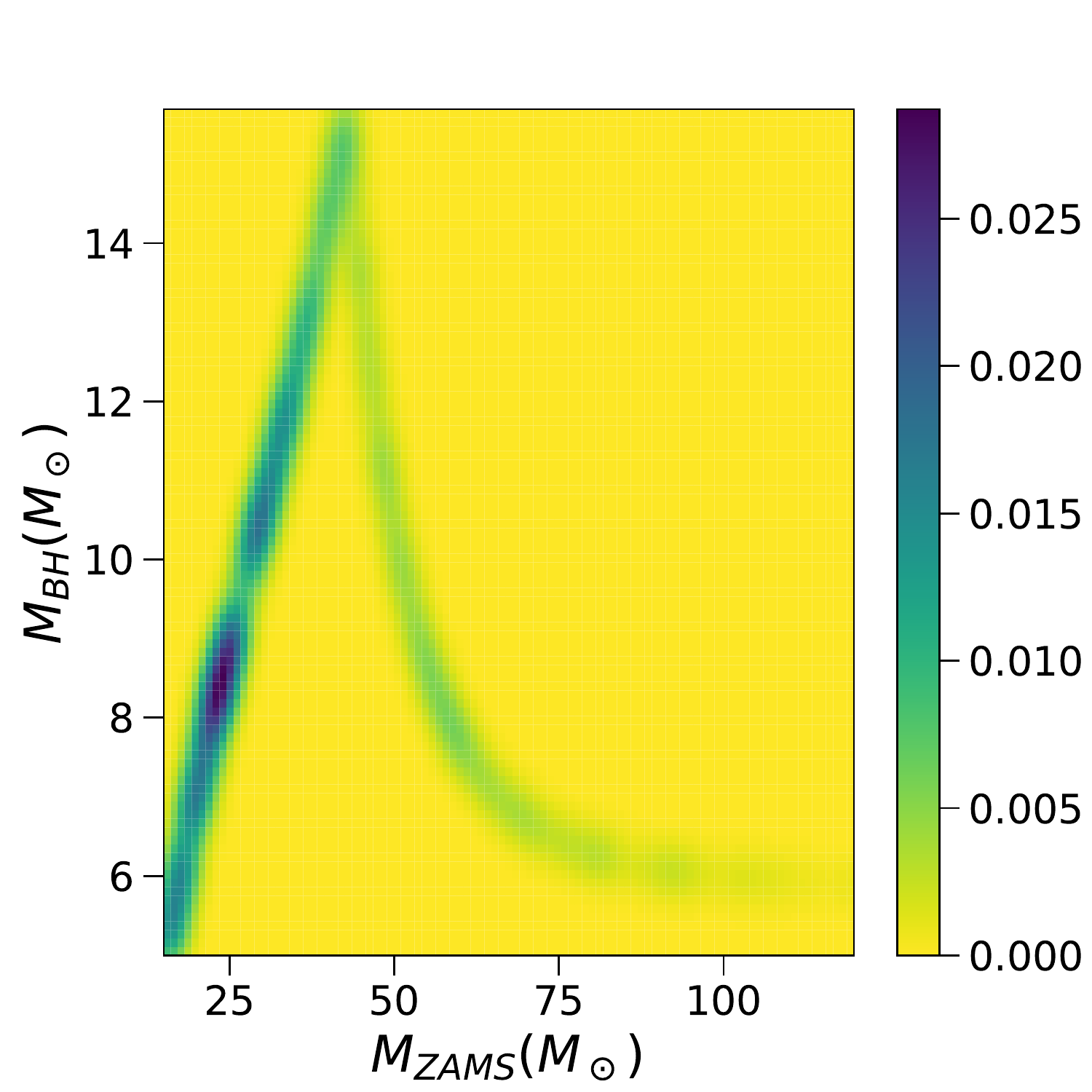}
    \caption{The black hole initial-final mass relation (BH IFMR), with the colorbar indicating the density of Milky Way stars at that part of the IFMR.
    For example, there is an overall trend where there are many more lower mass progenitors; this is mainly due to the IMF.
    However within that overall framework, there is more structure, which comes from the fact that some of the progenitor stars will form NSs and not BHs.
    The BH IFMR provides a mapping between the ZAMS mass of a star and the mass of the BH it forms, assuming it forms a BH and not a neutron star. 
    It is described by a piecewise function (Equation \ref{eq:bhifmr}, with ejection fraction $f_{ej} = 0.9$). 
    This value of $f_{ej}$ was selected as it most closely reproduced the observed distribution of black hole masses \citepalias{Raithel:2018}.}
\end{figure}
\label{fig:mzams_vs_mbh}

The \citetalias{Raithel:2018} IFMR also produces structures such as peaks and gaps in the BH PDMF.
The spike around $6 M_\odot$ is due to a combination of stars with ZAMS masses between $15-20 M_\odot$ and $70-120 M_\odot$. 
Although there are more stars of lower ZAMS masses due to the IMF, only 34\% of stars within $15-20 M_\odot$ form BHs, while 60\% of the $>70 M_\odot$ stars form BHs.
The paucity around $10 M_\odot$ is due to the fact that most stars between $25-27 M_\odot$ form NS and not BHs.
The general decrease in number of BHs greater than $8 M_\odot$ is simply due to the IMF; high mass stars are rarer.
These trends are shown visually in Figure \ref{fig:mzams_vs_mbh}, which combines the IFMR with the IMF and SFH.
We re-emphasize that this structure is specific to the \citetalias{Raithel:2018} IFMR; a different IFMR will produce a different BH PDMF. 
However, with this assumption, the structure in the BH PDMF may be detectable with a sample of $\sim$100 BHs.
This will become a possiblity with \emph{WFIRST}, as discussed in \S \ref{sec: BH Hunting with WFIRST}.

\subsection{Measuring Black Holes Masses with Individual Astrometric Follow-up}

Identifying a BH with microlensing requires showing that the lens mass exceeds $5 M_\odot$ and is not luminous.
Since $5 M_\odot$ stars are extremely bright, it is relatively easy to rule out massive stellar lenses once a lens mass is in hand.
However, as described in \S \ref{sec:Other Selection Parameters}, photometric microlensing alone is not sufficient to determine the mass of the lens in a microlensing event; astrometry is also needed to break degeneracies between lens mass and source and lens distances.
Thus, data from a photometric survey must be combined with astrometric follow-up of a smaller set of BH candidates. 
As an example, we investigate the strategy of \cite{Lu:2016} of selecting astrometric follow-up candidates with long Einstein crossing times $t_E \gtrsim 120$ days, high magnification $A_{max} \gtrsim 10 \equiv u_0 \lesssim 0.1$, and high source flux fraction $b_{SFF} \approx 1$ from photometric surveys updates such as the OGLE EWS (described in \S \ref{sec:PopSyCLE Simulations}). 

\subsubsection{Selecting BH Candidates with OGLE}
\label{sec:Selecting BH Candidates with OGLE}

\begin{figure*}[t]
    \centering
    \includegraphics[scale=0.5]{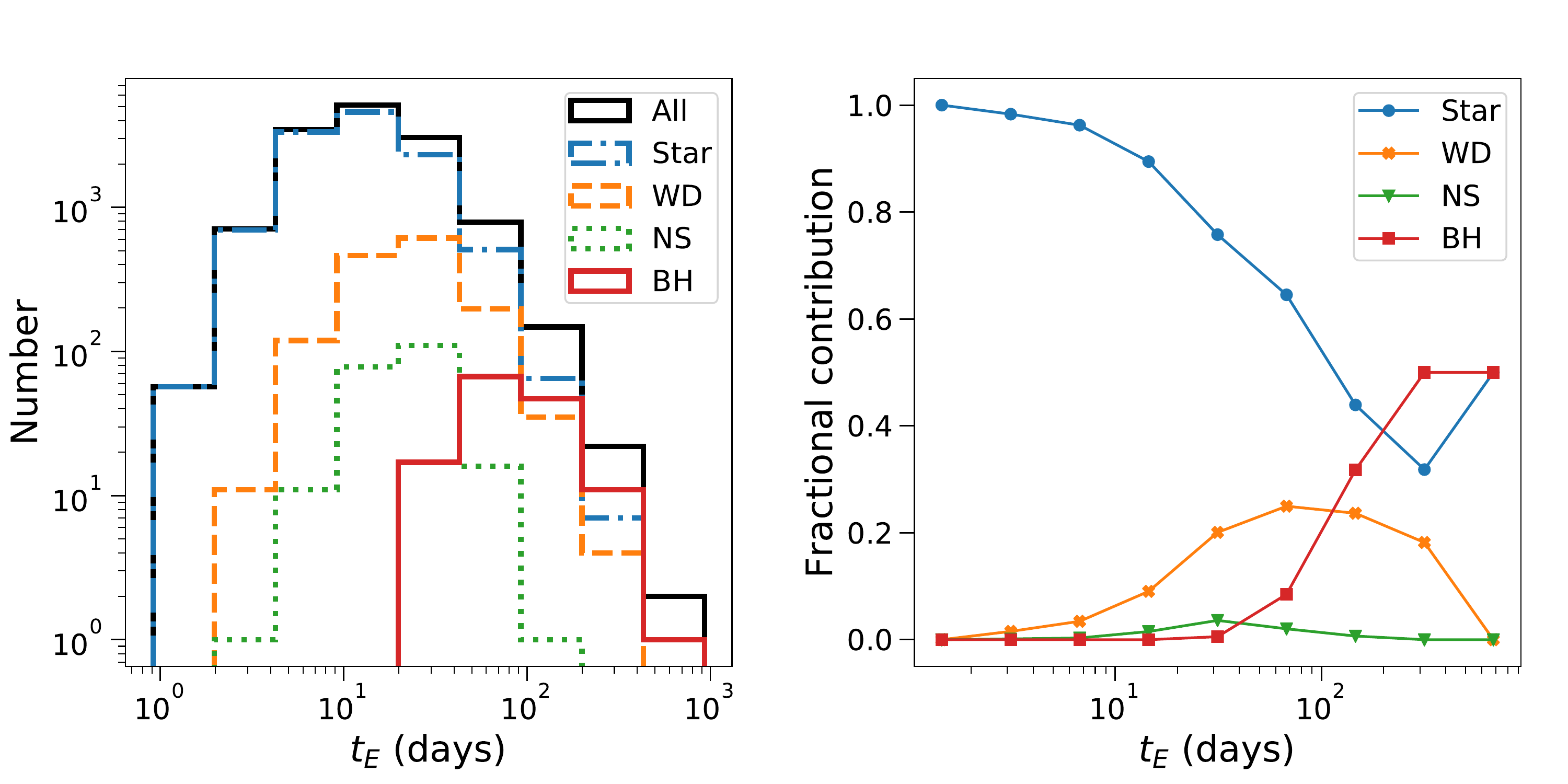} 
    \caption{\emph{Left:} $t_E$ distribution from the Mock EWS simulation, showing the different contributions due to different components.
    \emph{Right:} The fractional contribution from each type of object type at different $t_E$ times. 
    Note that the NS and BH contributions can be shifted to different $t_E$ depending on the adopted kick velocities.
    Also note that many of these events would be difficult to detect as they have parameters with low detection efficiency; however, it serves to illustrate the underlying contributions to the $t_E$ distribution.}
\label{fig:tE_contributions}
\end{figure*}

Out of all the Mock EWS events, $\sim$1\% of them have a BH lens. 
This is consistent with the fraction of BH lensing events in the unfiltered \texttt{PopSyCLE} event candidates list as well.
While BHs make up $\lesssim$0.1\% of objects in the Milky Way by number, their lensing rate is enhanced by their $\sim$10$\times$ larger Einstein radius. 
Thus, the observational selection criteria of the OGLE survey does not dramatically change the probability of detecting BH lensing events. 
In Figure \ref{fig:tE_contributions}, the various types of lenses contributing to the $t_E$ distribution is shown, which is comparable to the efficiency-corrected $t_E$ curves produced by surveys. 
Figure \ref{fig:ews} shows roughly the number of candidates that can be followed up based on EWS photometry per season.
Although there usually are nearly 100 events with $t_E > 120$ days, often only 10 or so of those have sufficiently high SNR data and good coverage of the event, which are necessary later when trying to fit the event and extract the microlensing parameters $t_E$ and $\pi_E$.

\begin{figure}[t]
    \centering
    \includegraphics[scale=0.5]{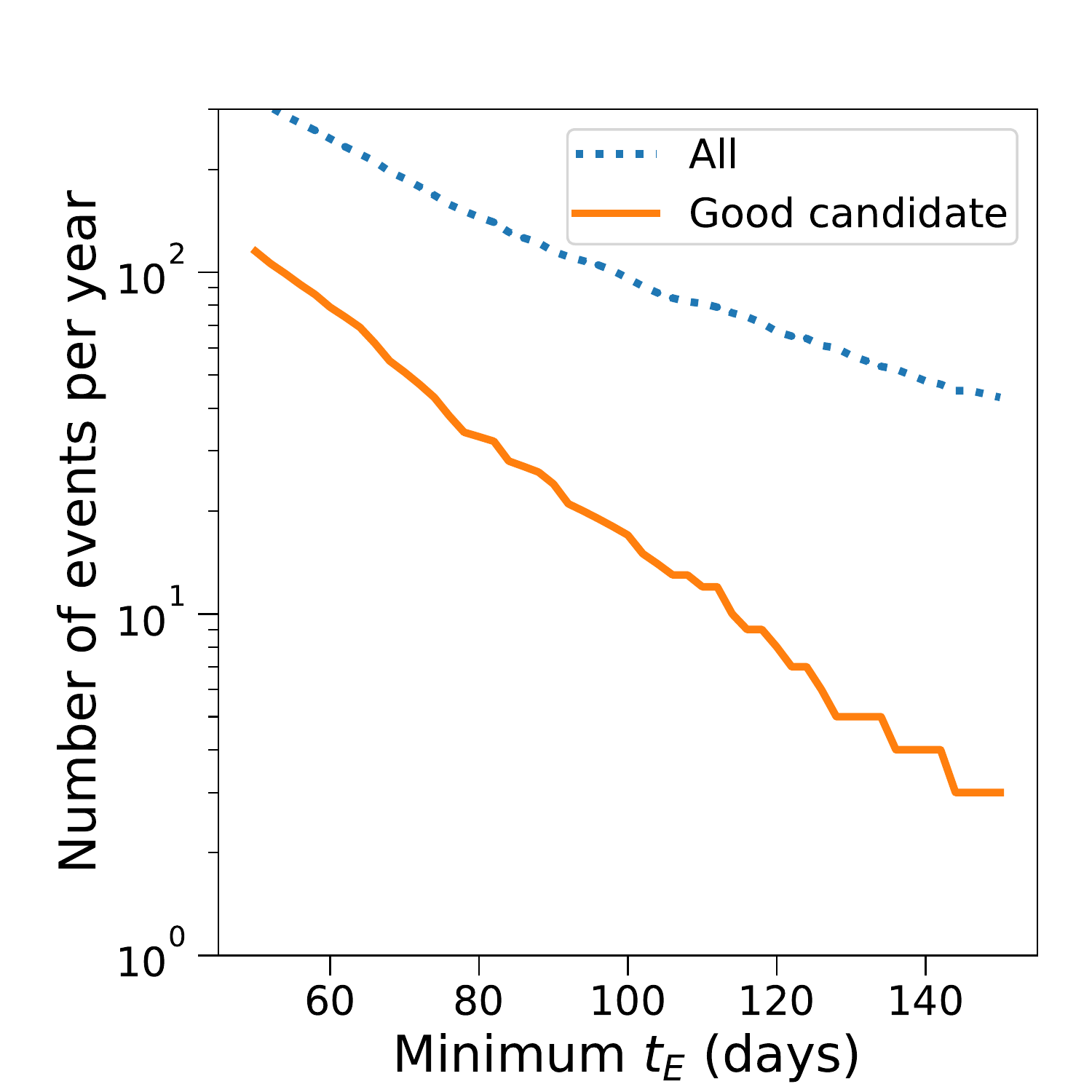}
    \caption{Number of events per season from the OGLE Early Warning System that have $t_E$ larger than some minimum value (dotted blue line).
    However, even events with $t_E > 120$ days are not necessarily good follow-up candidates, as they may not have sufficient coverage of the photometric peak or insufficient SNR.
    Thus, good candidates have 1) $\pm 0.5 t_E$ coverage of the peak, 2) $I_{base} < 19.5$ mag, 3) $\Delta I > 0.5$ mag (orange line).
    The numbers plotted are the averages over the 2016, 2017, and 2018 seasons reported on the OGLE EWS website.} 
    \end{figure}
\label{fig:ews}

\begin{figure}[t]
    \centering
    \includegraphics[scale=0.5]{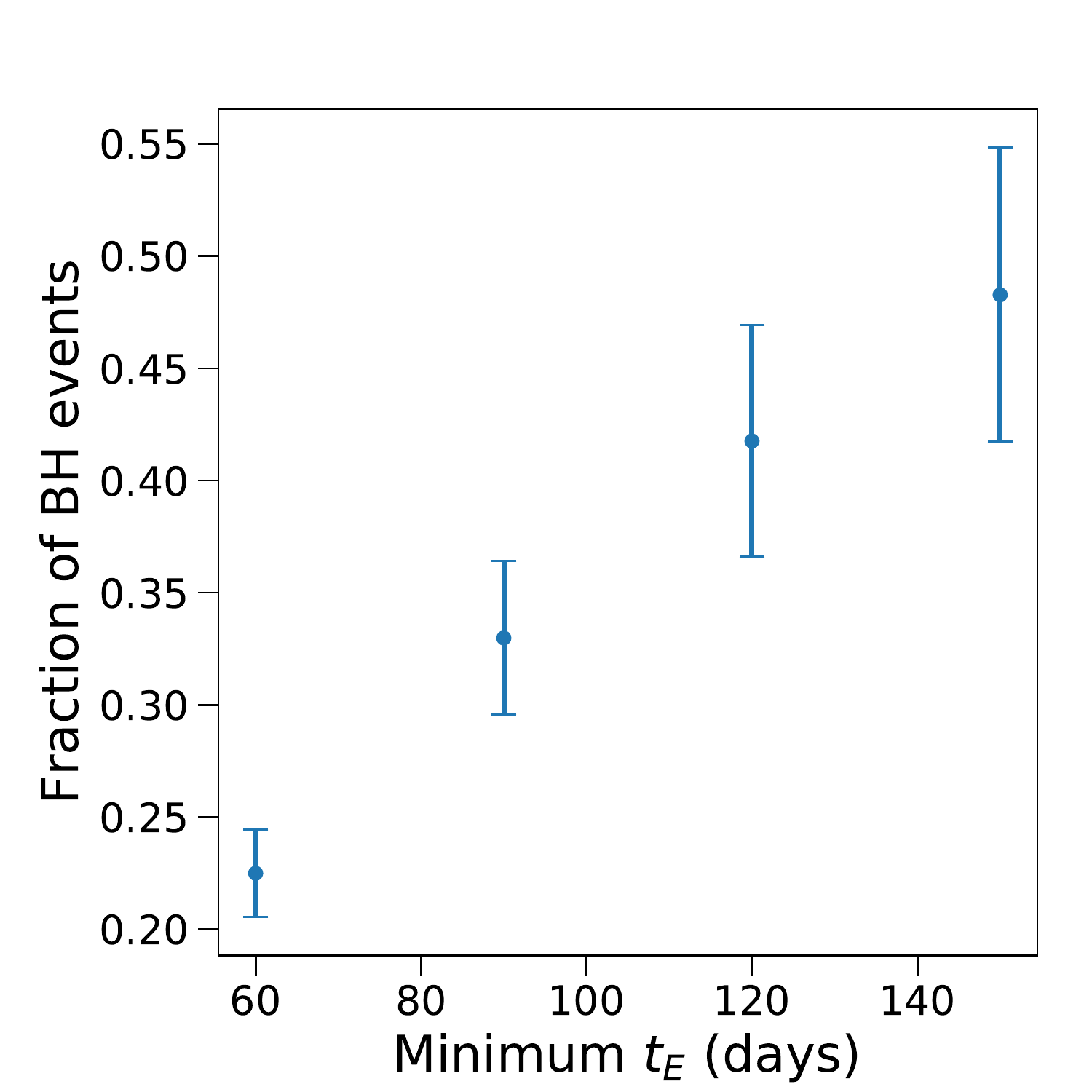}
    \caption{Fraction of long-duration Mock EWS events from \texttt{PopSyCLE} that have BH lenses.
    $t_E>120$ day selections are still recommended for maximizing the probability of finding a BH lens.} 
\label{fig:frac_bh}
\end{figure}

Selecting Mock EWS events with $t_E > 120$ days increases the probability of the lens being a BH to 40\% (Figure \ref{fig:frac_bh}). 
Note that a different Galactic model can increased this fraction by up to a factor of two (see Appendix \ref{Appendix:Galaxia Galaxtic Model}).
We then explored two additional selection criteria in addition to long events. 

First, we tried selecting on events from \texttt{PopSyCLE} simulations with $b_{SFF} \approx 1$.
Intuitively, one would expect a source to be less blended with a BH lens than with a stellar lens, as the BH does not contribute any flux.
However, for ground based surveys the seeing disk is so large that many background stars fall within the disk, and whether the lens is luminous or not does not significantly change $b_{SFF}$. 
In principle, additional high-resolution imaging from space or ground-based telescopes with adaptive optics provides a much smaller seeing disk, reducing contamination from neighboring stars.
This would then circumvent the aforementioned problem, meaning BH lensing events would truly have a high $b_{SFF}$.
However, this strategy fails, as OGLE microlensing events are observationally biased toward brighter sources and the average lens is quite faint ($I \sim 26$).
Thus, the distribution of $b_{SFF}$ for BH and stellar lenses are similar in a sample limited to I$\lesssim 21$.

Similarly, selecting events with small $u_0$ does not increase the probability of finding a BH lens.
Although the average BH is more massive than the average star and thus has a larger $\theta_E$, it does not have a correspondingly smaller $u_0$, as $u_0$ is independent of mass.

\subsubsection{Individual Candidate Follow-up}
\label{sec:Individual Candidate Follow-up}

As our ultimate goal is to measure the number of BHs in the Milky Way, we now consider how many BHs can be expected to be detected via astrometric follow-up. 
We estimate that $\sim$12 BH candidates (corresponding to 5 expected BH detections) are required to constrain the total number of BHs in the Milky Way to 50 percent, assuming Poisson statistics (Figure \ref{fig:nbh_sigma}). 
With an astrometric follow-up program, due to practicalities such as limited telescope time on facilities able to perform such measurements, 3-4 candidates can be observed each year
and the probability of the candidate being a BH is $\sim$40\%; thus about 1 - 2 BHs per year can be expected to be detected.
Over 5 years, this would result in a total of 5 - 10 BHs.
However, the BH PDMF would be difficult to constrain based on current sample sizes and sporadic astrometric follow-up, due to the inefficiency of the process. 
Dedicated astrometric surveys are needed to place useful constraints on the BH PDMF.

\begin{figure}[h!]
    \centering
    \includegraphics[scale=0.5]{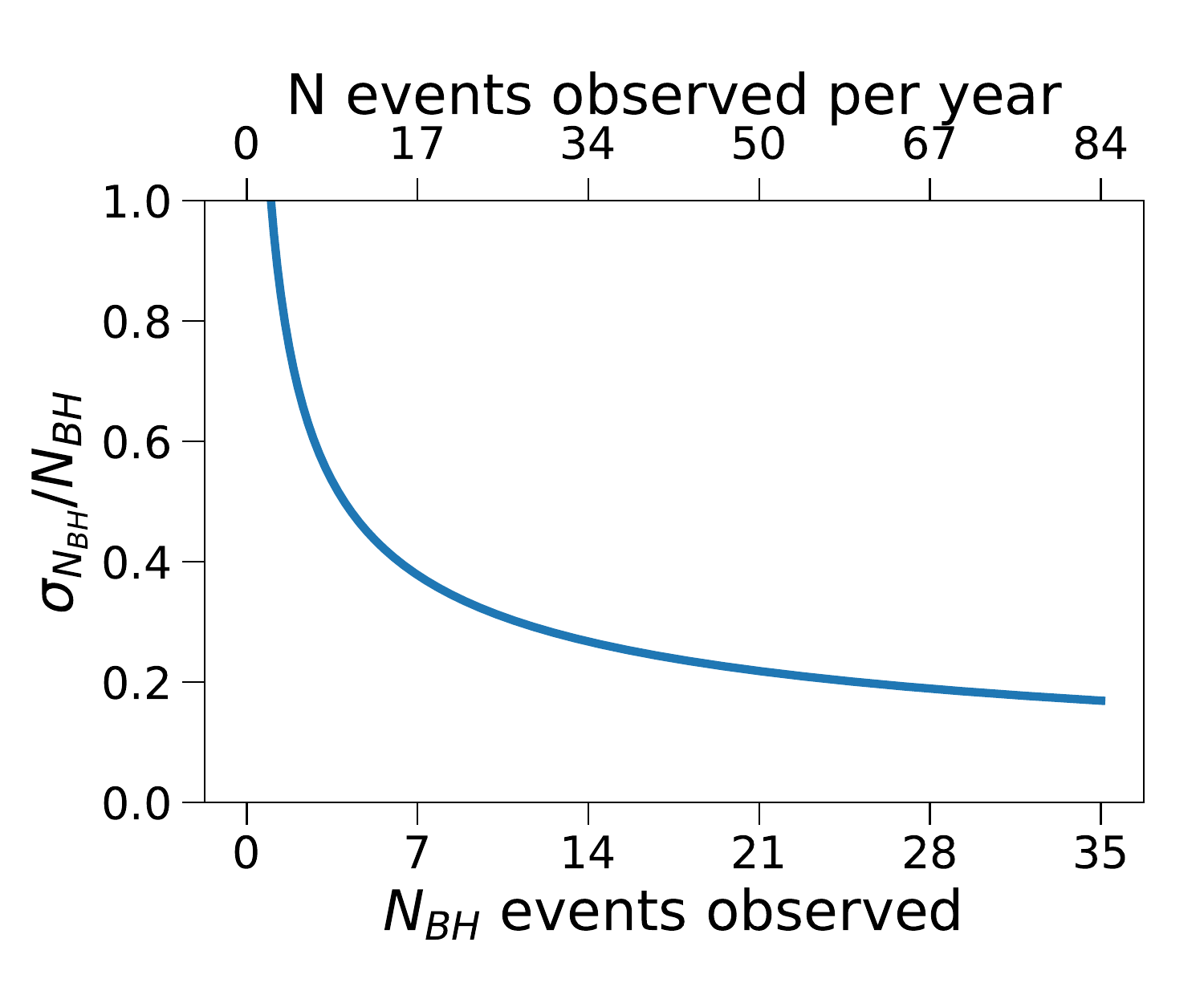}
    \caption{Number of BH candidates needed to constrain the total number of BHs in the Milky Way to some uncertainty (top horizontal axis), assuming Poisson uncertainty. 
    The number of actual BHs resulting from the sample of candidates is the bottom horizontal axis.
    This assumes we have selected candidates from long-duration ($t_E > 120$ days) OGLE EWS following the strategy outlined in \S \ref{sec:Selecting BH Candidates with OGLE}.}
\end{figure}
\label{fig:nbh_sigma}

\subsection{Confirming BH Lenses with Astrometry and Microlensing Parallax}
\label{sec:astro_parallax_BHconfirmation}

BH candidates identified from photometric microlensing surveys can be followed up with high-precision astrometric measurements in order to measure the astrometric microlensing shift, $\delta_{c,max}$. 
When combined with $t_E$, $\pi_E$, and $u_0$ from the photometric light-curve, the astrometric shift yields a constraint on the lens mass. 
In Figures \ref{fig:deltac_vs_piE_vs_tE} and \ref{fig:margin_deltac_vs_piE}, microlensing events are plotted in $\delta_{c,max}-\pi_E$ space.
In Figures \ref{fig:margin_piE_vs_tE} and \ref{fig:margin_deltac_vs_tE} the microlensing events are plotted in $\pi_E-t_E$ and $\delta_{c,max}-t_E$ space, respectively.
We note that blending due to the lens (if it is luminous) is incorporated into the astrometric shift calculations, assuming a $\sim$100 mas aperture size that is roughly equivalent to infrared imaging with JWST or with an 8--10 meter telescope and an adaptive optics system.
BHs occupy a very particular region in $\delta_{c,max}-\pi_E$  space (large $\delta_{c, max}$ and small $\pi_E$) as they are massive.
Although both a massive luminous stellar lens and equally massive black hole will have equal Einstein radii, all else equal, the astrometric shift will be larger for the BH, as it do not blend with the source images (see Equation \ref{eq:delta_c} vs. Equation \ref{eq:delta_c_LL}). 

\begin{figure}[t!]
    \centering
    \includegraphics[scale=0.55]{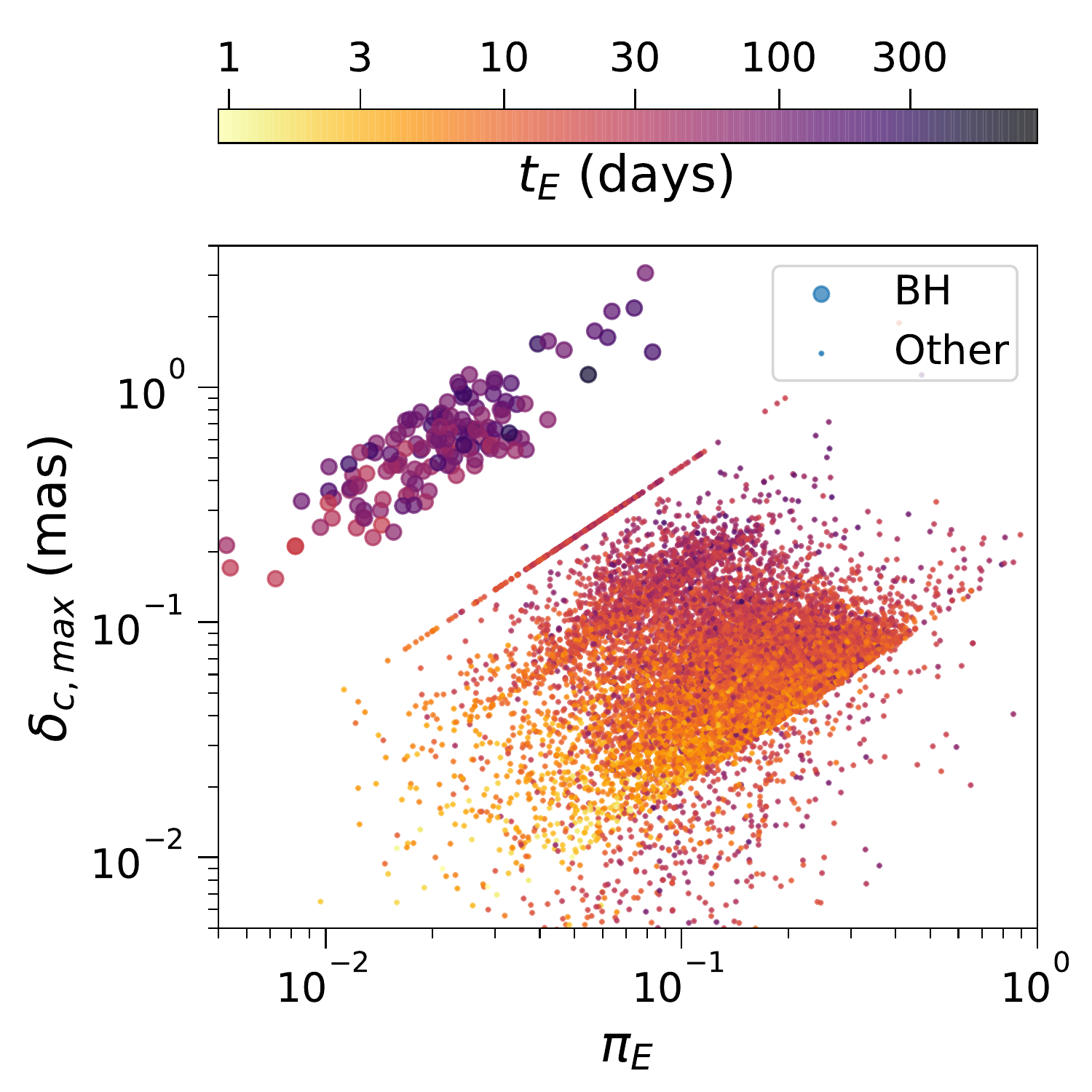}
    \caption{The maximum astrometric shift $\delta_{c,max}$ vs. the microlensing parallax $\pi_E$, with the color of the point indicating the Einstein crossing time $t_E$ of the event.
    We include blending between the lens and source when calculating $\delta_{c,max}$. 
    The BHs ({\em large points}) are easily seperable in this space.
    The points correspond to microlensing events from the Mock EWS simulation.}
\end{figure}
\label{fig:deltac_vs_piE_vs_tE}

\begin{figure}[t!]
    \centering
    \includegraphics[scale=0.4]{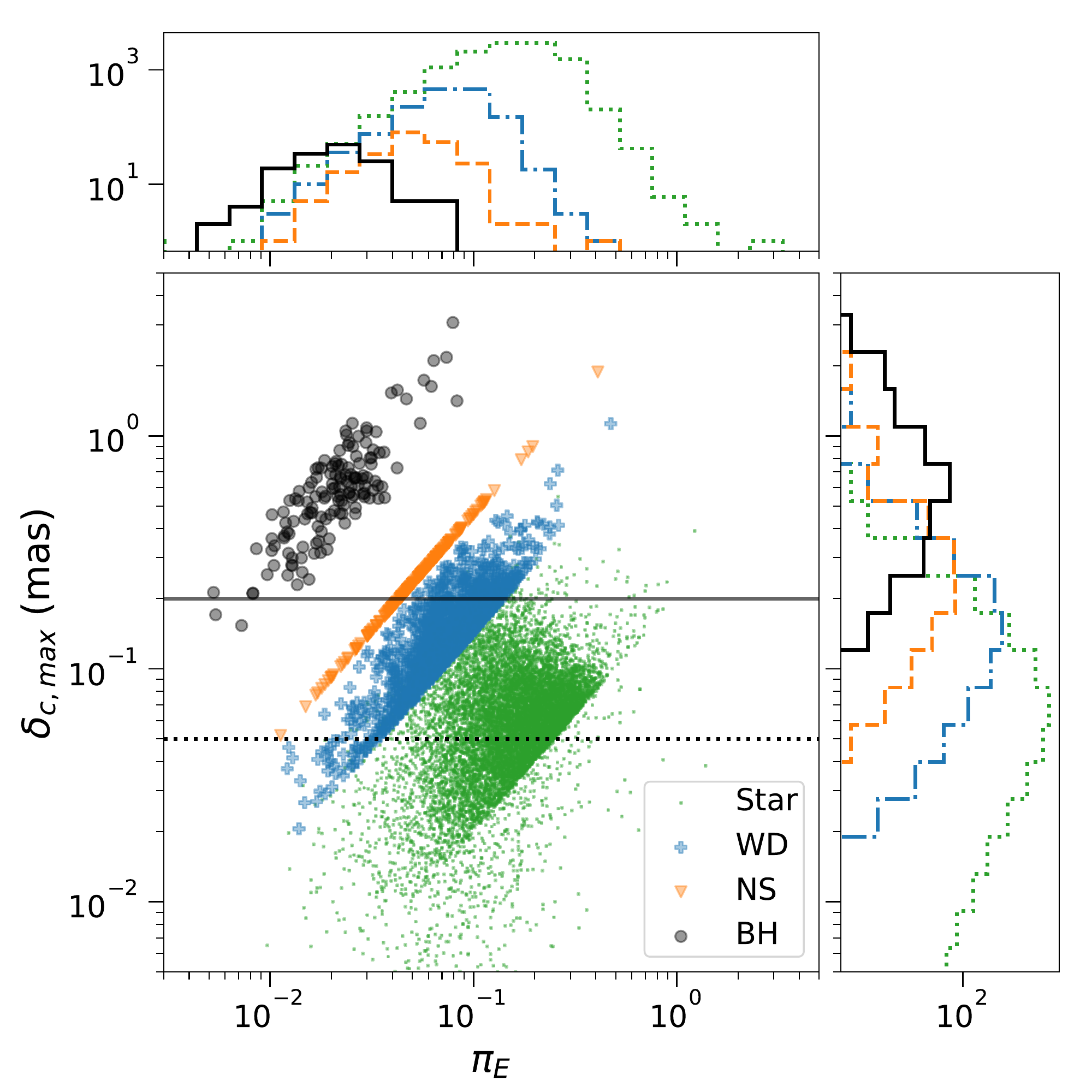}
    \caption{The maximum astrometric shift $\delta_{c,max}$ vs. the microlensing parallax $\pi_E$ for different lens types.
    Blending between the lens and source is included when calculating $\delta_{c,max}$.
    The solid line denotes the current astrometric precision ($\sim 0.2$ mas, using the Keck laser guide star adaptive optics system).
    The dotted line denotes anticipated astrometric precision achievable in the next decade ($\sim 0.05$ mas, using \emph{WFIRST} or the Thirty Meter Telescope). 
    The points correspond to microlensing events in the Mock EWS simulation.}
\end{figure}
\label{fig:margin_deltac_vs_piE}

\begin{figure}[t!]
    \centering
    \includegraphics[scale=0.4]{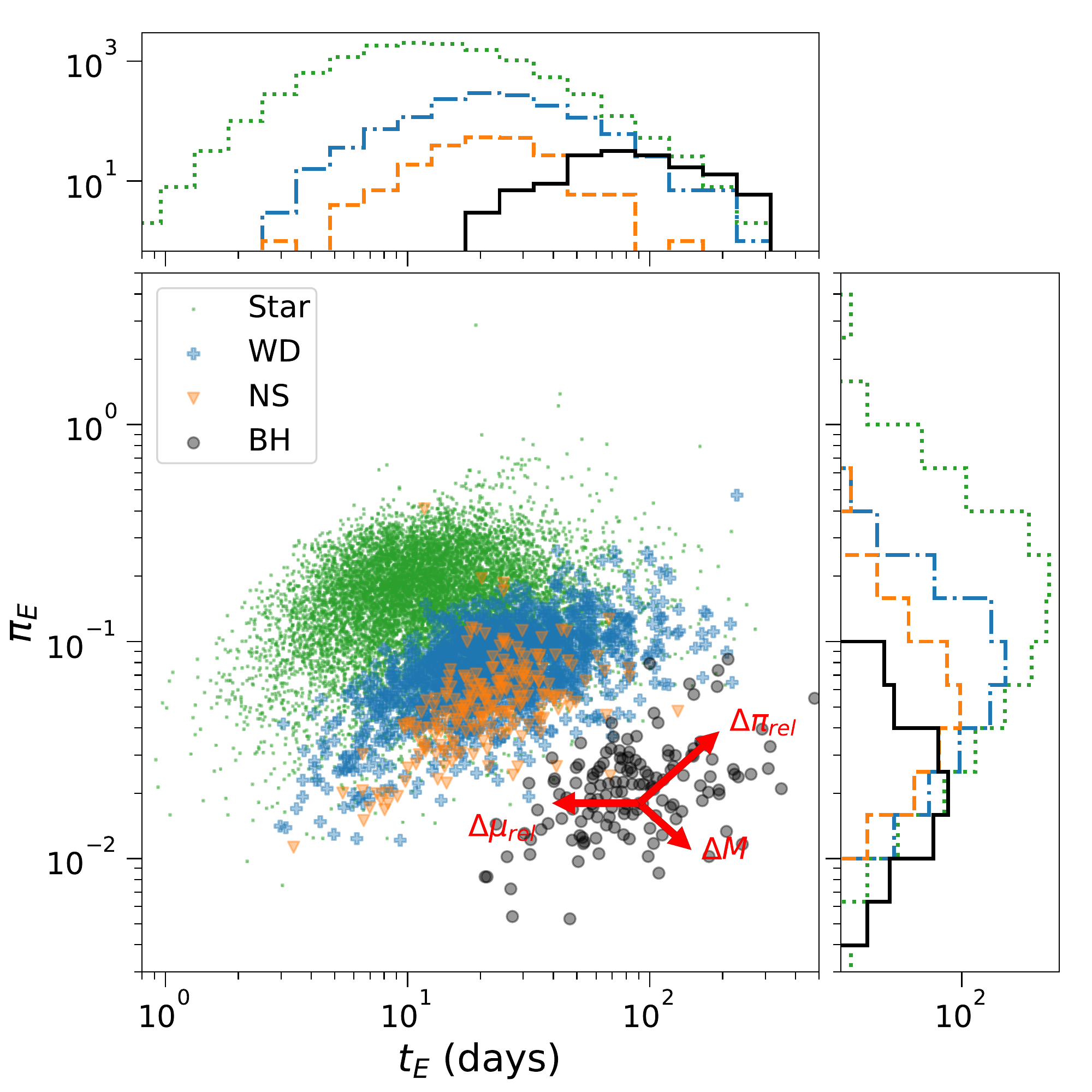}
    \caption{The microlensing parallax $\pi_E$ vs. the Einstein crossing time $t_E$.
    The points correspond to microlensing events in the Mock EWS simulation.
    The red arrows correspond to the effects of changing $\pi_{rel}$, $\mu_{rel}$, and $M$.
    Consider the fiducial parameters $d_L = 6.89$ kpc, $d_S = 8.62$ kpc, $\mu_{rel} = 6.55$ mas/yr, and $M = 11.12 M_\odot$; this corresponds to $t_E = 90.50$ days and $\pi_E = 0.018$.
    If the lens mass $M$ is increased to $30 M_\odot$ and all other parameters are held fixed, then $t_E = 148.64$ days and $\pi_E = 0.011$.
    If the relative proper motion $\mu_{rel}$ is increased to 15 mas/yr and all other parameters are held fixed, then $t_E = 39.53$ days and $\pi_E$ does not change.
    If the relative parallax $\pi_{rel}$ is increased from 0.029 mas to 0.13 mas by changing $d_L$ to 4 kpc and all other parameters are held fixed, then $t_E = 193.93$ days and $\pi_E = 0.039$.} 
\end{figure}
\label{fig:margin_piE_vs_tE}

\begin{figure}[t!]
    \centering
    \includegraphics[scale=0.4]{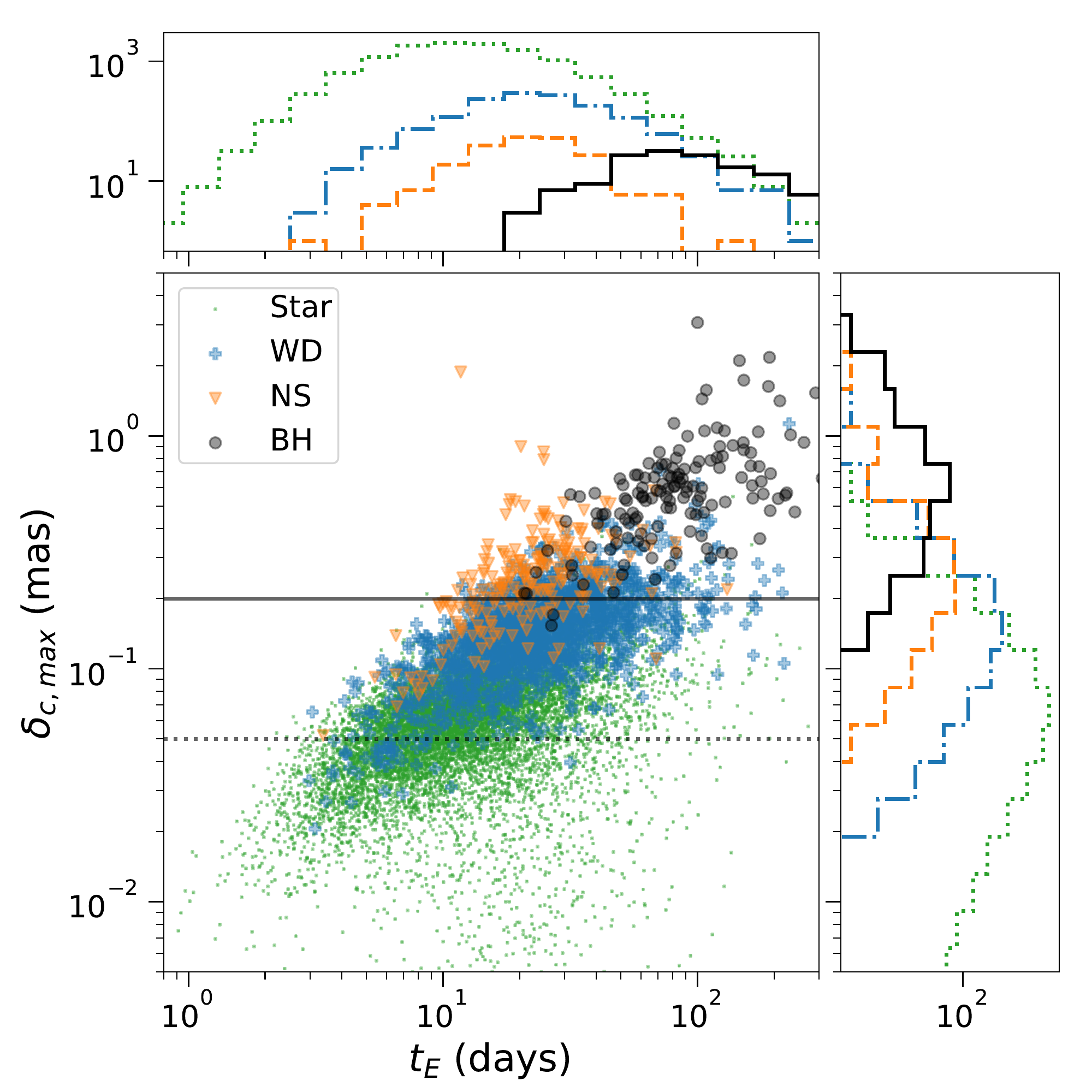}
    \caption{The maximum astrometric shift $\delta_{c,max}$ vs. the Einstein crossing time $t_E$.
    We assume blending between the lens and source when calculating $\delta_{c,max}$.
    The solid line denotes the achievable astrometric precision of $\sim 0.2$ mas using the Keck laser guide star adaptive optics system \citep{Lu:2016}.
    The dotted line denotes anticipated astrometric precision achievable in the next decade (e.g. $\sim 0.05$ mas, using \emph{WFIRST} or the Thirty Meter Telescope). 
    The points correspond to microlensing events in the Mock EWS simulation.}
\end{figure}
\label{fig:margin_deltac_vs_tE}

The very sharp delineation between the stars, WDs, NSs, and BHs in Figure \ref{fig:margin_deltac_vs_piE} can be simply explained.
Consider a lens of mass $M$.
Using Equations \ref{eq:theta_E}, \ref{eq:pi_E}, \ref{eq:pi_rel}, and the definition of $\kappa$ given in \S \ref{sec:Other Selection Parameters}, the microlens parallax will simply be
\begin{equation}
\label{eq:pi_E_2}
    \pi_E = \sqrt{\frac{\pi_{rel}}{\kappa M}}.
\end{equation}
Assuming an impact parameter of $u_0 \leq \sqrt{2}$, the maximum astrometric shift, assuming no blending, can be then written
\begin{equation}
\label{eq:delta_c_2}
    \delta_{c,max} = \sqrt{\frac{\kappa M \pi_{rel}}{8}}
\end{equation}
by combining Equations \ref{eq:theta_E}, \ref{eq:delta_c}, and \ref{eq:pi_rel}. 
Thus, for a given mass $M$, both $\pi_E$ and $\delta_{c,max}$ scale as $\sqrt{\pi_{rel}}$. 
Hence, when plotting $\delta_{c,max}$ against $\pi_E$ for a given mass lens, the slope is 1.
Since \texttt{PopSyCLE} currently assumes that all NSs are of the same mass, they lie artificially on a straight line.
Similarly, as there exists a minimum mass for BHs, WDs, and stars, there is a hard edge on the right diagonal side of those populations.
There is some downward scatter in $\delta_{c,max}$ due to some events with $\sqrt{2} < u_0 < 2$, as the maximum astrometric shift is less than what would be given in Equation \ref{eq:delta_c_2}.
Additionally when blending is included, $\delta_{c,max}$ scatters lower for some stars, depending on the lens luminosity.
The larger scatter from the slope of 1 in the BH, WD, and stellar populations are simply due to the fact that there are a range of masses.

Equations \ref{eq:pi_E_2} and \ref{eq:delta_c_2} can also be used to understand the $t_E$ gradient stretching from the bottom right (large $\pi_E$, small $\delta_{c,max}$) to the top left (small $\pi_E$, large $\delta_{c,max}$) shown in Figure \ref{fig:deltac_vs_piE_vs_tE}.
Since $t_E$ scales with the square root of the lens mass, heavier lenses will have smaller $\pi_E$, larger $\delta_{c,max}$, and longer $t_E$, on average.
Scatter in this relation is due to the fact that $t_E$ is also degenerate with the relative source-lens proper motion $\mu_{rel}$.

\subsection{Statistical Samples of Black Holes from Photometry Alone}

The microlensing parallax $\pi_E$ combined with a measurement of the Einstein crossing time $t_E$ appears to be a powerful means of picking out BHs (Figure \ref{fig:margin_piE_vs_tE}).
Unfortunately, $\pi_E$ can only be observed significantly after the photometric peak, and thus cannot be used as a BH-candidate selection criteria for astrometric follow-up.
However, the combination of $\pi_E$ and $t_E$ can still be used to obtain a sample of microlensing events with a very high fraction of BH lenses, as compared to looking at $t_E$ alone.
By selecting events with both $t_E > 120$ days and a microlensing parallax of $\pi_E < 0.08$, the detection rate of BHs is $\sim 85 \%$; by using an even lower value of $\pi_E$, the minimum value of $t_E$ could be shifted to lower values and still preserve the high fraction of BHs. 
This has the additional advantage that $t_E$ and $\pi_E$ are quantities that can be fit from photometry alone.
Thus, with a set of photometric lightcurves, events can be sorted into BH and non-BH lenses with high statistical confidence. 

This method is only useful if the $\pi_E$ distribution does not vary dramatically as the Galactic model changes.
It is not entirely clear how distinct this separation will be if different assumptions are made about the underlying distributions of $M$, $d_L$, $d_S$, and $\mu_{rel}$.
We evaluate the impact on the $\pi_E - t_E$ space by using the scaling relations,
\begin{align*}
    t_E &\propto \frac{\sqrt{\pi_{rel} M}}{\mu_{rel}} \\
    \pi_E &\propto \sqrt{\frac{\pi_{rel}}{M}}.
\end{align*}
as illustrated in the arrows on Figure \ref{fig:margin_piE_vs_tE}. 

Of all the populations in \texttt{PopSyCLE}, the most uncertainty lies in the BH spatial positions, velocities, and masses.
We consider the effects of changing these parameters.
The following trends are illustrated with specific values in Figure \ref{fig:margin_piE_vs_tE}.
\begin{itemize}
    \item Our BH mass distribution ranges from $5 M_\odot$ to 16 $M_\odot$; by including more high-mass BHs such as the ones discovered by LIGO into the distribution, the BH population would shift toward longer $t_E$ and smaller $\pi_E$.
    \item The kick velocities of BHs at birth, if there are any, is unknown. 
    Changes in kick velocity are manifested in changes in $\mu_{rel}$.
    If $\mu_{rel}$ increases (holding $d_L$ and $d_S$ fixed) $t_E$ becomes shorter; if $\mu_{rel}$ increases, $t_E$ becomes shorter. 
    \item The spatial distribution of BHs is also not known.
    Two extremes could be considered: a centrally concentrated population of BHs in the bulge, or a distribution spread throughout the stellar or dark matter halo.
    Bulge lenses have smaller $\pi_{rel}$, while disk lenses have a larger $\pi_{rel}$.
    Thus, having only BH bulge lenses would cause the average $\pi_{rel}$ to increase, causing both $\pi_E$ and $t_E$ to increase.
    On the other hand, with a non-centrally concentrated distribution of BHs, this would cause more lenses to fall closer to Earth, meaning the average $d_L$ and thus $\pi_{rel}$ would decrease, causing $\pi_E$ and $t_E$ to decrease.
\end{itemize}

Although this method of selecting BH lenses does not allow for direct confirmation nor mass measurement, it does allow for a statistical analysis of BH lensing events.
For example, the number of BHs in the Milky Way could be constrained; their masses could also be estimated by invoking some type of Galactic model, as is commonly done.
It's strength is that no additional data is necessary and the improved BH selection process can still lead to improved physical constraints.

The challenge associated with this method is constraining $\pi_E$ accurately down to the level which is necessary for this measurement.
However, with high cadence observations and high photometric precision this is possible.

\subsection{BH Hunting with WFIRST}
\label{sec: BH Hunting with WFIRST}

Lastly, we consider future microlensing surveys and the prospects they hold for measuring BH masses.
As described in \S \ref{sec:PopSyCLE Simulations}, the \emph{WFIRST} mission includes a microlensing survey designed for exoplanet detections. 
Some primary differences between the OGLE and \emph{WFIRST} surveys will be their filters, sensitivity, and resolution (seeing limited $I \lesssim 22$ vs. diffraction limited $H \lesssim 26$, respectively).
An additional important difference is that with \emph{WFIRST}, the astrometry will be obtained at the same time as the photometry.
Thus, it is not necessary to do individual follow-up of candidate BH events; all the information will be contained within the \emph{WFIRST} survey data.
Simultaneous astrometry also has the advantage that it is possible to go ``back in time" to look at astrometric data from before the photometric peak (i.e. before the microlensing event is recognized photometrically).

\citetalias{Penny:2019} presents thorough simulations and a detailed analysis to calculate the expected yield of bound planet detections with \emph{WFIRST} microlensing.
We present here a complementary analysis with \texttt{PopSyCLE} to determine the number of BHs we may expect a \emph{WFIRST}-like survey to find.
Detailed simulations of different season distributions and understanding how the BH yield changes is beyond the scope of this paper and is left as future work. 
However our continuous 1000 day survey simulation with Mock WFIRST parameters (Table \ref{tab:Simulation Parameters}) is a good order-of-magnitude estimation, considering the other uncertainties in the \texttt{PopSyCLE} simulation and the \emph{WFIRST} survey design itself.

To normalize the Mock WFIRST survey area to that of the actual \emph{WFIRST} microlensing survey, the number of events is multiplied by the ratio of the areas (a factor of 1.93).
Microlensing events from the Mock WFIRST simulation with BH lenses and an Einstein crossing time of $90 < t_E < 300$ days are defined to have ``measureable" BH masses.
Note that there will certainly be many BH microlensing events with $t_E$ that fall above or below this range; however, we do not consider those to have measureable masses, for the following reason.
As described in \S \ref{sec:Other Selection Parameters}, in addition to the photometric brightening, it is necessary to measure the astrometric shift and microlensing parallax to constrain the lens mass.
The choice for the lower bound on $t_E$ takes into the consideration that an event needs to have a somewhat long duration ($t_E \gtrsim 3$ months) to have a potentially measureable microlensing parallax $\pi_E$.
The choice for the upper bound on $t_E$ takes into consideration that the astrometric signal falls off much more slowly than the photometric signal; a rough guess is that to measure $\delta_{c,max}$ requires astrometric data for $\sim 5 - 10 t_E$.
Note the amount of astrometric data required also is dependent on when the astrometric measurements occur.
To properly determine this requires detailed simulation (e.g. \S 8 of \citet{Lu:2016}) and is beyond the scope of this paper.

With this definition, the \texttt{PopSyCLE} Mock WFIRST survey produces $\sim 1000$ BH lensing events with measureable masses, nearly 100 times more than with an individual astrometric follow-up program, as estimated in \S \ref{sec:Individual Candidate Follow-up}. 
With this number of mass measurements, we can constrain the present-day BH mass function (Figure \ref{fig:bhmf_detect}) and the results of supernova simulations.

\begin{figure}[t!]
    \centering
    \includegraphics[scale=0.5]{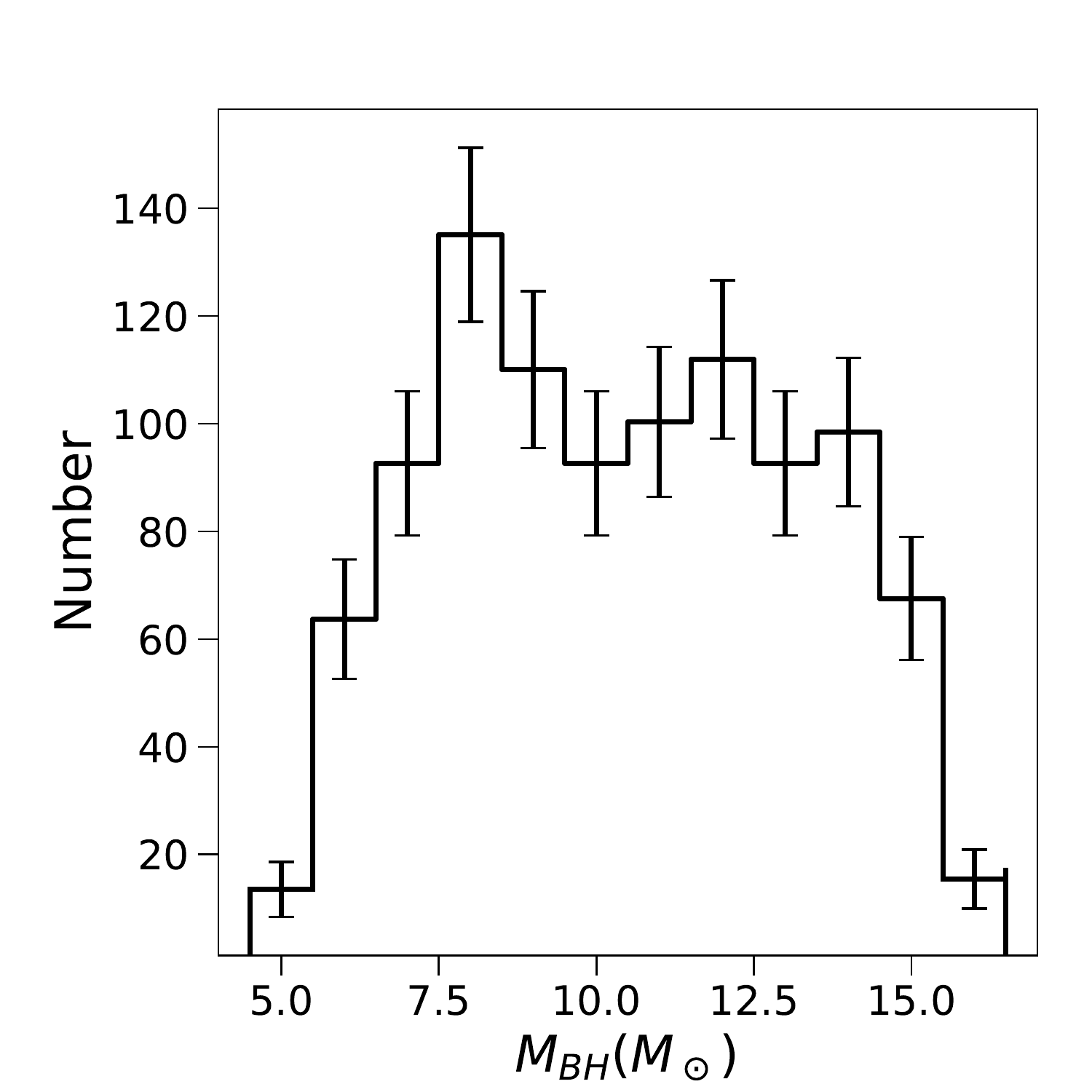} 
    \caption{Histogram of the BH masses that are measurable from the Mock WFIRST BH simulation, scaled to match the area of the actual \emph{WFIRST} survey. 
    To infer the underlying BH PDMF, an observational completeness correction will be required in order to account for the astrometric bias towards heavier lenses. 
    \texttt{PopSyCLE} is ideally suited for forward modeling populations, including completeness corrections.}
\end{figure}
\label{fig:bhmf_detect}

A major source of uncertainty in this estimate is due to the Galactic model, in both the stellar and compact object components.
As described in Appendix \ref{Appendix:Galaxia Galaxtic Model}, we consider two different angles ($\alpha = 11.1^\circ$ and $\alpha = 28^\circ$) for the Galactic bar/bulge; this modification significantly changes the number of stars along a given line of sight.
The more tilted bar with $\alpha = 11.1^\circ$ produces 4-5 times less microlensing event candidates than the less tilted bar with $\alpha = 28^\circ$.
By applying the Mock WFIRST criterion to the six fields in Table \ref{tab:Comparing PopSyCLE to surveys tilted bar}, the number of events with measureable BH masses is decreased by a factor of about 4. 
To take into account these uncertainties, we estimate the number of BHs with measureable masses is $\mathcal{O}(100-1000)$.

\section{Discussion}
\label{sec:Discussion}

\subsection{Comparison with other simulations}

\subsubsection{Microlensing Simulations}
Here we discuss several other recent microlensing simulations and compare them to \texttt{PopSyCLE}. 
\cite{Kerins:2009} generated synthetic maps of the microlensing optical depth, event rate, and average Einstein crossing time of the Galactic bulge, incorporating a 3-D extinction map. 
The \texttt{GULLS} code of \cite{Penny:2013, Penny:2019} improved upon this work by taking into account the effects of blending, while the \texttt{MaB$\mu$LS} code of \cite{Awiphan:2016} also included low-mass stars and brown dwarfs.  
All three of these simulations follow a similar method, as summarized in \citetalias{Penny:2019}, of drawing sources and lenses from a distribution described by the Besan\c{c}on model, then assigning weights proportional that source-lens pair's contribution to the total event rate along that sight line.
A constant correction factor to adjust for the number of sources and optical depth of the Galactic bulge as derived from the Besan\c{c}on model is also included to adjust the normalization of the event rate.
The Besan\c{c}on models used includes white dwarfs; however, neutron stars and black holes are not part of the models.

In particular, the \texttt{GULLS} code has been applied toward exoplanet microlensing survey designs for the \emph{Euclid} and \emph{WFIRST} missions.
Toward this end, it models single planets orbiting a single host star.
Properties of the detector and filters are included in the simulation, and \texttt{GULLS} can also generate images and lightcurves which includes Gaussian noise. 
Planetary detections were then evaluated on a $\Delta \chi^2$ criterion as is commonly done in other exoplanetary microlensing surveys and simulations.
However, there are discrepancies between the simulation and observed event rates, as there is substantial uncertainty in the Galactic models used.
As one of the purposes of \texttt{GULLS} is to estimate yields for microlensing survey missions, the results are rescaled by a factor to match observed star counts and optical depth \citep{Penny:2013} or star counts and event rates \citepalias{Penny:2019}.

In comparison, \texttt{PopSyCLE} is designed to understand how changes in the stellar and compact object population and imposition of observational selection criteria modify the underlying and observed microlensing event distribution.
It is modular in design to allow different initial-final mass relations, dust maps, observational cuts, etc.~to be used.
\texttt{PopSyCLE} is also unique in its emphasis for studying lensing by massive compact objects, specifically black holes, and its calculation of not just photometric, but also astrometric microlensing quantities.
The current version of \texttt{PopSyCLE} is not designed to generate lightcurves including complexities such as detector noise or atmospheric seeing, nor is it designed/optimized to probe the short-$t_E$ end of the distribution, as we are interested in the long-$t_E$ events.
Like \texttt{GULLS}, \texttt{PopSyCLE} also has a discrepancy in the event rates as compared to observations.
However, we choose to not rescale our event rates to match observation. 
Although this might limit \texttt{PopSyCLE}'s predictive power, we consider the discrepancies to clearly indicate limitations in our understanding of the physics of the simulations, corrections to our data, or both (e.g. Section 5.1 of \citet{Awiphan:2016}).
For this reason we present our \emph{WFIRST} BH yield as an order-of-magnitude estimate.
However, the \texttt{PopSyCLE} simulation results are available for download and can be rescaled as desired for survey yield predictions.

\subsubsection{Compact Object Population Synthesis}

There are several other compact object population synthesis packages. 
For example, a widely used package is the Stellar EVolution $N$-body (\texttt{SEVN}) code\footnote{http://web.pd.astro.it/mapelli/group.html}, which combines a single stellar evolution code along with several core-collapse supernovae models, pair-instability and pair-instability pulsational supernovae, along with many different binary evolution recipes \citep{Spera:2015}.
We note \texttt{SEVN} is purely a population synthesis model, that can (and has) been interfaced with N-body stellar dynamics codes, to study interactions in star clusters, for instance. 
It does not include a full model of the Milky Way, nor microlensing.
One thing to note is that the IFMRs in \texttt{SEVN} are heavily simplified analytic models that do not incorporate explosion physics.
Ultimately, it would still be worth adding support for \texttt{SEVN} inside \texttt{PopSyCLE} in addition to the current \texttt{PyPopStar} stellar evolution code; however, we chose \texttt{PyPopStar} initially due to its support for a larger range of models, flexibility, and Python implementation.

\subsection{Comparison with on-sky microlensing surveys}
\label{sec:Comparison with on-sky microlensing surveys}

We advocate for defining a microlensing event using quantities that are directly observable/measured rather than fit.
This is described in \cite{Dominik:2009} as a reparametrization when fitting microlensing events.
For example, in the survey papers discussed, cuts were made on events whose source magnitude, $m_S$ is fainter than some value, where $m_S$ is generally set by the magnitude limit of the telescope and camera.
However, the more easily observed quantity is the baseline magnitude $m_{base}$ since $m_S$ is generally a parameter derived from fitting and is subject to degeneracy with other parameters \citep{Dominik:2009}.
When fitting light curves, blending is degenerate with the observed timescale and peak magnification of the event, and if blending is not taken into account correctly the microlensing parameters derived will be incorrect; masses and timescales will be systematically underestimated \citep{DiStefano:1995, Wozniak:1997, Han:1999}.
In other words, events with small $t_E$, large $u_0$, and large $b_{SFF}$ are degenerate with events with large $t_E$, small $u_0$, and small $b_{SFF}$ \citep{Sumi:2011}.
In order to facilitate easier comparison between simulations and observations and explore this degeneracy, we recommend that future microlensing analyses adopt cuts using observable quantities, as advocated for in \cite{Dominik:2009}.
However, we also suggest that such observational cuts be applied for sample selection, as these are more easily reproduced and are less dependent on differences in fitting codes and prior assumptions.

Additionally, although the mean Einstein crossing time $\langle t_E \rangle$ is the most commonly reported parameter by surveys, the $t_E$ distribution is asymmetric in linear $t_E$ bins\footnote{$t_E$ is often plotted in bins of $\log t_E$ which makes the distribution look more symmetric.}; thus, the mean is easily skewed and depends on the range of $t_E$ used to estimate the mean. 
For future comparisons across observational surveys and simulations, the median is a better choice and is less impacted by particular cuts. 
It is also important to note the range of $t_E$ over which the distribution is made (e.g. as in \citet{Sumi:2013}).

\subsection{Hunting for BHs}
\label{sec:Hunting for BHs}

Over the years, many types of follow-up observations have been proposed and used to constrain lens properties. 
\cite{Agol:2002} proposed to constrain lens masses of BH microlensing candidates by searching for their X-ray emission due to accretion from the interstellar medium or stellar winds.
\cite{Nucita:2006} and \cite{Maeda:2005} used \textit{XMM-Newton} and \textit{Chandra} observations, respectively, to search for X-rays from the extremely long-duration BH candidates MACHO-96-BLG-5 \citep{Mao:2002}; however, no significant detections were made.

HST high-resolution imaging follow-up of BH candidates has also been used to measure the degree of blending \citep{Bennett:2002}. 
As discussed in \S \ref{sec:Selecting BH Candidates with OGLE}, this is not particularly useful for selecting photometric candidates for astrometric follow-up; however, it still may be a useful means of confirming that the lens is not a massive star.
\citet{Poindexter:2005} also suggested that HST observations could be used to measure the source proper motion; however they note that many degeneracies remain even with this measurement.
High-resolution images can be taken many years after an event, when it may be possible to resolve the source and lens \citep[][Abdurrahman et al.~in prep.]{Batista:2015, Bennett:2015, Alcock:2001b} and the absence of a lens may provide additional confirmation of a BH.

For nearby sources, it is sometimes also possible to measure the source distance via parallax, which would give complete event parameters. 
The combination of photometry and astrometry is powerful not only for BH lenses, but also for obtaining precise mass measurements of any type of lens, as was shown with the first astrometric microlensing signal detected outside our Solar System \citep{Sahu:2017}.
Another method of breaking degeneracies is to have the ability to resolve the images themselves, which is possible for stellar-mass lenses using interferometric techniques \citep{Dong:2019}.

We have not yet considered the number of BHs detectable with the \emph{Gaia} satellite, an astrometric space mission which has been operating since 2013 \citep{Gaia:2016}.
Although \emph{Gaia} has incredible astrometric precision for parallax measurement (down to $\sim 10\;\mu$as for some stars by the end of the mission), it does not perform as well for single-epoch astrometric measurements in the bulge due to the decreased observing cadence as well as significant stellar crowding ($> 1$ mas, \citet{Rybicki:2018}).
Additionally, \emph{Gaia} observes at green optical wavelengths, which cannot probe dusty, high extinction regions like the bulge.
However, there are estimates of the number of BHs that \emph{Gaia} will be able to detect over it's lifetime.
\citet{Mashian:2017} estimated that $\sim 2 \times 10^5$ BHs in astrometric binary systems can be detected over \emph{Gaia}'s 5 year mission; however, this is a severe overestimate, as this calculation has neglected extinction and crowding effects.
\citet{Rybicki:2018} estimated that a few isolated stellar mass BHs should be detectable astrometrically at the end of the \emph{Gaia} mission.
\citet{Wyrzykowski:2019} used \emph{Gaia} DR2 data to calculate distances and proper motions for sources in OGLE-III microlensing events from 2001 to 2009, providing additional information to perform a more careful reanalysis of the lens masses to determine whether they could be BHs.

In \S \ref{sec:Results}, two different methods to better understand BHs in the Milky Way are discussed.
The first involves using a combination of astrometry and photometry to measure $t_E$, $\pi_E$, and $\delta_{c,max}$, which allows lens masses to be measured for \emph{individual} microlensing events.
The second involves photometry only to measure $t_E$ and $\pi_E$, enabling statistical constraints on a \emph{population} of BHs.
Although the BH nature of individual lenses cannot be confirmed, ensemble information can still be gleaned.
To obtain masses with the second method would involve assuming some type of Galactic model or spatial distribution.
It is interesting to note that searching for a small/undetectable parallax signal to identify BHs is the opposite of previous approaches in the literature \citep{Poindexter:2005, Wyrzykowski:2016, Drlica-Wagner:2019}. 

\subsection{Future Work}
\label{sec:Future Work}

In the next version of \texttt{PopSyCLE}, we plan to add support for stellar and compact object binary systems, compact object mergers, metallicity-dependent IFMRs, different compact object spatial distributions, and primordial BHs. 
This will allow for a more realistic simulation and a larger exploration of parameter space.
The lack of these features at the present put some caveats on this work, which we discuss.

\subsubsection{Binarity and Mergers}
\label{sec:Binarity and Mergers}

Although all of the comprehensive catalogs of microlensing events in the literature from which event rates and optical depths are calculated are only for events which can be fit by point source point lens (PSPL) models \citep{Sumi:2016, Wyrzykowski:2015, Mroz:2017}, it is estimated that that binary lenses will consist of around 10\% of Galactic bulge stars lensing events \citep{Mao:1991}. 
It is difficult to ascertain the binary fraction from the observed distribution of microlensing events because binary lenses can produce lightcurves that resemble single lens events.
For a widely separated binary, only one of the lenses might be lensing the source star, while for a closely separated binary, both lenses act as a single more massive lens.
Moreover, the addition of another lens greatly increases the complexity and variety of lightcurve shapes, depending on how the source approaches and/or crosses the caustics, making binary lens events difficult to identify.
Additionally there is the issue of binary source stars.
Similar to binary lens events, binary source events can produce lightcurves that appear quite similar to PSPL events \citep{Dominik:1998, Han:1998}, although these degeneracies can be broken by the addition of astrometric information \citep{Nucita:2016}.

\texttt{Galaxia} does not have support for binaries; however, most massive stars are in binary systems \citep{Duchene:2013}.
Additionally, our current population synthesis method does not include compact object binaries (either compact-compact or compact-stellar).
\texttt{PyPopStar} currently has a heuristic prescription for stellar multiplicity, where stars are taken from single-stellar evolution models and combined into multiple systems based on multiplicity fraction.
However, this does not take into account the effect that mass exchange has on the evolution of stars in multiple systems.
For main sequence stars, such effects are likely negligible; however, in the later stages of stellar evolution, binary evolution cannot be neglected.
For a more rigorous treatment of stellar multiplicity, binary evolution models will be incorporated into \texttt{PyPopStar}.
Additionally, future work on binary IFMRs will also be added to \texttt{PyPopStar}. 

Adding support for binaries (stellar-stellar, compact-stellar, and compact-compact) into \texttt{PopSyCLE} would help us put further constraints on the number of BHs in the Milky Way and understand BH formation channels.
For example, both single and binary BHs may form from binary star systems; single star systems can be formed when the stellar binary is disrupted or merges.
If the binary is disrupted, this could impart large kick velocities to the BHs; the frequency at which disruption occurs depends on the assumptions made about the kicks (\citet{Wiktorowicz:2019} and references therein).
The spatial distribution will also be affected by disrupted binaries and the associated kicks; the scale height distribution of X-ray binaries can act as a proxy for different compact object formation mechanisms \citep{Repetto:2017}.

Another possibility for forming more massive stellar mass BHs is through the mergers of less massive BHs.
By adjusting the merger rate and fraction, we can adjust the mass spectrum and fraction of BHs in single and binary systems.
Adding support for compact-object mergers would allow us determine our ability to set constraints on the merger rate and fraction, which would again improve our understanding of BH formation channels and LIGO merger events.

\subsubsection{Metallicity}
\label{sec:Metallicity}

When performing population synthesis, metallicity has not been taken into account.
Although \texttt{Galaxia} and \texttt{PyPopStar} have metallicity support, the \citetalias{Raithel:2018} IFMR does not.
With the IFMR, metallicity can have a significant effect on the mass and type of the compact remnant.
In particular, having [Fe/H] $>$ [Fe/H]$_\odot$ will not particularly change the IFMR, while having [Fe/H] $<$ [Fe/H]$_\odot$ will produce more high-mass black holes.
New models to be used in the IFMR have been run to include a metallicity dependence (T. Sukhbold, private communication.)
A non-mutually exclusive option would be to add other IFMRs that have a dependence on both the progenitor mass and metallicity, such as the one in Appendix C of \citet{Spera:2015}.

In \texttt{Galaxia}, the metallicity of the bulge population is centered at solar metallicity with a spread $\sigma_{[Fe/H]} = 0.4$ (Table 2, \citet{Sharma:2011}), and with respect to the current iteration of \texttt{PopSyCLE}, about 80\% of stars within 10 kpc of Earth in the bulge direction have $-0.5 < $ [Fe/H] $ < 0.5$, where the IFMR likely does not change significantly.
Thus, the solar-metallicity approximation is reasonable for the bulk statistics; the largest effect would be a slight deficit of higher-mass BHs.

\subsubsection{Compact object distributions and kinematics}

In \texttt{PopSyCLE}, it is assumed that the positions of the compact objects followed the stellar spatial distribution. 
Additionally, the velocity of the object is assumed to be the stellar progenitor velocity and kick velocity added together; however, this is only true at the time the compact object is born.
This choice is roughly justified given that the number of objects leaving a region is the same as the number of objects entering that region, since the direction of the kick velocity is random.
However, effects like dynamical friction might cause black holes to settle toward the Galactic Center.
Conversely, supernovae may provide large enough kicks to unbind neutron stars from the Galactic disk.
In future iterations of \texttt{PopSyCLE}, we will allow for different compact-object spatial distributions.
Currently, the kick velocities for both NSs and BHs is tunable; however, it is only allowed to be a single value.
In the future, support for kick velocity distributions will be added.

The additions above are necessary to provide support for primordial black holes (PBHs) in \texttt{PopSyCLE}.
Ultimately, we plan to implement a PBH mass spectrum such that PBHs can be injected with their own position, velocity, and mass distributions that differ from the underlying stellar halo \citep{Carr:2016, Chapline:2018}.
Similar to the IFMR, a variety of different mass spectra could be implemented.

\section{Conclusions}
\label{Conclusions}

We have developed a Milky Way microlensing simulation, dubbed \texttt{PopSyCLE}, which is the first to consider both photometric and astrometric microlensing effects and perform compact object population synthesis in a realistic manner.
With \texttt{PopSyCLE}, we investigate different strategies to hunt for isolated stellar mass BHs in the Milky Way and measure their masses. 
We highlight the following results:
\begin{itemize}
    \item Assuming a state-of-the-art initial-final mass relations, the BH present-day mass function has structure (peaks and gaps) that can be measured with samples of $\mathcal{O}(100)$ BH mass measurements from both LIGO detections and BH microlensing surveys.
    \item The optimal isolated BH candidates to follow up astrometrically are long-duration microlensing events with Einstein crossing times $t_E \gtrsim 90 - 120$ days; additional selection criteria based on the source flux fraction $b_{SFF}$ or the impact parameter $u_0$ or obtaining high-resolution imaging does not significantly improve the outcome.
    Current photometric surveys and astrometric follow-up campaigns should yield a 40\% success rate for measuring BH masses.
    \item From photometry alone, BHs can be identified in a statistical manner with a combined measurement of $t_E$ and the microlensing parallax $\pi_E$.
    The BH detection rate can be raised to $\sim 85 \%$ by selecting events with both $t_E>120$ days and $\pi_E<0.08$. 
    \item BH lenses are easily distinguished from stellar lenses with a combined measurement of the microlensing parallax $\pi_E$ and the maximum astrometric shift $\delta_{c,max}$, providing a useful method for confirming the BH nature of the lens and measuring its mass. In particular, we note that BH lenses nearly always have $\pi_E<0.1$.
    \item The \emph{WFIRST} microlensing survey will be able to measure the masses of $\mathcal{O}(100-1000)$ isolated BHs over it's 5 year lifetime, which is at least an order of magnitude more than is possible with individual astrometric follow-up. 
    This is sufficient to constrain the BH IFMR, binary fraction, and kick velocity distribution.
\end{itemize}
\texttt{PopSyCLE} can also be used to forward model microlensing survey results and constrain the properties of compact objects, the initial-final mass relation, Galactic structure, the existence of primordial black holes, and more.
\texttt{PopSyCLE} can be downloaded from \url{https://github.com/jluastro/PopSyCLE} and community contributions are welcome.
We also provide the simulation files used for this work at \url{https://drive.google.com/open?id=12ALPCqMVOBN54fy1YhHfiJjI9Y1bnlQj}.

\section{Acknowledgements}

We thank the PALS collaboration and Will Clarkson for helpful conversations.
We also thank the anonymous referee for comments that improved this paper.
C.~Y.~L.~and J.~R.~L. acknowledge support by the National Aeronautics and Space Administration (NASA) under Contract No. NNG16PJ26C issued through the WFIRST Science Investigation Teams Program.
Portions of this work were performed under the auspices of the U.S. Department of Energy by Lawrence Livermore National Laboratory under Contract DE-AC52-07NA27344 and supported by the LLNL-LDRD Program under Project No. 17-ERD-120.

\software{Galaxia \citep{Sharma:2011}, astropy \citep{Astropy:2013, Astropy:2018}, MIST v1.2 \citep{Choi:2016, Dotter:2016}, Matplotlib \citep[][DOI:10.1109/MCSE.2007.55]{Hunter:2007}, NumPy \citep[][DOI:10.1109/MCSE.2011.37
]{vdWalt:2011}, SciPy \citep{Scipy:2019}, PyPopStar (Hosek et al. in prep, https://github.com/astropy/PyPopStar), PopSyCLE (this work, 10.5281/zenodo.3464129).}

\appendix

\section{\texttt{Galaxia} Galactic Model}
\label{Appendix:Galaxia Galaxtic Model}

There are several known issues in the Besan\c{c}on model, and by extension \texttt{Galaxia}, particularly with the bulge. 
Many of these are discussed in \S 6.2 of \citetalias{Penny:2019}, albeit for a slightly different version of the Besan\c{c}on model than that implemented in \texttt{Galaxia}.
We modify the bulge kinematics of \texttt{Galaxia}, specifically the pattern speed and velocity dispersions, in an attempt to ameliorate these issues.
We also validate these changes by performing some comparisons to star counts and event rates from \citet{Mroz:2019b} and \citet{Sumi:2016}.

\subsection{Galactic Models Comparison}

There are three specific properties of the bulge we consider modifying:
\begin{itemize}
    \item \emph{Pattern speed}: \texttt{Galaxia} implements a pattern speed of $\Omega = 71.62$ km/s/kpc (see \S 3.3 in \citet{Sharma:2011}).
    However, this is nearly twice as fast as the most recent values reported in the literature, which range from around 36 to 44 km/s/kpc \citep{Bovy:2019, Clarke:2019, Sanders:2019}, determined using combinations of Gaia DR2, VVV, and APOGEE.

    \item \emph{Velocity dispersion}: \texttt{Galaxia} implements velocity dispersions of $\sigma_R$ = $\sigma_\phi = 110$ km/s.
    However, this produces microlensing events with timescales that are too short, which suggests smaller velocity dispersions might be more appropriate.\footnote{This issue of too short microlensing events is also a problem described in \S 6.2.2 of \citet{Penny:2019}, who implement the Besan\c{c}on model to study microlensing, although it should be noted that the versions of the Besan\c{c}on model in \citet{Penny:2019} and \texttt{Galaxia} are slightly different.}
    In reality, $\sigma_R$ and $\sigma_\phi$ should vary with latitude and longitude \citep{Howard:2008, Howard:2009}, but to implement this in \texttt{Galaxia} would require significant rewriting of the code, which is far beyond the scope of this current work.

    \item \emph{Bar angle and length}: \texttt{Galaxia} implements a bar angle of $\alpha = 11.1^\circ$, where $\alpha$ is the angle from the Sun-Galactic Center line-of-sight, and a major axis scale length of $x_0 = 1.59$ kpc.
    It is suggested that the angle should be closer to $\alpha = 28^\circ$ \citep{Wegg:2013, Wegg:2015}, with a shorter scale length of 0.7 kpc.
    In fact, \cite{Portail:2017} performed dynamical modeling using $\alpha = 28^\circ$ to find a bar pattern speed of around 40 km/s/kpc.
    However, there is still considerable debate in the literature about the value of $\alpha$ and the scale length, with values for $\alpha$ ranging from $10^\circ - 45^\circ$ (see \citet{Robin:2012} for a summary and references).
\end{itemize}
We create two new variations of the original Galactic model implemented in \texttt{Galaxia}.
We dub ``v2" to be the version of \texttt{Galaxia} with a pattern speed $\Omega = 40$ km/s/kpc and bulge velocity dispersions $\sigma_R = \sigma_\phi = 100$ km/s.
We dub ``v3" to be the version where the bar is short and tilted, with $\Omega = 40$ km/s/kpc, bulge velocity dispersions $\sigma_R = \sigma_\phi = 100$ km/s, bar angle $\alpha = 28^\circ$ and major axis scale length $x_0 = 0.7$ kpc. 

Tables \ref{tab:Comparing PopSyCLE to surveys} and \ref{tab:Comparing PopSyCLE to surveys tilted bar} compare \texttt{PopSyCLE} stellar densities, event rates, and Einstein crossing times to results from \citetalias{Sumi:2016} and \citetalias{Mroz:2019b} for v2 and v3, respectively.
Figure \ref{fig:popsycle_vs_survey} summarizes the results of the two tables together.
The stellar densities of v2 match reasonably well with the observed number counts; v3 is consistently too low.
Note that although in projection the length of the bar is the same, as $\mathrm{sin}(11.1^\circ) \cdot 1.59 \, \mathrm{kpc} \approx \mathrm{sin}(28^\circ) \cdot 0.7 \, \mathrm{kpc}$ \citepalias{Penny:2019}, the number counts are quite different, due to extinction.
The event rates for v3 match reasonably well with the observed rates; v2 is consistently too high.
With respect to the average Einstein crossing time, it is not as clear whether v2 or v3 matches the observed values better.

\begin{figure}[h!]
    \centering
    \includegraphics[scale=0.4]{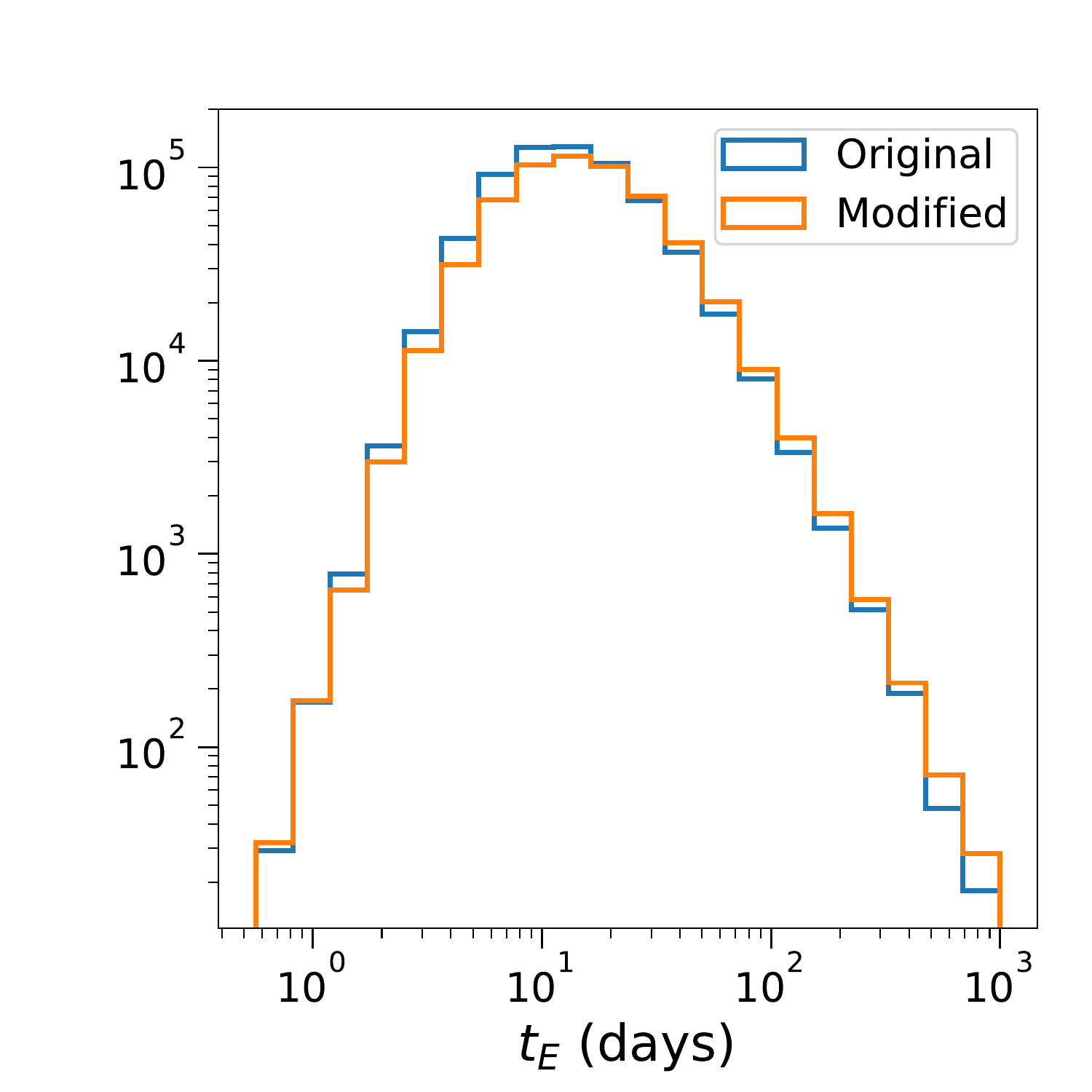}
    \caption{$t_E$ distribution of microlensing event candidates (i.e. without any selection cuts applied).
    The ``Original" curve was generated using \texttt{Galaxia} unmodified.
    The ``Modified" curve was generated using \texttt{Galaxia} v2 with $\Omega = 71.62$ km/s/kpc and $\sigma_R = \sigma_\phi = 110$ km/s.
    It can be seen that the modified \texttt{Galaxia} has more long $t_E$ events and fewer short $t_E$ events.}
\end{figure}
\label{fig:old_vs_new_tE}

Based on this analysis we chose to use v2 instead of v3 for the analysis in the main text of this paper.
Stellar density is the more fundamental observational quantity, hence we prefer the simulation reproduce this aspect accurately.
The event rate is microlensing-specific and dependent on many more factors (e.g., the detection efficiency correction).
However, it is curious in and of itself that the tilted bar creates agreement in one regime but not the other.
Additional modifications (such as the 3-D $E(B-V)$ map) may be necessary to bring the models into better agreement with observation.
We do not explore this further here, and leave detailed Galactic modeling to the investigation of future work.
However, it is worth noting how this uncertainty affects some of the results of this paper.
In \S \ref{sec:Selecting BH Candidates with OGLE}, the fraction of BH events at long times can be up to a factor of 2 higher for v3 than for v2.
In \S \ref{sec: BH Hunting with WFIRST} we consider the number of BH masses WFIRST can measure; the number is about a factor of 4 higher for v2 than for v3.

The rest of the analysis and validation performed in this appendix is also using v2.

\begin{deluxetable}{lCCCCCC}
\tablecaption{Comparing \texttt{PopSyCLE} to surveys (\texttt{Galaxia} v3) \label{tab:Comparing PopSyCLE to surveys tilted bar}}
\tablehead{
    \colhead{Field} &
    \multicolumn{2}{c}{n$_s$ (10$^6/\mathrm{deg}^2$)} &
    \multicolumn{2}{c}{$\Gamma$ (10$^{-6} \textrm{/star/yr)}$} &
    \multicolumn{2}{c}{$\langle t_E \rangle \textrm{(days)}$} \\
    \colhead{} &
    \colhead{Obs.} &
    \colhead{Mock} &
    \colhead{Obs.} &
    \colhead{Mock} &
    \colhead{Obs.} &
    \colhead{Mock}}
\startdata
F00$^{M19}$ & 2.62 & 1.48 & 1.3 \pm 0.8 & 4.35 & 32.6 & 14.8 \\
F01$^{M19}$ & 4.54 & 2.04 & 5.5 \pm 0.9 & 3.69 & 39.5 & 46.2 \\
F03$^S$ & 3.64 & 2.47 & 14.0$^{+2.9}_{-2.4}$ & 8.26 & 25.5 & 20.3 \\
F11$^{M19}$ & 4.95 & 3.66 & 16.2 \pm 1.3 & 17.88 & 21.8 & 19.7 \\
F12$^{M19}$ & 3.26 & 1.49 & 3.4 \pm 1.1 & 2.88 & 36.7 & 51.1 \\
F13$^{M19}$ & 4.51 & 2.19 & 5.2 \pm 1.1 & 4.9 & 30.8 & 28.2 \\
\enddata
\tablecomments{Identical analysis as presented in Table \ref{tab:Comparing PopSyCLE to surveys}, but using \texttt{Galaxia} v3 (with the tilted shorter bar) instead of v2.
Fields with $^{M19}$ indicate the observed values come from \citetalias{Mroz:2019b}, while those with $^S$ are from \citetalias{Sumi:2016}.}
\end{deluxetable}

\begin{figure}[h!]
    \centering
    \includegraphics[scale=0.4]{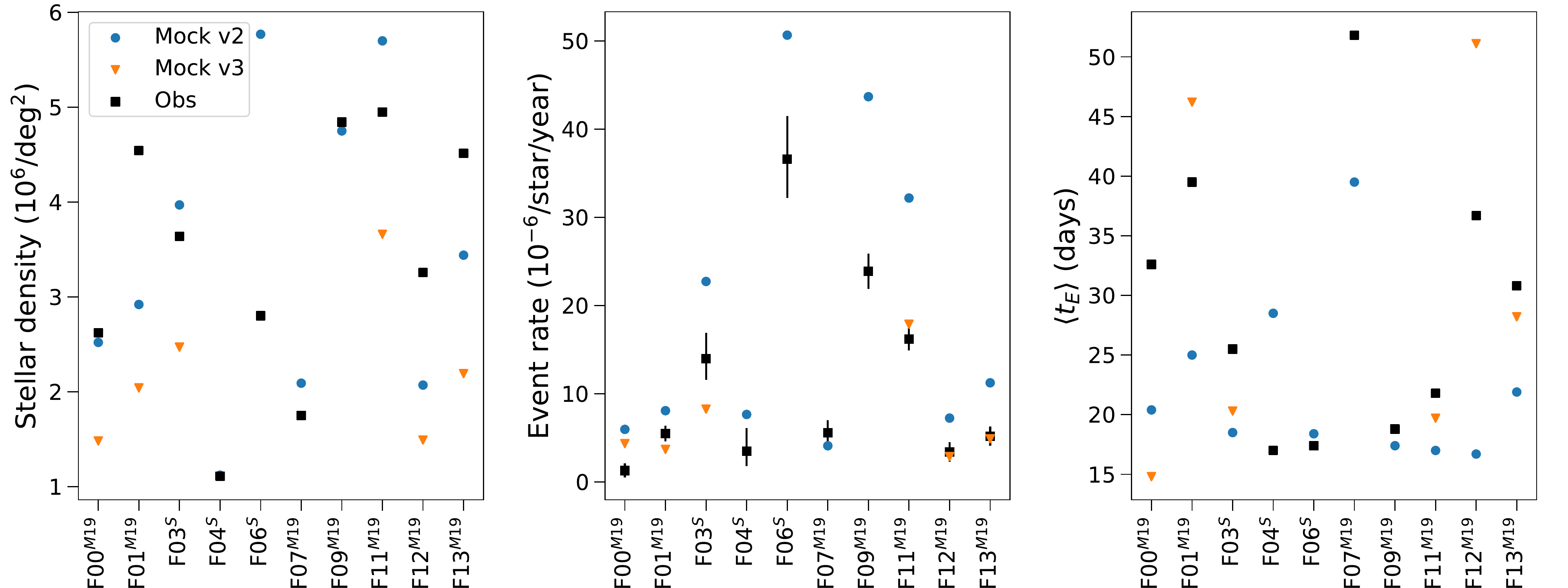}
    \caption{Comparison of the stellar density, event rate, and average Einstein crossing times for several fields.
    The labels F$XX$ correspond to the field used (see Table \ref{tab:PopSyCLE Fields}), and the superscript correspond to the paper from which observed values and selection criteria for the mock values were drawn (S from \citet{Sumi:2016}, M19 from \citet{Mroz:2019b}).
    Note that only a handful of fields were initially tested, which is why there are certain fields for v3 that have no points.}
\end{figure}
\label{fig:popsycle_vs_survey}

\subsection{Bulge Kinematics}

Following \S 6.2.2 of \citetalias{Penny:2019}, we compare the results of our simulation to observational studies of bulge kinematics.
In \citet{Clarkson:2008}, a study of proper motions in the Galactic Bulge is performed using observations of the HST SWEEPS field (centered at $(l,b) = (1.25, -2.65)$ covering an area of 11 arcmin$^2$) in the HST F606W and F814W bands.
To compare, we created a synthetic survey in the same direction of the same area using \texttt{Galaxia}; we use I and R band as these have the closest effective wavelength as the HST F606W and F814W bands.
First, we select stars in \texttt{Galaxia} photometrically (Figure \ref{fig:SWEEPS_CMD}) similarly to \citet{Clarkson:2008} to obtain a red population (bulge proxy) and a blue population (disk proxy).
We obtain 347 blue stars and 699 red stars (c.f. \citetalias{Penny:2019} 37 blue stars and 105 red stars, drawing from an area of 1.44 arcmin$^2$).

\begin{figure}[h!]
    \centering
    \includegraphics[scale=0.4]{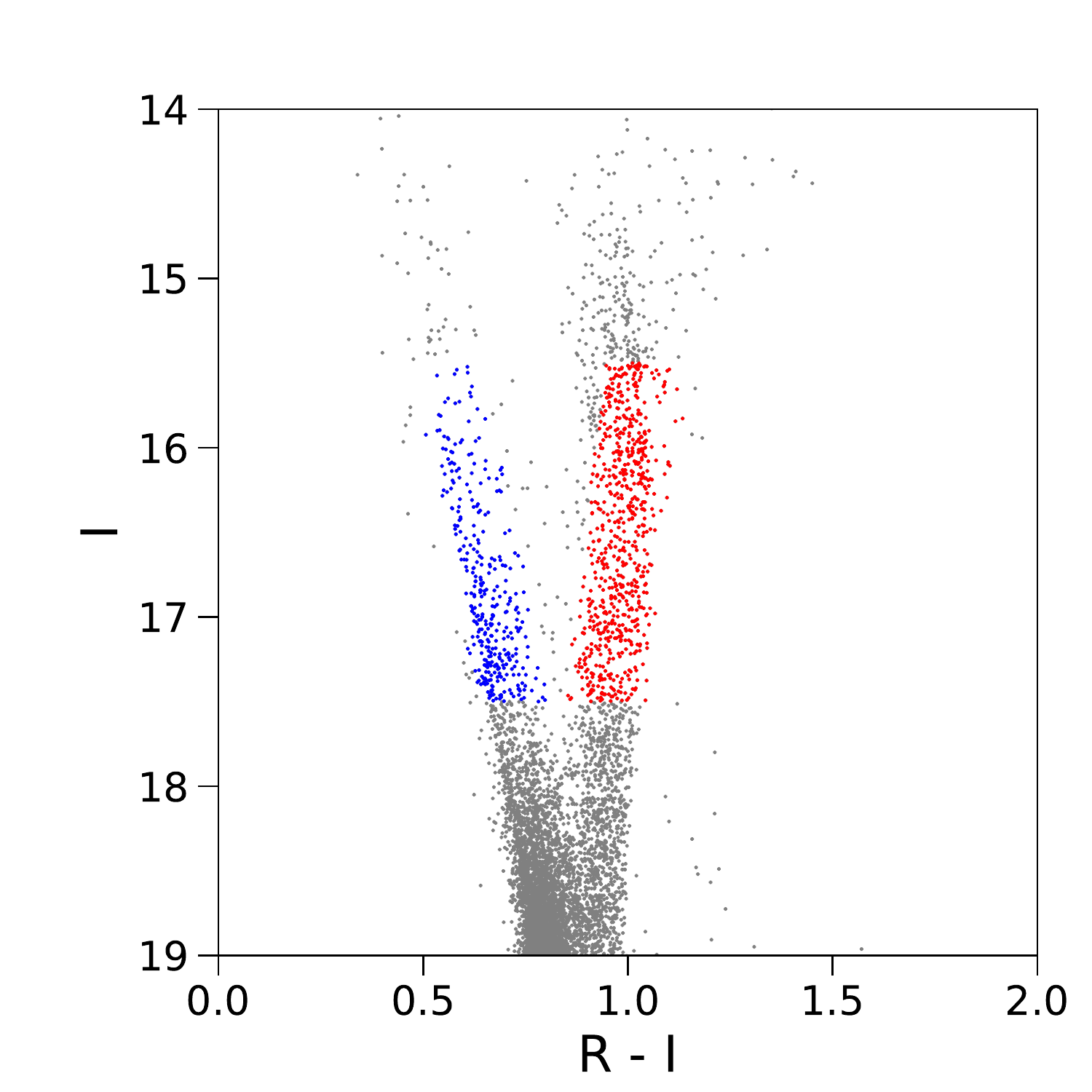}
    \caption{Photometrically selected red and blue populations, c.f. \citet{Clarkson:2008} Figures 8 and 9.
    The gray points correspond to all stars, while the red points correspond to the proxy bulge population and the blue points correspond to the proxy disk population.}
\end{figure}
\label{fig:SWEEPS_CMD}

We then compare our results to those of \citet{Clarkson:2008} and \citetalias{Penny:2019}, who use a different version of the Besan\c{c}on model, named BGM1106 for short.
The results are summarized in Table \ref{tab:PM compare}, and illustrated in Figures \ref{fig:SWEEPS_Penny} and \ref{fig:SWEEPS_Clarkson}.
In particular, the match between \citet{Clarkson:2008} is improved in $\sigma_{l,*}$ for the red population.

\begin{figure}[h!]
    \centering
    \includegraphics[scale=0.4]{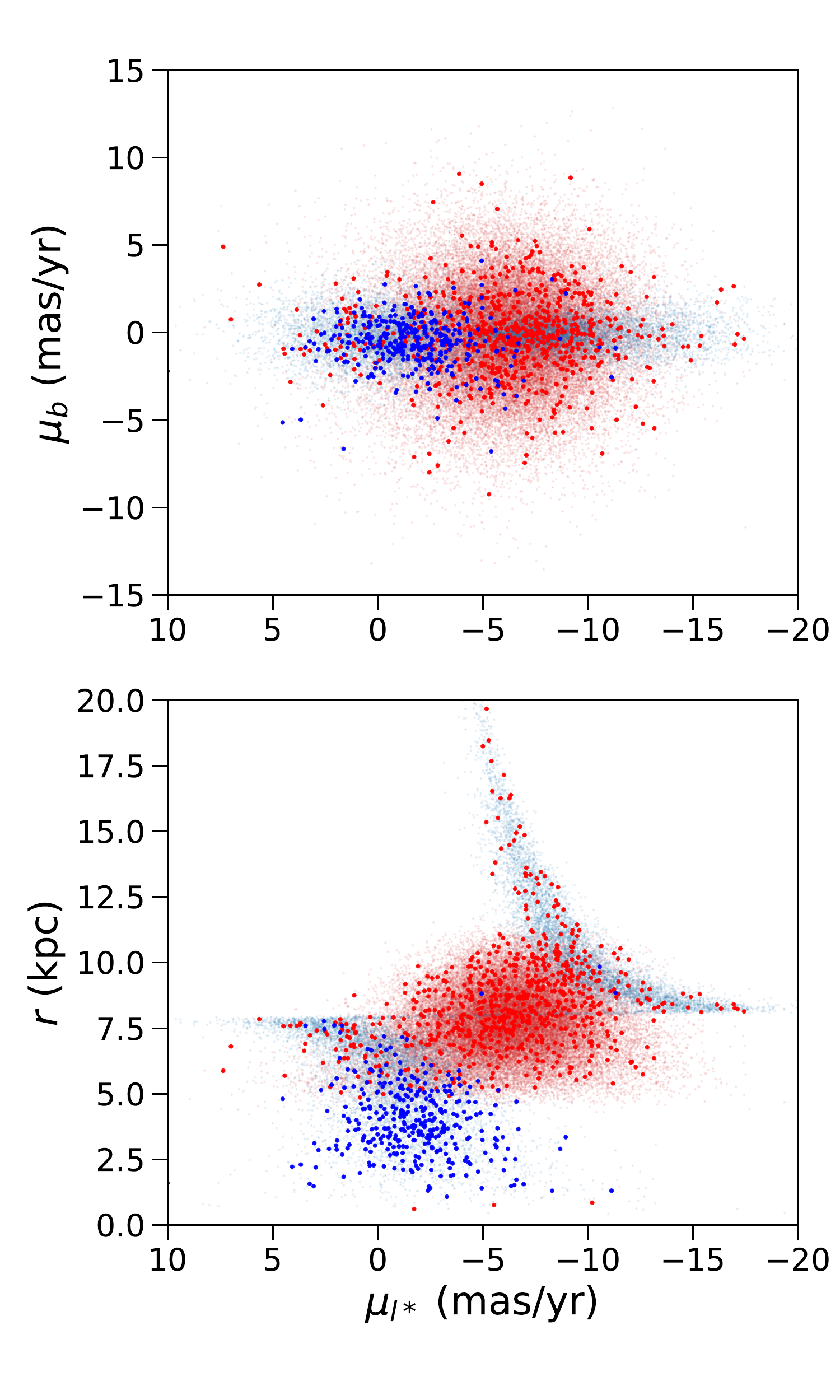}
    \caption{Large red points are the photometrically selected bulge stars (``red"/bulge proxy) while large blue points are the photometrically selected disk stars (``blue"/disk proxy).
    Small red points are the bulge stars and small blue points are the disk stars in \texttt{Galaxia}.
    \emph{Top}: Proper motion vector point diagram for the red and blue populations, c.f. \citetalias{Penny:2019} Figure 20.
    \emph{Bottom}: Distance vs. longitudinal proper motion diagram for the red and blue populations, c.f. \citetalias{Penny:2019} Figure 20.}
\end{figure}
\label{fig:SWEEPS_Penny}

\begin{figure}[h!]
    \centering
    \includegraphics[scale=0.4]{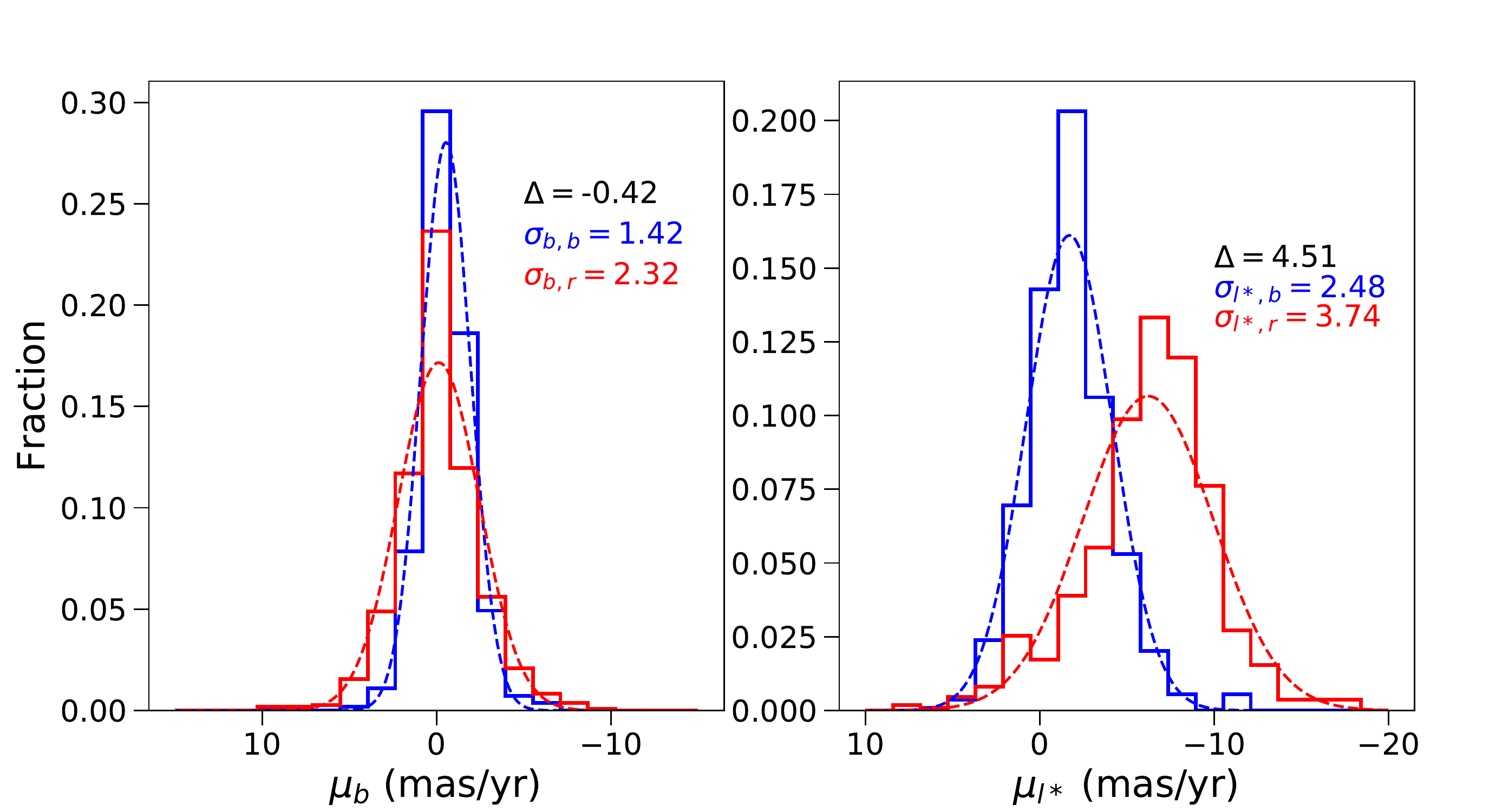}
    \caption{Histograms of proper motion for the red (bulge-proxy) and blue (disk-proxy) populations from Galaxia, with the dotted line being a Gaussian of the mean and standard deviation of each respective population.
    The $\Delta$ listed gives the differences between the means of the populations (blue $-$ red).
    Units of all inset numbers are mas/yr.}
\end{figure}
\label{fig:SWEEPS_Clarkson}

\begin{deluxetable}{lcccccc}
\tablecaption{\citet{Clarkson:2018}, BGM1106, \texttt{Galaxia} Proper Motion Comparison \label{tab:PM compare}}
\tablehead{
    \colhead{Model/Data} &
    \colhead{$\Delta \mu_{l*}$} &
    \colhead{$\Delta \mu_b$} & 
    \colhead{$\sigma_{l*}$, blue} & 
    \colhead{$\sigma_b$, blue} &
    \colhead{$\sigma_{l*}$, red} & 
    \colhead{$\sigma_b$, red} \\
}
\startdata
    \citet{Clarkson:2008} & 3.24 $\pm$ 0.15 & -0.81 $\pm$ 0.12 & 2.2 & 1.3 & 3.0 & 2.8 \\
    BGM1106 \citepalias{Penny:2019} & 3.53 $\pm$ 0.65 & -0.12 $\pm$ 0.32 & 2.47 $\pm$ 0.29 & 1.11 $\pm$ 0.13 & 5.19 $\pm$ 0.36 & 2.64 $\pm$ 0.18 \\ 
    \texttt{Galaxia} & 4.51 & -0.42 & 2.48 & 1.42 & 3.74 & 2.32 
\enddata
\tablecomments{All units are (mas/yr).
$\Delta$ is defined as blue $-$ red.}
\end{deluxetable} 

\section{Extinction}
\label{Appendix:Extinction}

The amount of extinction given in \texttt{Galaxia}, which uses the \citet{Schlegel:1998} dust map overestimates the extinction towards the Galactic Bulge. 
Thus, in \texttt{PopSyCLE} we have instead chosen to implement the reddening law of \citet{Damineli:16}, which is tailored for the direction toward the Galactic Plane.
We continue to use the color excess values $E(B-V)$ from \citet{Schlegel:1998}, which is not strictly correct; however, this is satisfactory, as we show below.

We compare the \texttt{Galaxia} model using the $E(B-V)$ from \citet{Schlegel:1998} with either the \citet{Schlegel:1998} or \citet{Damineli:16} reddening laws, to the data from the OGLE Early Warning System (EWS).
From the OGLE 2017 EWS, 4 different events were selected: OGLE-2017-BLG-0001 at (l,b) = (0.92, -1.63); OGLE-2017-BLG-0100 at (-2.07, 0.98); OGLE-2017-BLG-0150 at (1.74, -4.46); OGLE-2017-BLG-0921 at (-0.71, -1.79).
These 4 particular events were selected because their finding charts had different CMDs.
For these events, the $I$ and $V$ magnitudes of stars in a $2' \times 2'$ area centered on the event are provided.
With \texttt{Galaxia}, fields in those directions are generated with that equivalent area (0.0011 deg$^2$).
Since \texttt{Galaxia} returns the distance and absolute magnitude of the stars in various filters, along with and the Schlegel E(B-V) color excess, the apparent magnitude of these stars using either the Schlegel or Damineli reddening can be calculated and used to produce synthetic CMDs and luminosity and color functions.
The comparisons are plotted in Figure \ref{fig:compare_cmd}.

In general, \texttt{Galaxia} captures the various structures in the OGLE CMDs.
Although there is still some underestimation in the number of stars, the \citet{Damineli:16} reddening law does a significantly better job than \citet{Schlegel:1998} at capturing the total number of stars in the field.
The \citet{Damineli:16} law also does slightly better at replicating the CMD shape and luminosity and color functions.

\begin{figure}[h!]
    \centering
    \includegraphics[scale=0.18]{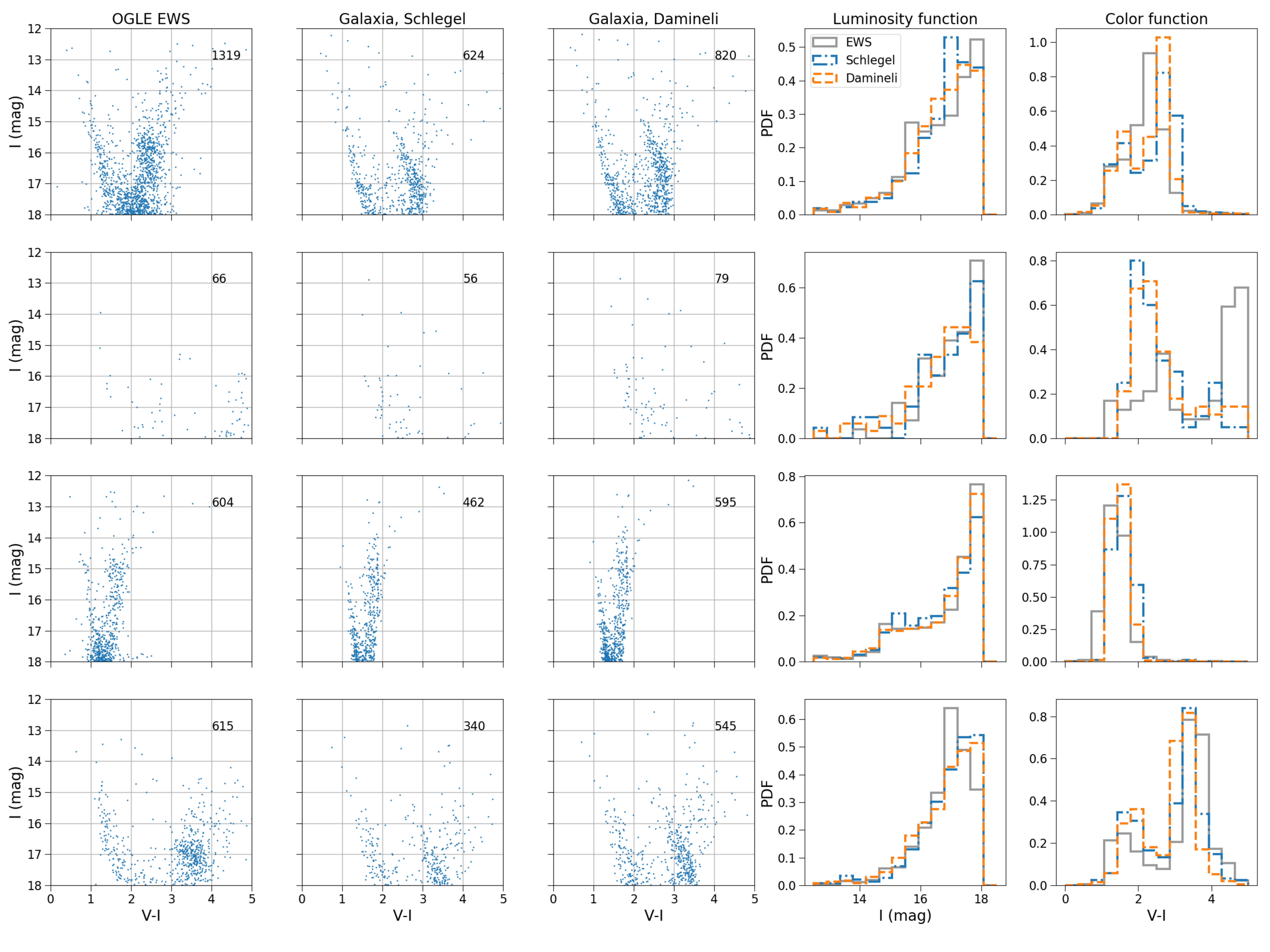} 
    \caption{Comparison of CMDs, luminosity, and color functions for different fields and extinction laws.
    The rows, from top to bottom, correspond to the fields OGLE-2017-BLG-0001 at (l,b) = (0.92, -1.63), OGLE-2017-BLG-0100 at (l,b) = (-2.07, 0.98), OGLE-2017-BLG-0150 at (l,b) = (1.74, -4.46), and OGLE-2017-BLG-0921 at (l,b) = (-0.71, -1.79).
    The numbers in the corners of the CMDs correpond to the number of stars in that CMD.}
 \end{figure}
 \label{fig:compare_cmd}
 
We also consider an extinction map produced by \citet{Nataf:2013} using data from the OGLE-III, VVV, and 2MASS surveys. 
However, for the fields available, the differences are random and small, hence they do not consistently skew our number counts in one direction.

\section{Initial-Final Mass Relation}
\label{Appendix:Initial-Final Mass Relation}

As defined in \citetalias{Raithel:2018} Equations (1) through (4), the black hole initial-final mass relation (IFMR) is a piecewise function, where the two pieces are dubbed Branches I and II. 
Branch I, which covers $15 \leq M_{ZAMS} \leq 40 M_\odot$, the remnant BH mass is given by
\begin{equation}
\label{eq:bhifmr}
    M_{BH}(M_{ZAMS}; f_{ej}) = f_{ej} M_{BH,core}(M_{ZAMS}) + (1 - f_{ej})M_{BH, all}(M_{ZAMS})
\end{equation}
where $f_{ej}$ is the ejection fraction describing how much of the envelope is ejected in a supernova explosion ($f_{ej} = 0$ is where the entire star collapses, $f_{ej} = 1$ is where only the star's He-core collapses), and $M_{BH, core}$ and $M_{BH, all}$ are defined as
\begin{align}
    M_{BH, core}(M_{ZAMS}) &= -2.049 + 0.4140 \; M_{ZAMS} \\
    M_{BH, all}(M_{ZAMS}) &= 15.52 - 0.3294(M_{ZAMS} - 25.97) - 0.02121(M_{ZAMS} - 25.97)^2 + 0.003120(M_{ZAMS} - 25.97)^3.
\end{align}
In \texttt{PopSyCLE} we use $f_{ej} = 0.9$ as this is the value reported by \citetalias{Raithel:2018} that most closely reproduces the observed distribution of BH masses.
The $M_{BH,core}$ term describes the BH IFMR where only the star's He-core collapses, while the $M_{BH,all}$ term describes the BH IFMR where the entire star collapses; $f_{ej}$ interpolates between the two.
The black hole IFMR for Branch II, which covers $45 \leq M_{ZAMS} \leq 120 M_\odot$, is
\begin{equation}
    M_{BH,core}(M_{ZAMS}) = 5.697 + 7.8598 \times 10^8 (M_{ZAMS})^{-4.858}.
\end{equation}
Similarly, there is a piecewise defined neutron star initial-final mass relation. 
Overall, there are seven branches, Branches I through VII, described in \citetalias{Raithel:2018} Equations (11) through (16). 
Each branch of the IFMR is defined over a particular range of ZAMS masses.
Five of these branches are described by third order polynomials, and two are described by Gaussian distributions.

We have made the following modifications to the original initial-final mass function/relation:
\begin{enumerate}
    \item Since the gap between $40 \leq M \leq 45 M_\odot$ between Branches I and II is due to the discrete sampling of the simulations from \cite{Sukhbold:2016}, we extend Branches I and II such that there isn't a gap.
    Specifically, we find where the function describing Branch I intersects Branch II, assuming $f_{ej} = 0.9$; this point is $M_{ZAMS} = 42.21 M_\odot$. 
    So for our purposes, Branch I goes from $15 \leq M_{ZAMS} \leq 42.21 M_\odot$ and Branch II goes from $42.21 \leq M_{ZAMS} \leq 120 M_\odot$.
    
    \item A distinction is made between black holes made by fallback, and those made by direct collapse. 
    As described by Equation (8), black holes only form from direct collapse immediately after the SNe. 
    However, for our purposes, a black hole formed by fallback vs. direct collapse is not relevant. 
    We modify the fraction of black holes to then be
    \begin{equation}
        X_{BH} = \frac{N_{BH} + N_{fb}}{N_{BH} + N_{NS} + N_{fb}} = 1 - X_{NS}
    \end{equation}
    We do this for completeness; however $N_{fb}$ is quite small compared to $N_{BH}$ and this will not substantially change the results.
    
    \item Since all the neutron stars fall in such a small mass range (between 1.3 and 1.9 $M_\odot$), for simplicity, we just assume the neutron star initial-final mass function is a constant, that is,
    \begin{equation}
        M_{NS}(M_{ZAMS}) = 1.6 M_\odot
    \end{equation}
    where $1.6 M_\odot$ was selected since is the mean of this mass range. 
    In future versions, we plan to release a more realistic NS IFMR.
\end{enumerate}

\section{Initial-Final Group Mass Ratio}
\label{Appendix:Initial-Final Group Mass Ratio}

\texttt{Galaxia} simulation output is divided into groups of stars with a similar age in order to determine the appropriate number, types, and masses of compact objects for that group. 
Each group is treated as a simple stellar population of a fixed age and solar metallicity. 
Currently, differences in metallicity within each group are ignored since the \citetalias{Raithel:2018} IFMR only has solar metallicity. 
The total stellar mass of the age group from \texttt{Galaxia} is the present-day group mass; however, the initial group mass is needed in order to determine how many black holes and other compact objects should be added to \texttt{Galaxia}.
As each group ages, the present-day mass decreases in a non-linear fashion and the conversion from present-day to the initial group mass must be derived through simulations. 
Ultimately, during \texttt{PopSyCLE} simulations, this relation between the initial-final group mass ratio and age is used to estimate the total initial group mass, generate a \texttt{PyPopStar} cluster of that mass and age, and insert the resulting white dwarfs, neutron stars, and compact objects back into \texttt{PopSyCLE}.

The relationship between the initial-final group mass ratio and age is calibrated using \texttt{PyPopStar} by simulating groups of stars of mass $10^7 M_\odot$ over a wide range of ages (Figure \ref{fig:ifgmr}).
The group mass is chosen to be large, to ensure that stochasticity does not dominate the result of the initial-final group mass ratio. 
For all age groups, we adopt the same Kroupa IMF and MIST evolutionary models described in Section \ref{sec:Population synthesis}.
The process is as follows:

A $10^7 M_\odot$ group of stars is generated using \texttt{PyPopStar} and evolved to the desired age.
As the MIST isochrones include some white dwarfs, while \texttt{Galaxia} does not include any white dwarfs, all stars beyond the main sequence (post-AGB and Wolf-Rayet stars) are discarded from the group.
The mass of the remaining stars constitute that age group's current stellar mass.
The current age group's stellar mass is then divided by the initial group's stellar mass ($10^7 M_\odot$) to obtain the initial-final group mass ratio.

Implicit in this process, it is assumed that the IMF shape and mass limits used to calculate the initial-current group mass ratio with \texttt{PyPopStar} are the same as the IMF used in \texttt{Galaxia}.

\begin{figure}[h!]
    \centering
    \includegraphics[scale=0.4]{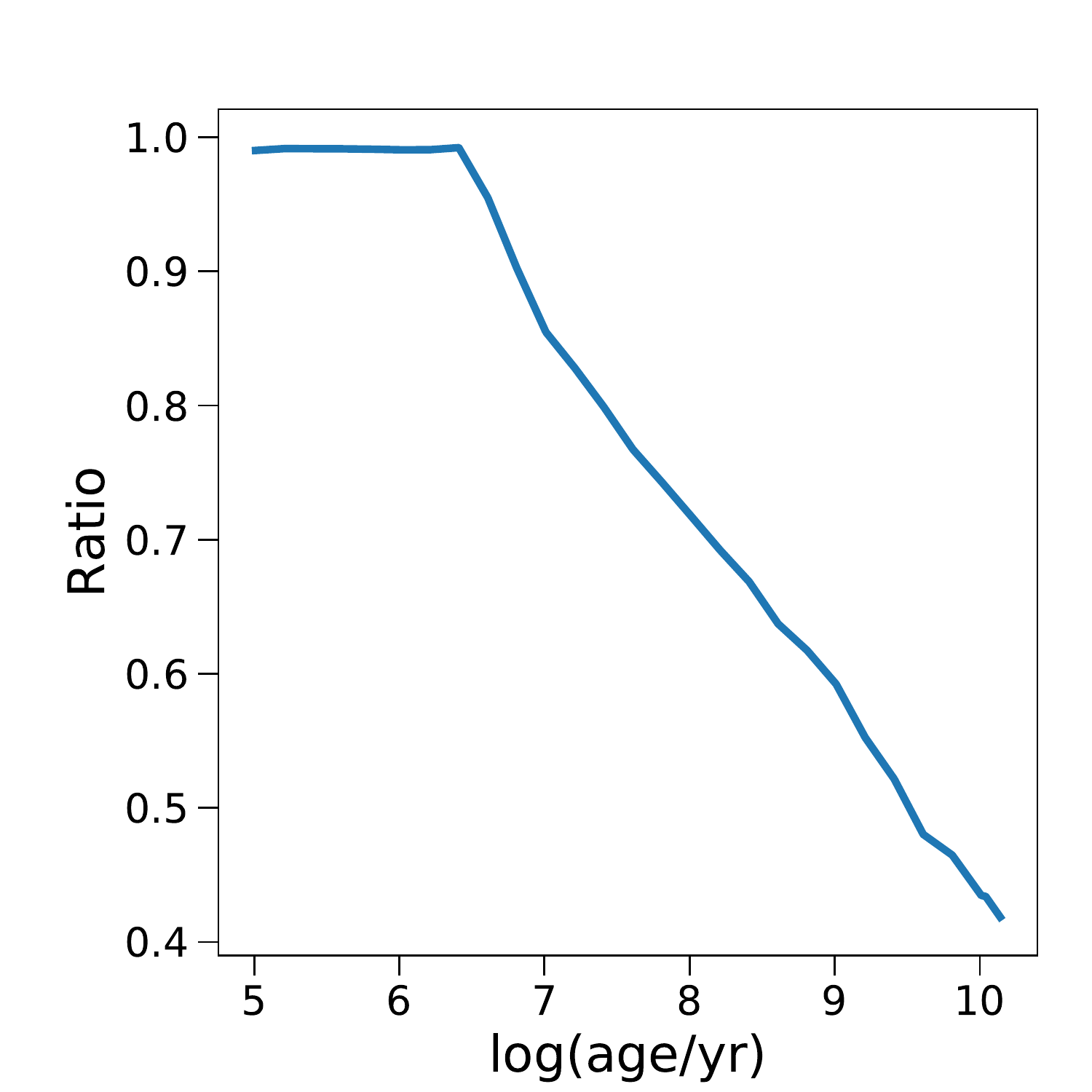}
    \caption{Ratio of current group mass to initial group mass as a function of group age.
    Most stars do not begin to evolve off the main sequence until after $10^6$ years.
    After that, stellar mass is converted to remnant mass in a roughly linear fashion as a function of log age.}
\end{figure}
\label{fig:ifgmr}

\bibliographystyle{aasjournal}
\bibliography{main.bib}


\end{document}